\documentstyle[preprint,eqsecnum,epsf,aps,tighten,floats]{revtex}%
\begin{document}
\preprint{ 
 \parbox{1.5in}{
                \leftline{WM-97-105} 
                \leftline{CFNUL-97-01}
                \leftline{JLAB-THY-97-09} }  
          }
%
\draft
\begin{title}
{\bf Covariant equations for the three-body bound state}
\end{title}
%
\author{Alfred Stadler$^{1,2}$, Franz Gross$^{1,3}$, and Michael 
Frank$^4$}
\address{$^1$College of William and Mary, Williamsburg, Virginia 23185}
\address{$^2$Centro de F\'\i sica Nuclear da Universidade de Lisboa, 1699
Lisboa Codex, Portugal$^*$}
\address{$^3$Thomas Jefferson National Accelerator Facility, Newport News, VA
23606}
\address{$^4$Institute for Nuclear Theory, University of Washington, 
Seattle WA 98195}
\maketitle
\begin{abstract}

The covariant spectator (or Gross) equations for the bound state
of three identical spin 1/2 particles, in which two of the three
interacting particles are always on shell, are developed and reduced to a
form suitable for numerical solution.  The equations are first written in
operator form and compared to the Bethe-Salpeter equation, then
expanded into plane wave momentum states, and finally expanded into
partial waves using the three-body helicity formalism first introduced by
Wick.  In order to solve the equations, the two-body scattering amplitudes
must be boosted from the overall three-body rest frame to their individual
two-body rest frames, and all effects which arise from these boosts,
including the Wigner rotations and
$\rho$-spin decomposition of the off-shell particle, are treated {\it
exactly\/}.  In their final form, the equations reduce to a coupled set 
of Faddeev-like double integral equations with additional channels
arising from the negative $\rho$-spin states of the off-shell particle.

\end{abstract}
\pacs{21.45+v, 11.10.St, 21.10.Dr}
\tableofcontents

\section{Introduction and Overview}

The three-body spectator (or Gross) equations were first introduced  and
applied to scalar particles in 1982\cite{R4}.  This original paper
included a treatment of non-identical particles and an introductory discussion
of the definition and role of three-body forces in a relativistic context.
Shortly afterward, in lectures  given at the University of Hannover\cite{R5},
the equations for three identical spin 1/2 particles were written down, but
many details needed for a practical solution of the equations were never
worked out.  In this paper we complete the development by expanding the
amplitudes into partial waves and reducing the equations to a compact form
suitable for numerical solution.  The development is carried out only
for the case when the three-body scattering amplitude can be obtained by
iterating successive two-body interactions, so that the three-body forces
of relativistic origin discussed in the original paper\cite{R4} are
neglected.  However, because our covariant equations include the
negative energy part of the Dirac propagator of the off-shell
nucleon, many contributions are automatically included which would arise from
three-body forces in a nonrelativistic context.

The bound state equations we present in this paper have already been solved
numerically for a variety of cases, and some results have already
been published\cite{conf,SG}.  From this
experience we know that the general development presented here is a suitable
basis for a practical solution of the covariant three-body problem.

In the remainder of this section we give a brief summary of the
current status of nonrelativistic calculations of the binding energy of the
three nucleon bound state, and a review of previous work on the
relativistic three-body problem.  Then we will give a brief summary of the
physics underlying our spectator equations, and present the final equations.
The derivation of these results is found in the subsequent sections.  In
Sec.~II we begin the development by writing the three-body equations in an
operator form which is independent of the basis states used to describe the
three-body system.  In Sec.~III we introduce basis states and
write the equations in momentum space.  In this representation the physical
content of the equations is clear, but the equations are not in a form
most convenient for numerical solution.  To solve the equations numerically it
is convenient to use a partial wave decomposition based on the helicity states
originally introduced (in a three-body context) by Wick \cite{R2} and this is
developed in detail in Sec.~IV.  The evaluation of the permutation operator,
which interchanges particles between interactions and permits us to express
the equation in terms of only one amplitude, is discussed in detail in
Sec.~V, and all of the results are collected together and the final equations
given in Sec.~VI.  There are three appendices which discuss some points in
detail.

\subsection{Brief history of the three-body bound state problem}

The first realistic nonrelativistic calculations of the triton binding energy
were
completed in the 1970's\cite{T1}.  Later it was
shown that different methods arrived at the same results, and that the
binding energy could be calculated to a numerical accuracy of a few keV by
considering all nucleon-nucleon ($NN$) partial waves up to $j=4$\cite{R1a}.
Today, if three-body
forces (3BFs) are not considered, a discrepancy of about 0.5-1.0 MeV
remains between the experimentally observed value of $-$8.48 MeV and values
obtained from realistic nonrelativistic $NN$ potentials.
Calculations of the contribution of the $\Delta$ resonance to the
3BF find that the net effect of the $\Delta$ is small\cite{Rxx,Rb}.
State-of-the art calculations often include in addition also 3BFs based on
meson-nucleon interaction processes other than $\Delta$
excitation \cite{Rxy}. When the strength of phenomenological 3BFs is
adjusted to give the correct triton binding energy, an excellent value is
also obtained for the
$^4$He binding energy (and to a lesser extent other light nuclei up to
$A\simeq7$)\cite{R2a}.

However, relativistic effects should  make a contribution to the binding energy
at the level of several hundred keV.  Using a mean momentum of about 200 MeV
(consistent with nonrelativistic estimates) we expect to see corrections of the
order of $(v/c)^2\simeq (p/m)^2 \simeq 4\%$.  If this is 4\% of the binding
energy, then it amounts to about 300 keV.  However, if relativity has a greater
effect on the attractive $\sigma$ exchange
part of the force (as it does in nuclear matter calculations using the
Walecka model\cite{walecka}) then we might obtain an effect 10 times larger.

Interest in relativistic three-body equations goes back to 1965, when
Alessandrini and Omnes\cite{Ra} used the Blankenbecler-Sugar
equation\cite{BBS} to describe the scattering of three particles, and
Basdevant and Kreps\cite{Rh} applied their ideas to a description of the
three pion system.  Taylor\cite{Rg} discussed the application of the
Bethe-Salpeter equation\cite{BS} to three-body systems in 1966. In 1968 Aaron,
Amado, and Young\cite{Rc} introduced three-body scattering equations in which
all the particles were on shell.  Later, Garcilazo and his
collaborators\cite{Rd} treated three-body bound states using the
Blankenbecler-Sugar equation, and Garcilazo\cite{Re} applied Wick's helicity
formalism to the three-body problem, and used it to treat the
$\pi NN$ system relativistically\cite{Rf}.  Recently, the size of relativistic
effects were estimated by Rupp and Tjon\cite{RT}   using a separable kernel in
the Bethe-Salpeter equation, by Sammarruca, Xu, and
Machleidt\cite{MSS} using minimal relativity and the Blankenbecler-Sugar
equation, and by the Urbana group\cite{Urbana} using the Schr\"odinger
equation with corrections of first order in $(v/c)^2$.  All of these
calculations include some contributions coming from relativistic kinematics,
but none treats the full Dirac structure of the nucleons, or investigates
effects which might arise from a realistic relativistic treatment of the $NN$
{\it dynamics\/}.

\subsection{The physics behind the spectator equations}

%
%
\begin{figure}[t]
\begin{center}
\mbox{
   \epsfysize=1.5in
\epsfbox{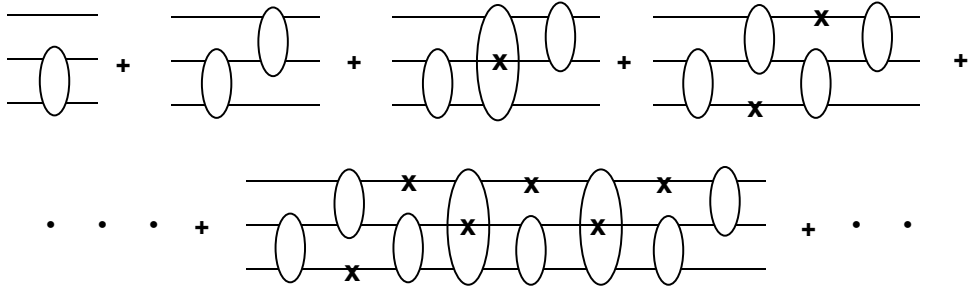}
}
\end{center}
\caption{Diagrams from the infinite class of successive two-body
scatterings (represented by the ovals) which contribute to scattering
of the three-body system.  We use the convention
that the initial state is on the right and the final state on the left in 
each diagram. The $\times$ labels internal spectators,
which are put on-shell in the spectator formalism. 
In this example, all of the diagrams but the first contribute to the
subamplitude  
${\cal T}^{13}$ where particle 1 (on the top) is the last spectator and
particle 3 (on the bottom) is the first spectator.  The first diagram
contributes to the subamplitude ${\cal T}^{11}$.}
\label{x1}
\end{figure}

In the absence of three-body forces, the three-body scattering amplitude
(and the three-body bound state vertex functions) can be obtained by summing
all successive two-body scatterings, as shown diagrammatically in Fig.~1.
This summation can be organized into Faddeev-like equations, shown
diagrammatically in Fig.~2.  When the three particles are identical, the
different Faddeev  subamplitudes can be obtained from each other by
interchange of variables, leading to a single equation for a single
subamplitude represented diagrammatically in Fig.~3.
It is necessary to know the two-body scattering amplitude before the
equation shown in Fig.~3 can be solved.  More specifically, the two-body
scattering amplitude must be known in the rest frame of the {\it three-body
system\/} (or any other frame independent of the internal variables).

%
%
\begin{figure}[t]
\begin{center}
\mbox{
   \epsfysize=2.5in
\epsfbox{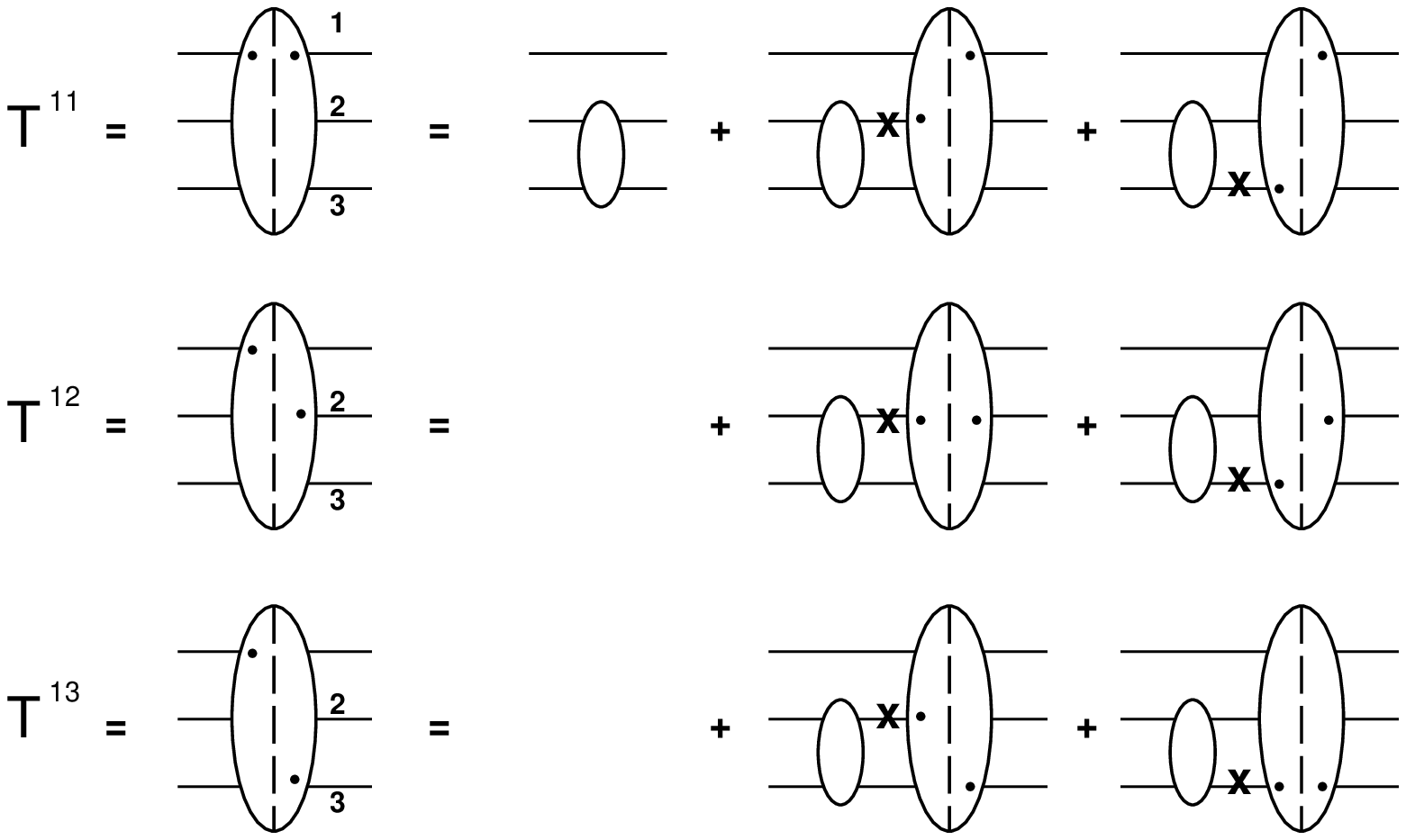} 
}
\end{center}
\caption{Diagrammatic representation of the Faddeev equations for the 
amplitudes ${\cal T}^{1i}$.  Note that the spectator is identified by
the solid dot.}
\label{x2}
\end{figure}

%
%
\begin{figure}[b]
\begin{center}
\vspace*{-0.6in}
\mbox{
   \epsfysize=1.5in
\epsfbox{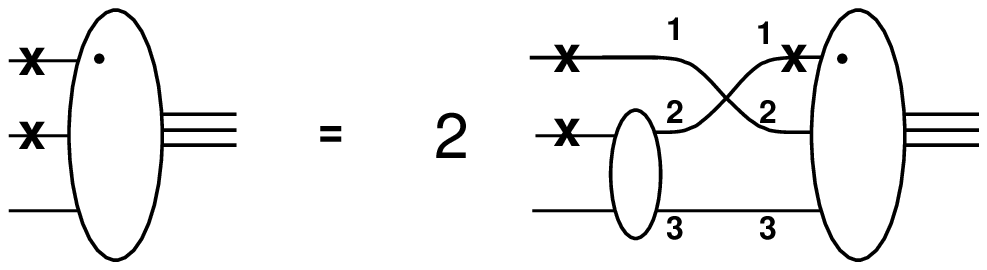} 
}
\end{center}
\caption{Diagrammatic representation of the bound state spectator equation
for three identical particles.  Spectators are identified by
the solid dots, and on-shell particles by the $\times$.  Note the
interchange of particles 1 and 2.}
\label{x3}
\end{figure}

The two-body amplitude is usually calculated in its own rest frame, so it
must be {\it boosted\/} to the three-body rest frame before it can be used
in the Faddeev equations.  The velocity of this boost depends on the
momentum of the spectator, which is one of the dynamical variables of the
problem, and hence the boost must be known for all velocities.  In the
nonrelativistic case this is trivial because the two-body amplitude is
invariant under Galilean boosts.  However, in the relativistic case this may
present a problem, depending on the type of formalism used.  Here, for the
purposes of discussion, we distinguish two fundamentally different ways to
approach relativistic calculations.  In one approach, which will be referred
to as Hamiltonian Dynamics (including light-cone methods)\cite{HD}, some of the
Poincar\'e generators include the interactions, and either boosts or rotations
cannot be carried out exactly.  In this method one must treat relativistic
effects approximately.  In a second method, which we will refer to as Manifestly
Covariant Dynamics\cite{MCD}, the generators are all kinematic, and the boosts
can be done {\it exactly\/}.  The spectator equations developed in this paper
are an example of the latter method; we will reduce the three-body equations
to a practical form by {\it exploiting our ability to boost the two-body
amplitudes to their rest frame\/}.

Of the methods discussed in the previous subsection, only the Bethe-Salpeter
(BS) formalism shares the property that the two-body amplitudes can be boosted
exactly to their rest frame, and we will therefore compare the spectator
equations with the corresponding BS equations.  Both approaches conserve total
four-momentum.  This leaves an integration over all independent internal
four-momenta, which are two for the three-body problem.  The three particle
Bethe-Salpeter equation does not restrict any of these eight independent
components, and after a partial wave decomposition there still remain four
integrations, leading to coupled {\it four\/}-dimensional Faddeev equations.
Furthermore, these equations contain singularities arising from the indefinite
nature of the Minkowsky metric.  In the spectator formalism the two time
components of the internal four-momenta are eliminated (or, more precisely,
expressed in terms of the other variables) by requiring that two of the three
particles be fixed to their positive energy mass shell.  This reduces the
number of independent variables to only {\it six\/}, and after a partial wave
decomposition one obtains coupled {\it two\/}-dimensional equations with a
Faddeev structure.  The three-body spectator equations therefore have the same
structure as nonrelativistic equations, and this is one of their most
significant advantages.

A particle is put on-shell when it is a {\it spectator\/} to the
interaction of two other particles.  When this is done systematically, 
{\it two of the three\/} particles are always on-shell.  The particle
which  is off shell is the (unique) particle which has just interacted
and is about to interact again (in a topological, not time-ordered,
sense), as illustrated in Fig.~1.

It is natural to assume that restricting particles
to their mass-shell represents an approximation to the BS equation, but it
can be shown that it is equivalent to a reorganization of the perturbation
series of all ladder and crossed ladder diagrams which, in some cases, sums
these diagrams more efficiently\cite{R4,Book}.

In summary, the spectator equation is used because:
\vspace*{0.2in}

\hang (i) it sums the infinite series of all ladder and crossed ladder
interactions efficiently,

\vspace*{0.2in}
\hang (ii) it  reduces the number of independent variables to a
minimum, making the covariant three-body problem tractable,
and

\vspace*{0.2in}
\hang (iii) it permits us to boost the two-body amplitudes to their rest
frame and calculate relativistic effects exactly.

\vspace*{0.2in}

Before we turn to the details of the derivation of the spectator equations,
we present the equations in their final form in the next subsection.

\subsection{Spectator equations for three spin 1/2 particles}

In the absence of 3BFs the three-body scattering amplitude is obtained from a
sum of all successive two-body scatterings.  Because the three particles are
identical,
each two-body scattering differs from the others only by a permutation, and
they
can therefore all be summed by one operator equation of the form
\begin{equation}
| \Gamma^1 \rangle  =2 M^1 G^1  {\cal P}_{12}  | \Gamma^1 \rangle\, ,
\label{Eq1x}
\end{equation}
where $|\Gamma^1\rangle$ is a vertex function describing the contribution
to the bound state from all processes in which the 23 pair was the last to
interact (with particle 1 a spectator), the two-body amplitude $M^1$ describes
the scattering of the 23 pair,
$G^1$ is the propagator for the 23 pair, and ${\cal P}_{12}$ is a 
permutation operator interchanging particles 1 and 2.  These are labeled
in Fig.~3.  The factor of 2 comes from the contribution of ${\cal
P}_{13}$ which equals the one of ${\cal P}_{12}$.  The permutation 
operator rearranges the particles so that the same equation sums up the
scattering of {\it all\/} pairs: 12, 23, and 13.

The three-body spectator equations have the same structure as
(\ref{Eq1x}), but incorporate the additional feature that the spectator
is restricted to its positive energy mass-shell in all intermediate
states.  With the conventions implied above,
consistency also requires that particle 2 be on-shell, so that two
particles are always on-shell.  As already stated above, we think of
these constraints as a reorganization of Eq.~(\ref{Eq1x}) which will, in some
cases, improve its convergence.
The constraints are manifestly covariant, and lead to the following equation
\begin{equation}
| \Gamma^1_2 \rangle  =  2 M^1_{22} G^1_2 {\cal P}_{12}
| \Gamma^1_2 \rangle  \, , \label{Eq2x}
\end{equation}
where the lower index labels the second on-shell particle.  Hence
only particle 3, the (unique) particle which has just left one interaction and
is about to enter another one, is off-shell in Eq.~(\ref{Eq2x}).

%
%
\begin{figure}[t]
\begin{center}
\mbox{
   \epsfysize=2in
\epsfbox{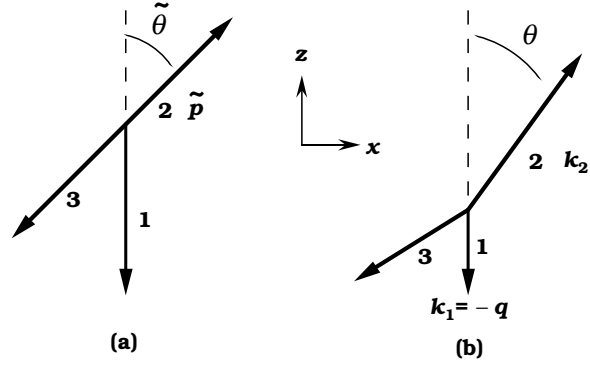}
}
\end{center}
\caption{Diagrams showing the momenta in the two-body rest frame (left
panel) and in the three-body rest frame (right panel).  We chose the
momenta of particle 1 to be in the $-\hat z$ direction, so the boost from
the two to three-body rest frames is $+\hat z$ direction.}
\label{x4}
\end{figure}

To prepare Eq.~(\ref{Eq2x}) numerical evaluation, we take matrix 
elements of the operators using three-particle states.  Both $\rho$-spin 
states (where $\rho=+$ is the $u$ spinor positive energy state and
$\rho=-$ is the $v$ spinor negative energy state) of the off-shell
particle must be treated.   First we  reduce the equation using states
with definite particle helicities, similar to those defined by
Wick\cite{R2}.  These three-body states will be written in the
abbreviated form $| J 1 (23)\rho \rangle$, where $J$ is the total angular
momentum of the state, $\rho$ the $\rho$-spin of the off-shell particle,
$1=\{q,\lambda_1\}$ (where $q$ and $\lambda_1$ are the magnitude of the
three-momentum and the helicity of the spectator in the three-body c.m.), and
$(23)=\{\tilde{p},j,m,\lambda_2,\lambda_3\}$ (where $\tilde{p}$ is the
magnitude of the relative three-momentum of the 23 system, $j$ and 
$m$ are the angular momentum of the pair and its projection in the
direction of ${\bf q}$, and $\lambda_2$ and $\lambda_3$ are the
helicities  of particles 2 and 3, {\it all defined in the rest frame of
the 23 pair}).  The momenta are defined in Fig.~4, which also shows
the relation between the rest frames of the two and three-body systems.
The three particles have mass $m$, and the total mass of the three-body
bound state is denoted by $M_t$. [We use the symbol $m$ to denote both the
projection of the momentum and the particle mass, but the difference between them
should be clear from the context.]  Using this notation, and suppressing isospin,
the final form of the three-body spectator equation  for
$\Gamma^1$ is given in  Eq.~(\ref{Eq6.1}). It can be written 
%
\begin{eqnarray}
\langle J 1 (23)\rho | \Gamma^1 \rangle =&&
\sum_{j'm'}\!
\sum_{{\lambda''_2 \lambda''_3\,\rho''}
\atop {\lambda'_1 \lambda'_2 \lambda'_3\,\rho'} }
\int_0^{q_{\hbox{{\tiny crit}}}} {q'^2 dq'}{m\over
E_{q'}}
\int_0^\pi d\chi \sin\chi \;  \nonumber\\
&&\quad\times
\langle j (23)\rho | M^1| j (2''3'')\rho''\rangle
\,{m\over E_{\tilde{p}''}}\, g^{\rho''}(q,{\tilde p}'')\nonumber\\
&&\quad\times{\cal P}_{12}^{\rho''\rho'}[1(2''3''),1'(2'3')]
\,{m \over E_{\tilde{p}'}}\,
\langle J' 1' (2'3')\rho' |\Gamma^1 \rangle \, ,  \label{Eq6.1x}
\end{eqnarray}
%
%
where $\langle j (23)\rho | M^1| j (2''3'')\rho''\rangle$ is the $j$th
two-body partial wave amplitude for the scattering of particles 2 and 3 in
their own rest frame (precisely the amplitude obtained from the two
body spectator theory as described in Ref.~\cite{R1}),
${\cal P}_{12}^{\rho''\rho'}[1(2''3''),1'(2'3')]$ is the matrix element
of the
permutation operator, given in Eq.~(\ref{Eqxx}) below, and $g^{\rho}(q,{\tilde
p})$ the propagator of the off-shell particle in different $\rho$-spin states
\begin{equation}
g^+(q,{\tilde p}) = {\displaystyle{1\over 2E_{{\tilde p}} -W_q}} \, ,
\qquad
g^-(q,{\tilde p}) = -{\displaystyle{1\over W_q }} \, '
\label{Eq6.2x}
\end{equation}
where $W_q$ is the mass of the 23 pair, and depends on $q$,
\begin{equation}
W_{q}^2=M_t^2+m^2-2M_tE_q\, , \label{wq}
\end{equation}
with $E_q=\sqrt{m^2+{\bf q}^2}$.
Note that Eq.~(\ref{Eq6.1x}) includes a sum
over intermediate helicities and angular momentum quantum numbers, and an
integration over the internal spectator momentum $q'$ and the angle
$\chi$ between the directions of ${\bf q}'$ and ${\bf q}$.  The momenta
$\tilde{p}'$ and $\tilde{p}''$ depend on $q$, $q'$, and $\chi$, as given in
Eq.~(\ref{Eq4.15}).

The integration over $q'$ has been limited to the finite
interval $[0,q_{\hbox{{\scriptsize crit}}}]$, where
$q_{\hbox{{\scriptsize crit}}}$ is the root of the equation
$W_{q_{\hbox{{\scriptsize crit}}}}=0$.  At this critical spectator momentum
(equal to $\simeq4m/3\simeq1200$ MeV), the two-body subsystem is
recoiling at the speed of light and the relativistic effects are enormous!
One consequence of this is that the solutions of the three-body equations go
smoothly to zero as $q\to q_{\hbox{{\scriptsize crit}}}$ (this is discussed in 
detail in Sec.~VI).  Contributions from  $q'>q_{\hbox{{\scriptsize
crit}}}$, which come from two-body states with {\it spacelike\/} four-momenta,
are suppressed both because of this zero and because the propagators for large
$q$ are small.  Hence, even if the spacelike two-body scattering amplitude is
not small, we expect spacelike contributions to the overall three-body
amplitudes to be very much suppressed, and it  seems sensible to simply neglect
the region $q'\ge q_{\hbox{{\scriptsize crit}}}$ and set the three-body
amplitudes to zero there. This also removes the need to calculate two-body
amplitudes for spacelike total four-momenta.

Exchanging particles 1 and 2 implies that particle 2 becomes the
spectator and now its momentum and helicity must be expressed in
the c.m.\ frame of the three-body system, while the variables of
particles 1 and 3 must be expressed in the rest frame of the 13 pair.
Boosting from one frame to another introduces Wigner rotations of both 
the single particle and two-body helicities.  In the helicity basis,
this exchange operator is
%
\begin{eqnarray}
{\cal P}_{12}^{\rho''\rho'}[1(2''3''),1'(2'3')]=&&
(-1)^{m-\lambda_1+ \lambda'_3}
\sqrt{2j+1}\sqrt{2j'+1} \nonumber\\
&&\times d^{(J)}_{m-\lambda_1,m'-\lambda'_2}(\chi)
d^{(j)}_{m,\lambda''_2-\lambda''_3}(\tilde{\theta}'')
\,d^{(j')}_{m',\lambda'_1-\lambda'_3}(\tilde{\theta}')
\nonumber\\
&&\times
\,d^{(1/2)}_{\lambda_1\lambda'_1}(\beta_1)\,
d^{(1/2)}_{\lambda''_2\lambda'_2}(-\beta_2)\,
{\cal N}^{\rho''\rho'}_{\lambda''_3 \lambda'_3}
(q,q',\chi)\, , \label{Eqxx}
\end{eqnarray}
where the functions $d^{(1/2)}_{m_1,m_2}(\beta)$ are the Wigner rotation
matrices, and
${\cal N}^{\rho''\rho'}_{\lambda''_3\lambda'_3}(q,q',\chi)$ describes
{\em exactly}
the Wigner rotations of the off-shell particle 3, as well as the
nontrivial
matrix elements between the different $\rho$-spinors
$u$ and $v$ of particle 3 as they appear in the rest frames of the 23 pair and
the 13 pair.  The matrix ${\cal N}$ is defined in Eq.~(\ref{Eq5.40}), the
angles $\tilde{\theta}'$ and $\tilde{\theta}''$ in Eq.~(\ref{Eq4.15}), and
the Wigner rotation angles $\beta_1$ and $\beta_2$ in Eqs.~(\ref{Eq5.8}) and
(\ref{Eq5.10}).

For practical calculations it is more convenient to express
Eq.~(\ref{Eq6.1x}) in terms of states with definite isospin and
parity. These states will be denoted $| T j^{\;r} (m
\lambda) \rho \rangle$, where we suppress reference to the total
angular momentum and parity $J^\Pi=1/2^+$,
$T=0$ or 1 and $r=\pm1$ are the isospin and parity of the 23 pair, and
$\lambda=\lambda_2-\lambda_3 = 1/2-\lambda_3$.  As discussed in
Sec.~VID, the states of good parity are superpositions of positive and
negative helicity states [see Eq.~(\ref{eqpar2})], so that the two-body
subspace is fully described by adopting the convention
$\lambda_2=+1/2$, and identifying the states by their parity $r$ and
helicity difference $\lambda=0$ or 1.  In this basis
Eq.~(\ref{Eq2x}) becomes   
\begin{eqnarray}
\langle T j^{\;r} (m \lambda) \rho |\Gamma^1_T\rangle =&&
\sum_{{j'r'}\atop{m'T'}}
\sum_{{\lambda''\,\rho''}
\atop {\lambda'\,\rho'}}
\int_0^{q_{\hbox{{\tiny crit}}}} {q'^2 dq'}{m\over
E_{q'}}
\int_0^\pi d\chi \sin\chi \;
 \nonumber\\
&&\quad\times \langle T j^{\;r} (m
\lambda) \rho | M^{1\,T} |T j^{\;r} (m \lambda'') \rho''\rangle 
\,{m\over E_{\tilde{p}''}}\, g^{\rho''}(q,{\tilde p}'')\nonumber\\
&&\quad\times \overline{{\cal P}}_{12}^{\rho''\rho'}
[T j^{\;r} (m \lambda'') \rho'' ,T' j'^{\;r'} 
(m' \lambda') \rho' ] \,{m \over E_{\tilde{p}'}}\,
\langle   T' j'^{\;r'} (m'\lambda') \rho' | \Gamma^1_T \rangle
\, , \label{Ifinalx}
\end{eqnarray}
where the permutation operator $\overline{{\cal P}}_{12}^{\rho''\rho'}$
is given in Eqs.~(\ref{I6}) and (\ref{EqZZ8}).  Note that
Eq.~(\ref{Ifinalx}) includes a sum over the intermediate isospin $T'$. 

This concludes our brief introduction; we now turn to a detailed
derivation of the three-body equations (\ref{Eq6.1x}) and
(\ref{Ifinalx}) given above.

\section{Three-body equations in operator form}

We start with a derivation of Faddeev-type Bethe-Salpeter equations and
introduce the spectator equations afterwards by substituting a new propagator
and repeating the derivation with all necessary modifications.

\subsection{Bethe-Salpeter Equations}

The total scattering amplitude for the three-nucleon system
$\cal T$ can be decomposed into three parts
${\cal T}^i$,
\begin{equation}
{\cal T} = \sum_{i=1}^3 {\cal T}^i \, .
\end{equation}
The partial amplitude ${\cal T}^i$ sums up all diagrams in which 
particle $i$ is the spectator during the ``last'' interaction (in the
sense of  ``leftmost'' in the  diagrams of Fig.~1). Each amplitude
${\cal T}^i$ is further split into sub-amplitudes ${\cal T}^{ij}$, this
time according to which particle does not participate in the ``first''
(or ``rightmost'') two-body interaction,
\begin{equation}
{\cal T}^i = \sum_{j=1}^3 {\cal T}^{ij} \, .
\end{equation}
The amplitudes ${\cal T}^{ij}$ satisfy the integral equation
\begin{equation}
{\cal T}^{ij} = i\,\delta_{ij} {\cal M}^i G_i^{-1} - {\cal M}^i G_{BS}^i
  \sum_{k \not= i}  {\cal T}^{kj} \, , \label{EqMij}
\end{equation}
where $G_i$ is the propagator of a {\it single\/} off-shell particle $i$,
$G_{BS}^i=G_{BS}^i\otimes 1_i=-i\,G_j\otimes G_k\otimes 1_i$ is
the free two-body propagator for the $\{j,k\}$ pair, and
${\cal M}^i={\cal M}^i\otimes 1_i$ is the two-body scattering operator
acting in the two-body subspace of particles $j$ and $k$, with $1_i$ the
identity  operator for the spectator particle $i$.  In our notation $G_i$
is real, and any  overall factor of $i$ which emerges when the operator
expressions are represented by Feynman diagrams is included in the
propagator $G^i_{BS}$.  If
$V^i=V^i\otimes 1_i$ represents the sum of all irreducible diagrams
describing the interaction of the two particles $j$ and $k$ with
particle $i$ a spectator, the Bethe-Salpeter equation
\begin{equation}
{\cal M}^i = V^i - V^i  G_{BS}^i {\cal M}^i
\end{equation}
yields the scattering operator ${\cal M}^i$.

A bound state of the three-body system can be defined as the
residue of a pole of the three-body scattering amplitude ${\cal T}$. For
the triton we denote the position of the pole as
$P^2 = M_t^2$, where $P=k_1+k_2+k_3$ is the total four-momentum of the
system and the $k_i$ are single-particle four-momenta.
One can write ${\cal T}^{ij}$ as the sum of a pole term and a part
${\cal R}^{ij}$ regular at $P^2 = M_t^2$:
\begin{equation}
{\cal T}^{ij} = -\frac{| \Gamma^i \rangle \langle \Gamma^j |}
                     {M_t^2-P^2} + {\cal R}^{ij}  \, ,
\end{equation}
where $|\Gamma^i\rangle$ are the partial vertex amplitudes for the bound state.
Insertion into Eq.~(\ref{EqMij}), multiplication by $(M_t^2-P^2)$, and
performing the limit $P^2 \rightarrow M_t^2$ yields
\begin{equation}
| \Gamma^i \rangle = - {\cal M}^i G_{BS}^i \sum_{j \not= i} | \Gamma^j
\rangle \, .
\label{EqBS1}
\end{equation}
These are the Bethe-Salpeter equations for the partial bound state
vertex amplitudes.

Up to this point the equations are very general and apply to systems of
any three distinguishable particles. Now we want to specialize to the
case of three identical particles. We define the transpositions
${\cal P}_{ij}$  of two particles
$i$ and $j$ as follows
\begin{eqnarray}
{\cal P}_{12} | abc \rangle &=&  |bac\rangle \nonumber\\
{\cal P}_{13}|abc \rangle &=&  |cba\rangle \, , \label{EqP123}
\end{eqnarray}
Note that ${\cal
P}_{ij}$ interchanges
the quantum numbers of the particles in the $i$th and $j$th locations in
the state ket.
The symmetry of the scattering
amplitude under particle interchange can be expressed as
\begin{eqnarray}
{\cal P}_{ij} {T}  &&= \zeta {T}  \nonumber\\
{T} {\cal P}_{ij}  &&= \zeta {T} \, , \label{EqSymM}
\end{eqnarray}
where $\zeta=+1$ for bosons and $-1$ for fermions, and $T$ is the
symmetrized version of ${\cal T}$.  If we introduce the combined amplitude
\begin{equation}
| \Gamma \rangle = \sum_{i=1}^3 | \Gamma^i \rangle \, , \label{Combamp}
\end{equation}
then the symmetry  (\ref{EqSymM}) of ${T}$ carries over to $ | 
\Gamma\rangle $, i.e.,
\begin{eqnarray}
{\cal P}_{ij}  | \Gamma \rangle  &&= \zeta | \Gamma \rangle \nonumber\\
\langle   \Gamma | {\cal P}_{ij} &&= \zeta \langle  \Gamma | \, .
\end{eqnarray}

These relations can be used to derive the permutation properties
of the individual vertex factors $| \Gamma^i \rangle $.  If the particles 
are identical, then the two-body scattering operators and propagators
acting in each two-body
subspace are identical, and this is expressed formally by the relations
\begin{eqnarray}
 {\cal P}_{ij} M^i {\cal P}_{ij} & = & M^j \,  \nonumber\\
 {\cal P}_{ij} G_{BS}^i {\cal P}_{ij} & = & G_{BS}^j \, ,
\end{eqnarray}
where $M$ is a symmetrized version of ${\cal M}$.
Using these, and the fact that ${\cal P}_{ij}^2=1$, we obtain
\begin{eqnarray}
{\cal P}_{ij} | \Gamma^i \rangle & = & - {\cal P}_{ij} M^i
  {\cal P}_{ij} {\cal P}_{ij} G_{BS}^i \, \sum_{k \not= i}| \Gamma^k
\rangle    \nonumber \\
 & = & - M^j G_{BS}^j {\cal P}_{ij} \left(
 | \Gamma \rangle - | \Gamma^i \rangle \right) \nonumber \\
 & = & - M^j G_{BS}^j \left(
 \zeta |\Gamma \rangle -  {\cal P}_{ij} | \Gamma^i \rangle \right) \, ,
\end{eqnarray}
Comparing with
\begin{equation}
\zeta| \Gamma^j \rangle  =  - M^j G_{BS}^j \left(
 \zeta | \Gamma \rangle - \zeta| \Gamma^j \rangle \right)
\end{equation}
one obtains immediately
\begin{equation}
{\cal P}_{ij} | \Gamma^i \rangle  =\zeta | \Gamma^j \rangle \, . \label{EqPerm1}
\end{equation}
Thus the three-body equations for identical particles can be written
\begin{equation}
| \Gamma^i \rangle = -\zeta M^i G_{BS}^i \left( {\cal P}_{ij} + {\cal P}_{ik}
 \right) | \Gamma^i \rangle \, . \label{EqFad1}
\end{equation}
The three equations for the three possible choices of $i$ are
equivalent. It is therefore sufficient to solve Eq.~(\ref{EqFad1})
for, say, $i=1$, and calculate $| \Gamma^2 \rangle $ and
$ | \Gamma^3 \rangle $ by means of Eq.~(\ref{EqPerm1}).

Eq.~(\ref{EqFad1}) can be simplified further if we take into
account the fact that the two-body amplitude $M^1$ is symmetric
or antisymmetric under exchange of particles 2 and 3 for the
case of identical bosons or fermions, respectively.  Thus
\begin{equation}
{\cal P}_{23} M^1 = M^1 {\cal P}_{23} = \zeta M^1 \, . \label{EqSymT1}
\end{equation}
Using this relation, Eq.~(\ref{EqFad1}) (with $i=1$) can be written
\begin{eqnarray}
{\cal P}_{23} | \Gamma^1 \rangle  & = & -\zeta {\cal P}_{23} M^1 G_{BS}^1
\left( {\cal P}_{12} + {\cal P}_{13} \right) | \Gamma^1 \rangle  \nonumber \\
& = & -\zeta^2 M^1 G_{BS}^1
\left( {\cal P}_{12} + {\cal P}_{13} \right) | \Gamma^1 \rangle  \nonumber \\
& = & \phantom{-}\zeta | \Gamma^1 \rangle  \, . \label{EqSymG1}
\end{eqnarray}
Next, using the definitions Eq.~(\ref{EqP123}) note that
\begin{eqnarray}
{\cal P}_{23} {\cal P}_{12} {\cal P}_{23}|abc\rangle&&= {\cal P}_{23} 
{\cal P}_{12}
|acb\rangle={\cal P}_{23} |cab\rangle \nonumber\\
&&=|cba\rangle=  {\cal P}_{13}|abc\rangle \, .
\end{eqnarray}
Hence, the operator ${\cal P}_{13}$ can
be written
\begin{equation}
{\cal P}_{13} = {\cal P}_{23} {\cal P}_{12} {\cal P}_{23} \, .
\label{EqSymP1}
\end{equation}
Using the relations (\ref{EqSymT1}) -- (\ref{EqSymP1}) together
with the fact that $G_{BS}^1$ commutes with ${\cal P}_{23}$ we can write
the Faddeev
equations (\ref{EqFad1}) in the following simple form
\begin{eqnarray}
| \Gamma^1 \rangle  & = & -\zeta M^1 G_{BS}^1
 \left( {\cal P}_{12} + {\cal P}_{23} {\cal P}_{12} {\cal P}_{23}
\right) | \Gamma^1 \rangle  \nonumber \\
& = & -\zeta M^1 G_{BS}^1 \left( 1 +\zeta {\cal P}_{23} \right)
{\cal P}_{12}
| \Gamma^1 \rangle \nonumber \\
& = & -2\zeta M^1 G_{BS}^1  {\cal P}_{12}  | \Gamma^1 \rangle \, .
\label{EqFad2}
\end{eqnarray}
To reduce these equations to a practical form, it is sufficient to
evaluate the
permutation operator ${\cal P}_{12}$.

\subsection{Spectator Equations}

Now we turn to the spectator equations.
We begin by replacing the two-body propagator $G_{BS}^i\otimes 1_i$,
which describes the
propagation of particles $j$ and $k$ (both not equal to $i$) in
Eq.~(\ref{EqBS1}), by a new propagator,
\begin{equation}
G_{BS}^i\otimes 1_i \to G_j{\cal Q}_k \otimes 1_i \,  , \label{EE1a}
\end{equation}
where ${\cal Q}_k$ is a {\it projection operator\/} which
places particle $k$ on the positive energy mass-shell, and, as in the BS case,
$G_j$ is the propagator of a single off-shell particle $j$.
Choosing particle $k$ to be the spectator during the ``previous''
interaction gives the unclosed form of the spectator Faddeev equations
\begin{equation}
| \Gamma^i \rangle = - \sum_{k \not= i \not= j}
M^i G_j{\cal Q}_k | \Gamma^k \rangle \, , \label{EE1}
\end{equation}
where the sum is over $k$ and $j$ with $i$ fixed and no two indices equal.
Explicitly, Eq.~(\ref{EE1}) is shorthand for the following three 
equations:
\begin{eqnarray}
| \Gamma^1 \rangle =&& - \left\{
M^1 G_2{\cal Q}_3 | \Gamma^3 \rangle + M^1 G_3{\cal Q}_2 | \Gamma^2 \rangle
\right\} \nonumber\\
| \Gamma^2 \rangle =&& - \left\{
M^2 G_3{\cal Q}_1 | \Gamma^1 \rangle + M^2 G_1{\cal Q}_3 | \Gamma^3 \rangle
\right\} \nonumber\\
| \Gamma^3 \rangle =&& - \left\{
M^3 G_1{\cal Q}_2 | \Gamma^2 \rangle + M^3 G_2{\cal Q}_1 | \Gamma^1 \rangle
\right\} \, . \label{EE2}
\end{eqnarray}
Note that the projection operator ${\cal Q}_k$ insures that
particle $k$
is on-shell both as it leaves the partial amplitude
$|\Gamma^k\rangle$ and
as it enters the two-body scattering amplitude $M^i$.

To make a closed set of equations from Eq.~(\ref{EE1}), it is necessary
to place the  final spectator particle $i$ on shell, which then also 
forces one of the two interacting particles in the final state
(denoted by $k'$) to be on shell.  The spectator scattering equations
are shown diagrammatically in Fig.~2.  The final bound state equations can
be written algebraically in the following form
\begin{equation}
{\cal Q}_i{\cal Q}_{k'}| \Gamma^i \rangle = -
\sum_{k \not= i \not= j} {\cal Q}_{k'}  M^i
 {\cal Q}_k \; G_j\; {\cal Q}_k {\cal Q}_i | \Gamma^k \rangle
\, , \label{EqSpec1a}
\end{equation}
where no summation over the index $i$ is implied, and we used the
projection property ${\cal Q}_k {\cal Q}_k ={\cal Q}_k $ and
$G_j {\cal Q}_k = {\cal Q}_k G_j$.

Alternatively, we may introduce the notation
\begin{eqnarray}
| \Gamma^i_j \rangle =&& {\cal Q}_i{\cal Q}_j| \Gamma^i \rangle
\nonumber\\
M^i_{k'k}= && {\cal Q}_{k'}  M^i{\cal Q}_k \otimes 1_i \nonumber\\
G^i_k = && G_j {\cal Q}_k \otimes 1_i \, , \label{EE3}
\end{eqnarray}
where the indices $i$, $j$, and $k$ are all different, and
the lower indices on $M$ and $\Gamma$ label which particles, apart from
the spectator, are on mass shell.  In this notation Eq.~(\ref{EqSpec1a})
becomes
\begin{equation}
| \Gamma^i_{k'} \rangle = - \sum_{k \not= i}   M^i_{k'k} G^i_k
| \Gamma^k_i \rangle
 \, , \label{EqSpec1}
\end{equation}
We will use this notation in most of the remainder of this section, but will
return to the definitions (\ref{EE3}) later in the paper.
As an example, consider the case $i=1$ and $k'=2$:
\begin{equation}
| \Gamma^1_2 \rangle = - M^1_{22} G^1_2 | \Gamma^2_1 \rangle
                 - M^1_{23} G^1_3 | \Gamma^3_1 \rangle   \, .
\end{equation}
As discussed in Ref.~\cite{R4}, for distinguishable particles without
three-body forces Eq.~(\ref{EqSpec1})
becomes a coupled set of {\it six\/} equations for the six
amplitudes $| \Gamma^i_j \rangle$, instead of only three equations for three
$| \Gamma^i \rangle$, as in the Bethe-Salpeter case.

We emphasize that an important difference between the spectator 
subamplitudes
$| \Gamma^i_j \rangle$ and the Bethe-Salpeter subamplitudes $| \Gamma^i
\rangle$
is that it is no longer possible to add the spectator subamplitudes
together in order
to construct a total amplitude, as we did in Eq.~(\ref{Combamp}).  This is
because
the amplitude $|\Gamma^1_2\rangle$, for example, restricts particles 1 and 2 to
the mass shell, while the amplitude $|\Gamma^1_3\rangle$ restricts
particles 1 and
3 to the mass shell, and hence they are defined for different regions of
phase space.
Only operators or amplitudes which satisfy identical constraints, such as
$|\Gamma^1_2\rangle$ and $|\Gamma^2_1\rangle$, for example, can be combined.
{\it No total three-body amplitude exists in the spectator formalism\/}.

For identical particles, Eqs.~(\ref{EqSpec1}) can be further reduced by
using permutation
operators. Using the fact that the operator $M^1$ is symmetric under
particle interchange, Eq.~(\ref{EqSymT1}), and the relation 
\begin{equation}
{\cal P}_{32}{\cal Q}_2={\cal Q}_3{\cal P}_{32} \, , \label{EqZZ3}
\end{equation} 
we obtain
\begin{eqnarray}
{\cal P}_{32} M^1_{22} && = \zeta M^1_{32} = M^1_{33}{\cal P}_{32} \nonumber\\
{\cal P}_{23} M^1_{33} && = \zeta M^1_{23} = M^1_{22}{\cal P}_{23} \,  ,
\label{SpecSymm1}
\end{eqnarray}
where $\zeta=+1$ for bosons and $-1$ for fermions, as before.  (These
are the operator form of the symmetry relations discussed in
Ref.~\cite{R1}; note that, in the spectator formalism, the exchange
operator does not relate an amplitude to itself, but to another
amplitude with a different particle off-shell.)  We will  find it
convenient to exploit the fact that
${\cal P}_{jk}={\cal P}_{kj}$, and always write relations like those above
so that the initial and final indices on both sides of the equation 
match.  The spectator and on shell interacting particle can also be
interchanged, leading to the
following relations for the operators $G^i_j$
\begin{eqnarray}
{\cal P}_{12} G^1_2  {\cal P}_{21} & = & G^2_1  \,   \nonumber \\
{\cal P}_{23} G^1_3  {\cal P}_{32} & = & G^1_2  \, .
\end{eqnarray}
The two-body amplitudes exhibit a similar symmetry
\begin{eqnarray}
{\cal P}_{12}M^1_{22} {\cal P}_{21} & = & M^2_{11} \,   \nonumber \\
{\cal P}_{23}M^1_{33} {\cal P}_{32} &=& M^1_{22} \,  . \label{SpecSymm2}
\end{eqnarray}
Further relations can be found by combining the relations (\ref{SpecSymm1}) and
(\ref{SpecSymm2}).  One relation we will use below is
\begin{equation}
{\cal P}_{12}M^1_{23}{\cal P}_{32}{\cal P}_{21}=\zeta
M^2_{11}=M^2_{13}{\cal P}_{31}
\, .  \label{SymmSpecial}
\end{equation}

It is now easy to derive the effect of permutations on the spectator
subamplitudes.
For example, under the interchange of two particles in the interacting pair,
\begin{eqnarray}
{\cal P}_{32}| \Gamma^1_2 \rangle  & = &
- {\cal P}_{32}M^1_{22} G^1_2 | \Gamma^2_1 \rangle
                 -{\cal P}_{32} M^1_{23} G^1_3 | \Gamma^3_1 \rangle \nonumber \\
& = & -\zeta M^1_{32} G^1_2 | \Gamma^2_1 \rangle
                 -\zeta M^1_{33} G^1_3 | \Gamma^3_1 \rangle  \nonumber \\
& = & \zeta  | \Gamma^1_3 \rangle  \, . \label{Symm1}
\end{eqnarray}
Using this, the interchange of the spectator with the on-shell particle in
the interacting pair is
\begin{eqnarray}
{\cal P}_{12}| \Gamma^1_2 \rangle & = & - {\cal P}_{12} M^1_{22} {\cal P}_{21}
{\cal P}_{12} G^1_2 {\cal P}_{21} {\cal P}_{21} | \Gamma^2_1 \rangle
- {\cal P}_{12}M^1_{23} {\cal P}_{32}{\cal P}_{21}{\cal P}_{12}
{\cal P}_{23} G^1_3 |
\Gamma^3_1 \rangle \nonumber \\
& = &
- M^2_{11}G^2_1 {\cal P}_{21} | \Gamma^2_1 \rangle
- M^2_{13}{\cal P}_{31}{\cal P}_{12}G^1_2 {\cal P}_{23} | \Gamma^3_1
\rangle \nonumber\\
& = &
- M^2_{11}G^2_1 {\cal P}_{21} | \Gamma^2_1 \rangle
- M^2_{13}G^2_3 {\cal P}_{31}{\cal P}_{12} {\cal P}_{23} | \Gamma^3_1
\rangle \nonumber\\
& = &
- M^2_{11}G^2_1 {\cal P}_{21} | \Gamma^2_1 \rangle
- M^2_{13}G^2_3 {\cal P}_{21} | \Gamma^3_1 \rangle \nonumber\\
& = &
- M^2_{11}G^2_1 {\cal P}_{21} | \Gamma^2_1 \rangle
- \zeta M^2_{13}G^2_3  | \Gamma^3_2 \rangle   \, ,
\end{eqnarray}
where Eq.~(\ref{EqSymP1}) (with $1 \leftrightarrow 3$) was used in the next to
last step.  Comparison with the equation for $| \Gamma^2_1 \rangle$,
\begin{equation}
\zeta | \Gamma^2_1 \rangle = -\zeta M^2_{11} G^2_1 | \Gamma^1_2 \rangle
                 - \zeta M^2_{13} G^2_3 | \Gamma^3_2 \rangle  \, ,
\end{equation}
implies
\begin{equation}
{\cal P}_{21}| \Gamma^2_1 \rangle  = \zeta | \Gamma^1_2 \rangle   \, .
\end{equation}

Using these relations, we can obtain a single equation for $ | \Gamma^1_2
\rangle $
\begin{eqnarray}
| \Gamma^1_2 \rangle  & = & -\zeta M^1_{22} G^1_2 {\cal P}_{21}
                             | \Gamma^1_2 \rangle
-\zeta^2 M^1_{23} G^1_3 {\cal P}_{13}{\cal P}_{32} | \Gamma^1_2 \rangle
\nonumber \\
& = &
-\zeta M^1_{22} G^1_2 {\cal P}_{21} | \Gamma^1_2 \rangle
-\zeta^2 M^1_{23} {\cal P}_{32}{\cal P}_{23}G^1_3 {\cal P}_{32} {\cal P}_{21}
| \Gamma^1_2 \rangle
\nonumber \\
& = &
-2\zeta  M^1_{22} G^1_2 {\cal P}_{21} | \Gamma^1_2 \rangle  \, ,
\end{eqnarray}
where Eq.~(\ref{EqSymP1}) was used in the next to last step.
From now on we will only consider fermions, so that the three-body
equation for the vertex function
$\Gamma^1_2$, which singles out particle 1 as spectator in the
``last'' interaction and particle 2 as the interacting particle
to be put on mass shell, becomes
\begin{equation}
| \Gamma^1_2 \rangle  =  2 M^1_{22} G^1_2 {\cal P}_{21}
| \Gamma^1_2 \rangle  \, . \label{EqSpec2}
\end{equation}
This equation is illustrated diagrammatically in Fig.~3.

A more explicit form of Eq.~(\ref{EqSpec2}), which expresses $M^1_{22}$ and
$G^1_2$ as operators, and $| \Gamma^1_2 \rangle$ as a vector in Dirac
space, is
\begin{equation}
| \Gamma^1_2 \rangle_{\alpha\beta\gamma}
= 2 [M^1_{22}]_{\beta \beta_1,\gamma \gamma_1} \;
[G^1_2]_{\beta_1 \beta_2, \gamma_1\gamma_2} \;
\left[{\cal P}_{12}| \Gamma^1_2 \rangle\right]_{\alpha \beta_2 \gamma_2} \, ,
\end{equation}
where $\alpha$, $\beta$, and $\gamma$ are Dirac indices for particles
1, 2, and 3 respectively, and summation over repeated Dirac indices is
implied.  Note that $M$ and $G$ operate on a two-body space only; the third
particle (the spectator) is unaffected by these operators.

In the next section we will give a momentum space representation of these
equations.

\section{Momentum Space Representation}

We specialize to three identical particles with mass $m$, spin $1/2$,
and four-momenta $k_1$, $k_2$, and $k_3$. The total momentum
\begin{equation}
P=k_1 + k_2 + k_3
\end{equation}
is conserved.
The Gross equation restricts two of the three particles to be on mass
shell, which
for the choice (\ref{EqSpec2}) are particles 1 and 2, with particle 3 off
mass shell.
In the three-body c.m. system
\begin{equation}
P=( M_t,{\bf 0}) \, ,
\end{equation}
where $M_t$ is the mass of the three-body system, the momenta are
%
\begin{eqnarray}
k_1 & = & ( E_{k_1}, {\bf k}_1) \nonumber\\
k_2 & = & ( E_{k_2}, {\bf k}_2) \nonumber\\
k_3 & = & ( k_{30}, {\bf k}_3) = (M_t - E_{k_1} - E_{k_2}, - {\bf k}_1 -
{\bf k}_2) \, . \label{Momxxx}
\end{eqnarray}
In Eq.~(\ref{Momxxx}) the four-momentum of the off-shell particle, $k_3$,
is fixed by four-momentum conservation.
It is obvious that the problem has only 6 independent momentum variables, just
as in the nonrelativistic case.

The three-body basis states are direct products of a single particle state
and a
two-particle state,
\begin{equation}
| k_1 (k_2 \overline{k}_3) \rangle =
 | k_1 \rangle \otimes | k_2  \overline{k}_3 \rangle \, , \label{Eq17a}
\end{equation}
where, by convention, the off-shell particle has a bar over its momentum.

Completeness and orthogonality relations are:
\begin{eqnarray}
\langle k_1 (k_2 \overline{k}_3 ) | k'_1 (k'_2 \overline{k'_3}) \rangle &&=
2 E_{k_1} \delta^{3}(k_1-k'_1)
2 E_{k_2} \delta^{3}(k_2-k'_2) \delta^4(P-P') \label{Eq22}\\
1 && = \int \frac{d^3k_1}{2 E_{k_1}}\frac{d^3k_2}{2 E_{k_2}} d^4P
|k_1( k_2 \overline{k}_3 )\rangle \langle k_1( k_2 \overline{k}_3 ) |   \, .
\label{Comp}
\end{eqnarray}

Next, we specify the matrix elements of all operators in
this momentum space basis.  The propagator is
\begin{eqnarray}
\langle k_1 (k_2 \overline{k}_3 ) |
[G^1_2]_{\beta \beta',\gamma \gamma'}
|k'_1 (k'_2 \overline{k'_3} )\rangle = & & 2 E_{k_1} \delta^{3}(k_1-k'_1)
\, 2E_{k_2} \delta^{3}(k_2-k'_2) \, \delta^4(P-P') \nonumber\\
&&\qquad \times 2m \left[ \Lambda_+(k_2) \right]_{\beta \beta'}
\frac{(m+ \rlap/ k_3)_{\gamma\gamma'}}{m^2-k^2_3-i \epsilon} \, ,
\end{eqnarray}
where $\Lambda_\pm(k)=(m\pm \rlap/ k)/2m$ are the positive and negative energy
projection operators.
The two-body $M$ matrix is
\begin{eqnarray}
\langle k_1 (k_2 \overline{k}_3 ) |
[M^1_{22}]_{\beta\gamma, \beta'\gamma'}
|k'_1 (k'_2 \overline{k'_3} ) \rangle = &&
2 E_{k_1} \delta^{3}(k_1-k'_1) \,\delta^4(P-P') \nonumber\\
&& \qquad\times M_{\beta \beta', \gamma \gamma'}
(k_{23},k'_{23};P-k_1) \, , \label{TwobodyT}
\end{eqnarray}
where $P-k_1$ is the total two-body four-momentum, and the relative
momenta in the two-body space are denoted by
\begin{equation}
k_{ij}={\textstyle{1\over2}}\left(k_i-k_j\right) \,  .
\end{equation}
Note that $k_{ij}=-k_{ji}$.  The two-body amplitudes in
Eq.~(\ref{TwobodyT}) are identical to those discussed in Sec.~IIA of
Ref.~\cite{R1}.
The partial vertex amplitudes will be written
\begin{equation}
\langle k_1 (k_2 \overline{k}_3 ) |\Gamma^1_2 \rangle_{\alpha\beta\gamma}
= \Gamma_{\alpha \beta \gamma}(k_1, k_2, k_3) \, ,
\end{equation}
where, by convention, it is understood that the last momentum is the one which
is off shell.  Therefore
\begin{equation}
\langle k_1 (k_2 \overline{k}_3)  |{\cal P}_{12} |\Gamma^1_2
\rangle_{\alpha \beta \gamma}
=- \langle k_2 (k_1 \overline{k}_3) |\Gamma^1_2 \rangle_{\beta\alpha  \gamma}
=-\Gamma_{\beta \alpha \gamma}(k_2, k_1, k_3) \, .
\end{equation}

We can now obtain the momentum space representation of
Eq.~(\ref{EqSpec2}).  Inserting the completeness relation (\ref{Comp})
gives
\begin{eqnarray}
\langle k_1 (k_2 \overline{k}_3 ) | \Gamma^1_2
\rangle_{\alpha \beta \gamma} & = &
2 \int \frac{d^3k'_1}{2 E_{k'_1}}\frac{d^3k'_2}{2 E_{k'_2}} d^4P'
\frac{d^3k''_1}{2 E_{k''_1}}\frac{d^3k''_2}{2 E_{k''_2}} d^4P''
\langle k_1( k_2 \overline{k}_3 ) |
[M^1_{22}]_{\beta\beta',\gamma\gamma'}
|k'_1 (k'_2 \overline{k'_3}) \rangle \nonumber \\
& &\qquad\times
\langle k'_1 (k'_2 \overline{k'_3})  |
[G^1_2]_{\beta,\beta'',\gamma'\gamma''}
|k''_1 (k''_2 \overline{k''_3}) \rangle \;
\langle k''_1 (k''_2 \overline{k''_3})  |
 {\cal P}_{12} |\Gamma^1_2 \rangle_{\alpha \beta'' \gamma''}
  \, .
\end{eqnarray}
Inserting the above expressions for $M$ and $G$, and
carrying out all integrals gives
\begin{eqnarray}
\Gamma_{\alpha \beta\gamma}(k_1, k_2, k_3)
& = &- 2\int d^3k'_2 \frac{m}{E_{k'_2}}
M_{\beta \beta', \gamma \gamma'}(k_{23},k'_{23};P-k_1)
\nonumber \\
& & \qquad\times
\left[ \Lambda_+(k'_2) \right]_{\beta' \beta''}
\frac{(m+  \rlap/ k'_3)_{\gamma' \gamma''}}
{m^2-k'^2_3-i \epsilon}
\Gamma_{\beta'' \alpha \gamma''}(k'_2, k_1,k'_3) \, , \label{Reduced1}
\end{eqnarray}
where $k'_3 = P - k'_2 - k_1$.  This equation is manifestly covariant.

These equations may be further reduced by multiplying the $M$ matrix and
the three-body vertex functions $\Gamma$ by the on-shell spinors $u$ (for
on-shell particles in the initial state) and $\bar{u}$ (for on-shell
particles in the final state):
\begin{eqnarray}
\Gamma_{\lambda_1 \lambda_2 \gamma}(k_1, k_2, k_3)
&&= \bar{u}_{\alpha} ({\bf k_1}, \lambda_1) \bar{u}_\beta ({\bf k_2},
\lambda_2)
\Gamma_{\alpha \beta \gamma}(k_1, k_2, k_3) \nonumber\\
M_{\lambda_2 \lambda'_2, \gamma \gamma'}(k_{23},k'_{23};P-k_1)
&&= \bar{u}_\beta ({\bf k_2},\lambda_2)
M_{\beta \beta', \gamma \gamma'}(k_{23},k'_{23};P-k_1)
u_{\beta'} ({\bf k'_2}, \lambda'_2)  \, ,
\end{eqnarray}
where $u_{\alpha} ({\bf k_1}, \lambda_1)$ is an on-shell Dirac spinor with
three-momentum ${\bf k_1}$ and helicity $\lambda_1$.  This gives us quantities
with ``mixed indices''; a Dirac index on a
matrix element is replaced by a helicity index when it is contracted with
a $u$-spinor of that helicity and with matching momentum.  These amplitudes
are still covariant, and simpler because the four-dimensional Dirac space
is replaced by a two-dimensional helicity space.  If we then replace
the on-shell projection operator by a sum over on-shell $u$-spinors
\begin{equation}
\left[ \Lambda_+(k_2) \right]_{\beta \beta'} = \sum_{\lambda_2} u({\bf
k_2},\lambda_2)
\bar{u}({\bf k_2},\lambda_2) \, , \label{Eq26a}
\end{equation}
and multiply Eq.~(\ref{Reduced1}) from the left by
$\bar{u}_{\alpha} ({\bf k_1}, \lambda_1) \bar{u}_{\beta} ({\bf k_2}, \lambda_2)$
we get
\begin{eqnarray}
\Gamma_{\lambda_1 \lambda_2 \gamma}(k_1, k_2, k_3) & = &- 2
\sum_{\lambda'_2}
\int d^3k'_2 \frac{m}{E_{k'_2}}
M_{\lambda_2 \lambda'_2, \gamma \gamma'}(k_{23},k'_{23};P-k_1)
\nonumber \\
& & \qquad\times
\frac{(m+ \rlap/ k'_3)_{\gamma' \gamma''}}
{m^2-k'^2_3-i \epsilon}
\Gamma_{\lambda'_2 \lambda_1 \gamma''}(k'_2, k_1,k'_3) \, .
\label{Reduced2}
\end{eqnarray}

Equation (\ref{Reduced2}) is still manifestly covariant, but
is not suitable for a numerical solution.  The main reason is that
the two-body $M$ matrix is given as a partial wave expansion in the
{\it two-body rest frame\/}, and not the three-body c.m.\ system, as
needed in the above equation.  A related problem is that the propagator
for particle 3
depends on the angle between the vectors ${\bf k_1}$ and ${\bf k'_2}$ and
is therefore not diagonal with respect to all angular momenta after a
partial wave decomposition.

In the nonrelativistic case, the first problem does not occur because
the partial wave expansion is invariant under a Galilean boost, and
the second is solved by introducing Jacobi coordinates. Because of the
different energy-momentum relations in special relativity, neither of
these problems can be handled so simply here.

However, we can eliminate these problems here by exploiting the covariance
of the formalism, and by explicitly boosting the two-body subsystem to its
rest frame.
To prepare the way, introduce the
total four-momentum of the two-body subsystem,
\begin{equation}
P_{23} = k_2 + k_3 = P- k_1 = P+q \, ,
\end{equation}
where here it is convenient to introduce the momentum $q = -k_1$.
Next, the boost operator $\Lambda_{k_1}$ is defined by the requirement
\begin{equation}
\Lambda_{k_1} P_{23} = \tilde{P}_{23} = (W_q, {\bf 0}) \, .
\end{equation}
The square of the mass of the (23) pair is then
\begin{eqnarray}
W^2_q & = &\tilde{P}^2_{23} = P^2_{23} = (M_t - E_q)^2-{\bf q}^2 \, ,
\nonumber \\
 & = & M^2_t + m^2 -2 M_t E_q \, .  \label{EqWq}
\end{eqnarray}
A tilde (``\~{}'') on top of a variable always indicates that it is defined in the
two-body rest frame. We have, e.g.,
\begin{eqnarray}
{\tilde k}_2 & = & \Lambda_{k_1} k_2  \, , \nonumber\\
{\tilde k}_3 & = & \Lambda_{k_1} k_3  \, .
\end{eqnarray}
We now define the relative momentum ${\tilde p}$ through
\begin{eqnarray}
{\tilde k}_2 & = & {\textstyle{1 \over 2}} \tilde{P}_{23} + {\tilde p}
= ( E_{\tilde p}\, , \,{\tilde {\bf p}}) \nonumber\\
{\tilde k}_3 & = & {\textstyle{1 \over 2}} \tilde{P}_{23} - {\tilde p}
= (W_q - E_{\tilde p}\, , \, -{\tilde {\bf p}})\, ,
\end{eqnarray}
and therefore
\begin{equation}
{\tilde p} = {\tilde k}_{23} = {\textstyle{1 \over 2}}
( {\tilde k}_2 - {\tilde k}_3 )
=(E_{\tilde p} - {\textstyle{1 \over 2}}W_q \, , \, {\tilde {\bf p}}) \, .
\end{equation}

Next, we introduce the representation on the Dirac space, $S(\Lambda)$,
of a Lorentz boost $\Lambda$.  These transform Dirac matrices
and spinors according to the following rules
%
\begin{eqnarray}
&& S^{-1}(\Lambda)\gamma^\mu S(\Lambda) = \Lambda^\mu{}_\nu\gamma^\nu
\label{Trans1}\\
&& S(\Lambda) u({\bf k},\lambda) =
\sum_\mu {\cal D}^{(1/2)}_{\mu \lambda}(R_{\Lambda k})
u(\Lambda {\bf k}, \mu ) \label{Trans2} \\
&&{\bar u} ({\bf k},\lambda) S^{-1}(\Lambda)  =
\sum_\mu {\bar u}(\Lambda {\bf k}, \mu )
{\cal D}^{(1/2)*}_{\lambda \mu }(R_{\Lambda k}) \, , \label{Trans3}
\end{eqnarray}
%
where $R_{\Lambda k}$ is the Wigner rotation accompanying the boost that
connects the momenta ${\bf k}$ and $\Lambda {\bf k}$ (see Appendix C),
and the Dirac indices have been suppressed.  The propagator
of the off-shell particle 3 in the three-body rest frame can therefore be 
expressed in terms of its form in the two-body rest frame using (\ref{Trans1})
\begin{equation}
\frac{(m+ \rlap/ k_3)}{m^2-k^2_3-i \epsilon}= S^{-1}(\Lambda_{k_1})
\frac{(m+ \rlap/ {\tilde k}_3)}{m^2-{\tilde k}^2_3-i \epsilon}
S(\Lambda_{k_1}) \, .
\end{equation}
Similarly, the full two-body $M$ matrix in the three-body system can be
written
\begin{equation}
M_{2,3}(k_{23}, k'_{23};P_{23})=
S_{2}^{-1}(\Lambda_{k_1}) S_{3}^{-1}(\Lambda_{k_1})
M_{2,3}({\tilde p},{\tilde p}';{\tilde P}_{23})
S_{2}(\Lambda_{k_1}) S_{3}(\Lambda_{k_1}) \, ,
\end{equation}
where the subscripts 2 and 3 are shorthand for pairs of Dirac indices on
particle 2 ($\beta\beta'$, etc.) and on particle 3 ($\gamma\gamma'$, etc.),
and the two-body scattering amplitude
$M({\tilde p},{\tilde p}';{\tilde P}_{23})$  is a solution
of the two-body Gross equations in the two-body c.m.\ frame.
[Do not confuse $M_{2,3}$
with amplitudes like $M^1_{23}$ used in the last subsection; here the
subscripts refer to
the Dirac indices, and in the previous subsection they referred to which
of the interacting
particles was on shell.  From now on we have made the choice that particle 2
is on shell, and
in the language of the previous subsection, all two-body amplitudes are
$M^1_{22}$.]
Using (\ref{Trans2}) and (\ref{Trans3}), we obtain the following expression for
the mixed index $M$ matrix
\begin{eqnarray}
M_{\lambda_2 \lambda'_2, 3}&&(k_{23}, k'_{23};P_{23})
\nonumber\\&&=
S_{3}^{-1}(\Lambda_{k_1})
{\cal D}^{(1/2)*}_{\lambda_2 \mu_2 }(R_{\Lambda_{k_1} k_{2}})
M_{\mu_2 \mu'_2, 3}({\tilde p},{\tilde p}';{\tilde P}_{23})
{\cal D}^{(1/2)}_{\mu'_2 \lambda'_2 }(R_{\Lambda_{k_1} k'_{2}})
S_{3}(\Lambda_{k_1}) \, ,
\end{eqnarray}
where summation over all repeated indices (including helicities) is
implied.  Substituting these relations into Eq.~(\ref{Reduced2}) gives
\begin{eqnarray}
\Gamma_{\lambda_1 \lambda_2 \gamma}(k_1, k_2, k_3) & = & -2
\int d^3k'_2 \frac{m}{E_{k'_2}} \;S_{\gamma \gamma_1}^{-1}(\Lambda_{k_1})
{\cal D}^{(1/2)*}_{\lambda_2 \mu_2 }(R_{\Lambda_{k_1} k_{2}})
M_{\mu_2 \mu'_2, \gamma_1 \gamma_2}({\tilde p},{\tilde p}';{\tilde P}_{23})
{\cal D}^{(1/2)}_{\mu'_2 \lambda'_2 }(R_{\Lambda_{k_1} k'_{2}})
\nonumber \\
& & \qquad\times
\frac{(m+ \rlap/ {\tilde k}'_3)_{\gamma_2 \gamma_3}}
{m^2-{\tilde k}'^2_3-i \epsilon}
S_{\gamma_3 \gamma'}(\Lambda_{k_1})
\Gamma_{\lambda'_2 \lambda_1 \gamma'}(k'_2, k_1,k'_3) \, .
\label{Reduced3}
\end{eqnarray}

This equation can be further reduced if we decompose of the propagator
of the off-shell particle 3 into positive and
negative energy parts
\begin{eqnarray}
\frac{(m+ \rlap/ {\tilde k}_3)_{\gamma \gamma'}}
{m^2-{\tilde k}^2_3-i \epsilon} &&=
{m \over E_{\tilde p}} \sum_{\lambda_3} \left[
\frac{u_{\gamma}({\bf {\tilde k}_3},\lambda_3) {\bar u}_{\gamma'}
({\bf {\tilde k}_3},\lambda_3)}
     {2 E_{\tilde p} - W_q - i \epsilon}
- \frac{ v_{\gamma} (-{\bf {\tilde k}_3},\lambda_3 )
\bar{v}_{\gamma'}
(-{\bf {\tilde k}_3},\lambda_3)}
     {W_q - i \epsilon} \right]  \, \nonumber\\
&&={m \over E_{\tilde p}} \sum_{\lambda_3} \left[
\frac{u_{\gamma}(-{\bf {\tilde p}},\lambda_3) {\bar u}_{\gamma'}
(-{\bf {\tilde p}},\lambda_3)}
     {2 E_{\tilde p} - W_q - i \epsilon}
- \frac{ v_{\gamma} ({\bf {\tilde p}},\lambda_3 )
\bar{v}_{\gamma'}
({\bf {\tilde p}},\lambda_3)}{W_q - i \epsilon} \right]  \, , \label{Decomp}
\end{eqnarray}
where the second expression can be obtained from the first using the
fact that the spinors depend only on the three-momentum and
${\bf {\tilde k}_3}= - {\bf {\tilde p}}$.  At this point it is
convenient to introduce ``$\rho$''-spin by letting
\begin{equation}
u^\rho({\bf p},\lambda)=\cases{u({\bf p},\lambda)\phantom{-}
\qquad\hbox{if } \rho=+\cr
	v(-{\bf p},\lambda) \qquad\hbox{if } \rho=- \; . \cr} \label{Eq20a}
\end{equation}
Then the decomposition (\ref{Decomp}) becomes
\begin{equation}
\frac{(m+ \rlap/ {\tilde k}_3)_{\gamma \gamma'}}
{m^2-{\tilde k}^2_3-i \epsilon} =
{m \over E_{\tilde p}}
u^\rho_{\gamma}(-{\bf {\tilde p}},\lambda_3) \, g^\rho(q,{\tilde p})
\, {\bar u}^\rho_{\gamma'}  (-{\bf {\tilde p}},\lambda_3)
 \, , \label{RhoDecomp}
\end{equation}
where summation over $\rho$ and $\lambda_3$ is implied, and
\begin{eqnarray}
g^+(q,{\tilde p}) &&= {\displaystyle{1\over 2E_{{\tilde p}} -W_q -i\epsilon}}
\nonumber\\
g^-(q,{\tilde p}) &&= -{\displaystyle{1\over W_q -i\epsilon}} \, .
\label{Eq20b}
\end{eqnarray}
Substituting (\ref{RhoDecomp}) into Eq.~(\ref{Trans3}), multiplying
from the left by $\bar{u}^\rho({\bf k_3},\lambda_3)$, and using (\ref{Trans2})
and (\ref{Trans3}), give the following reduced three-body equations
\begin{eqnarray}
\Gamma^{\rho}_{\lambda_1 \lambda_2 \lambda_3}  (k_1, k_2, k_3 )
 =&& -2  \int d^3{\tilde k}'_2
\left( \frac{m}{E_{{\tilde k}'_2}} \right) ^2
{\cal D}^{(1/2)*}_{\lambda_2 \mu_2 }(R_{\Lambda_{k_1} k_2})
{\cal D}^{(1/2) *}_{\lambda_3 \mu_3 }(R^\rho_{\Lambda_{k_1} k_3})
\nonumber \\
& & \times
M^{\rho \rho'}_{\mu_2\mu'_2, \mu_3 \mu'_3}({\tilde p},{\tilde p}';W_q)
{\cal D}^{(1/2)}_{\mu'_2\lambda'_2 }(R_{\Lambda_{k_1} k'_2})
{\cal D}^{(1/2)}_{\mu'_3 \lambda'_3 }(R^{\rho'}_{\Lambda_{k_1} k'_3})
\nonumber\\
&&\qquad\times g^{\rho'}(\tilde{p}',q) \;
\Gamma^{\rho'}_{\lambda'_2 \lambda'_1 \lambda'_3}(k'_2, k'_1,k'_3 )
\, , \label{Reduced4}
\end{eqnarray}
where $R^\rho_{\Lambda k}$ is the Wigner rotation for the spinor $u^\rho$,
and
\begin{eqnarray}
\Gamma^{\rho}_{\lambda_1 \lambda_2 \lambda_3}  (k_1, k_2, k_3 )
 =&& \bar{u}^\rho_\gamma({\bf k_3},\lambda_3)
\Gamma_{\lambda_1 \lambda_2 \gamma}  (k_1, k_2, k_3 )
\nonumber \\
M^{\rho \rho'}_{\mu_2\mu'_2, \mu_3 \mu'_3}({\tilde p},{\tilde
p}';W_q)  =&& \bar{u}^\rho_\gamma({\bf {\tilde k}_3},\mu_3)
M_{\mu_2\mu'_2, \gamma \gamma'}({\tilde p},{\tilde p}';W_q)
{u}^{\rho'}_{\gamma'}({\bf {\tilde k}'_3},\mu'_3)
\, .
\end{eqnarray}

We have reduced the three-body equations to
six-dimensional integral equations for the coupled set of
$2^4=16$ amplitudes
$\Gamma^{\rho}_{\lambda_1 \lambda_2 \lambda_3}$, which can be
written
\begin{equation}
\Gamma^{\rho}_{\lambda_1 \lambda_2 \lambda_3} (k_1, k_2, k_3 )
 =
\langle k_1 \lambda_1 ( k_2 \lambda_2 \overline{k}_3 \lambda_3) \rho|
\Gamma \rangle
\, . \label{Eq16a}
\end{equation}
The new states $| k_1 \lambda_1 ( k_2 \lambda_2 \overline{k}_3 \lambda_3)
\rho \rangle $ have simple completeness and orthogonality relations
(developed in
the next section) which make them a useful starting point for further
discussion.

This form (\ref{Reduced4}) for the three-body equations
displays the Wigner rotations which appear when the two-body scattering
amplitude is boosted from the overall three-body rest frame to its
two-body rest frame.  For practical calculations the equations will be
further reduced by decomposing the amplitudes into partial waves,
which will be discussed in the next section.

\section{Angular momentum states}

In this section we follow the conventions of Wick \cite{R2} and
define  a basis of three-body partial-wave helicity states.
Completeness and orthogonality relations are defined and the matrix
elements of the propagator and two-body scattering amplitude are
obtained. Using these states, the operator equations
(\ref{EqSpec2}) are written directly in terms of the partial
wave states.  To obtain the final equations, the matrix elements of
the permutation operator must be evaluated, and this is done in the
following section.

\subsection{Construction of the states}

The three-body states are constructed in three stages.  First,  we
construct the state of particle 2 and 3 in its rest system, choosing the
momenta so that
${\bf k}_2$ lies in the $xz$ plane with $k_{2x}$ positive, as shown  in
Fig.~4. By convention, particle three is off shell, and requires both
$u$ ($\rho=+$) and
$v$ ($\rho=-$) spinors
to describe its Dirac structure.  This degree of freedom is referred to
as the ``$\rho$-spin'' of the off-shell particle.  Next, we boost
the (23) system to a frame with three-momentum
${\bf q}=-{\bf k}_1$
in the positive $z$ direction, and take the direct product of this state
with the state
of particle one with its three-momentum ${\bf k}_1$  in the negative
$z$ direction.  Finally, we obtain the partial wave states by an angular average
over the Euler angles $\{\Phi,\Theta,\phi\}$, as
defined below.  In shorthand, this three-body state is denoted
$|1(23)\rangle$, to
remind us that particles 2 and 3 are the pair which was boosted from their
rest system.

Begin with the construction of the state for particle 2 with momentum
$|\tilde{\bf k}_2| = \tilde{p}$ pointing in the
positive z-direction, and with helicity $\lambda_2$.  This state will be
denoted by
$|(\tilde{p},0,0),\lambda_2 \rangle$, where the second two arguments in
the parentheses are the polar and azimuthal angles of the momentum.
The state with momentum pointing in an arbitrary direction can be obtained
by applying a rotation operator $R_{\phi,\tilde \theta,\gamma}
=e^{-i\phi J_z}e^{-i\theta J_y} e^{-i\gamma J_z}$ through Euler
angles $\phi$, $\theta$, and $\gamma$.  For vectors without
internal structure, we need only two angles, and following Wick characterize
the states by the polar angles $\tilde \theta$ and $\phi$, and
represent the rotations by $R_{\phi,\tilde \theta,0}$, so that
\begin{equation}
|(\tilde{p},\tilde \theta,\phi),\lambda_2 \rangle  =
R_{\phi,\tilde \theta,0} |(\tilde{p},0,0),\lambda_2 \rangle\, .
\label{Eq1}
\end{equation}
Note that this differs by a phase from the convention adopted in
Jacob and Wick \cite{R3} and used in Ref.~\cite{R1}, where the
rotation was defined to be $R_{\phi,\tilde \theta,-\phi}$ instead of
$R_{\phi,\tilde \theta,0}$.  The phase difference is
\begin{equation}
R_{\phi,\tilde \theta,-\phi} |(\tilde{p},0,0),\lambda_2 \rangle  =
e^{i\phi\lambda_2}R_{\phi,\tilde \theta,0} |(\tilde{p},0,0),
\lambda_2 \rangle\, .
\end{equation}
As discussed in Wick \cite{R2}, the new phase convention turns out
to have significant advantages for the treatment of the three-body
system, and {\it gives identical results if\/} $\phi=0$, where
the two-body states were previously defined \cite{R1}.

The state for particle 3, which {\it in the rest system of the pair\/} has
a momentum of the same magnitude but opposite in direction, is defined
\begin{equation}
|(\tilde{p},\tilde \theta,\phi),\lambda_3 \;\rho\rangle  =
R_{\phi,\tilde \theta,0} |(\tilde{p},\pi,\pi),\lambda_3 \;\rho\rangle\, ,
\label{Eq3}
\end{equation}
where $\rho=\pm$ is the $\rho$-spin of the state (for more details, see the
discussion below), and
the following phase convention is incorporated into the definition of the
state $|(\tilde{p},\pi,\pi),\lambda_3 \;\rho \rangle$:
\begin{equation}
|(\tilde{p},\pi,\pi),\lambda_3 \;\rho \rangle  = e^{-i\pi s_3}
R_{\pi,\pi,0} |(\tilde{p},0,0),\lambda_3 \;\rho \rangle\, ,
\label{Eq2}
\end{equation}
where $s_3$ is the spin of particle 3 (in our case $s_3=1/2$).
The phase factor $e^{-i\pi s_3}$ is precisely what is needed for the
definition Eq.~(\ref{Eq3}) to agree with the phase convention of
Jacob and Wick \cite{R3} for ``particle 2'', which was
used previously in Ref.~\cite{R1}.  To see this, recall that
\begin{equation}
R_{\pi,\pi,0} = e^{-i\pi J_z} e^{-i\pi J_y} = e^{-i\pi J_y} e^{i\pi J_z}
\label{Eq4}
\end{equation}
and hence
\begin{eqnarray}
e^{-i\pi s_3} R_{\pi,\pi,0} |(\tilde{p},0,0),\lambda_3 \;\rho \rangle
& = & e^{-i\pi s_3}  e^{-i\pi J_y} e^{i\pi J_z}
|(\tilde{p},0,0),\lambda_3 \;\rho \rangle
\nonumber \\
& = & e^{-i\pi (s_3-\lambda_3)} R_{0,\pi,0}|(\tilde{p},0,0),\lambda_3
\;\rho\rangle
\nonumber\\
& = &(-1)^{s_3-\lambda_3} R_{0,\pi,0}|(\tilde{p},0,0),\lambda_3 \;\rho
\rangle\, ,
\label{Eq5}
\end{eqnarray}
as used in Ref.~\cite{R1}.  (According to our phase conventions, the value of
$J_z$ is independent of $\rho$; see Eq.~(A9) of Ref.~\cite{R1}.)
The two-particle state (in its rest system) is now written
\begin{eqnarray}
|(\tilde{p},\tilde{\theta},\phi),\lambda_2 \lambda_3 \;\rho \rangle
&&=  R_{\phi,\tilde \theta,0}
|(\tilde{p},0,0),\lambda_2 \lambda_3 \;\rho \rangle \nonumber\\
&&=  R_{\phi,\tilde \theta,0}\, \left(
|(\tilde{p},0,0),\lambda_2 \rangle \otimes
|(\tilde{p},\pi,\pi),\lambda_3 \;\rho \rangle \;\right)\, ,
\label{Eq6}
\end{eqnarray}
where we emphasize that the phase $e^{-i\pi s_3}$ is included in the
definition of
$|(\tilde{p},\pi,\pi),\lambda_3 \;\rho \rangle $, as given in Eq.~(\ref{Eq2}).
Two particle states of definite total angular momentum and total helicity
can be
projected from these general two-particle states by integrating over
the polar and azimuthal angles
\begin{equation}
| \tilde{p} j m, \lambda_2 \lambda_3 \;\rho \rangle
= \eta_j
\int_{0}^{2\pi} d \phi
\int_{0}^{\pi} d \tilde{\theta} \sin \tilde{\theta}\;
{\cal D}^{(j)*}_{m,\lambda_2 -\lambda_3}(\phi,\tilde \theta,0)
R_{\phi,\tilde \theta,0}
|(\tilde{p},0,0),\lambda_2 \lambda_3 \;\rho \rangle \, ,
\label{Eq7}
\end{equation}
where we use the abbreviation
\begin{equation}
\eta_j  =  \left( \frac{2j+1}{4\pi} \right)^{1/2} \, .\label{Eq7a}
\end{equation}

The next step is to boost the two-particle state in the direction of the
positive z-axis such that its total three-momentum becomes $q$.
The required boost operator will be denoted $Z_q$ (it is equal to
$\Lambda^{-1}_q$ of the last section), and the (23) pair can  be treated
like an elementary particle with momentum $q$ in the positive
$z$-direction, with
``spin'' $j$ and
``helicity'' $m$, and a mass $W_{q}$ given by $W_{q}^2=(P-k_1)^2$.
The boosted state of the pair is no
longer an eigenstate of the single-particle helicities, because a boost
which is not in the direction of a particle's momentum mixes helicities.  The three
body helicity state is then constructed by taking a direct product of the
boosted
(23) pair and the state of the single particle 1 with a momentum of
magnitude $q$ in
the negative z-direction.  For consistency, the same phase convention is
used to define
the state of particle 1 that was used before to define particle 3, i.e..
\begin{equation}
|(\tilde{q},\pi,\pi),\lambda_1 \rangle  = e^{-i\pi s_1}
R_{\pi,\pi,0} |({q},0,0),\lambda_1 \rangle\, .
\label{Eq8}
\end{equation}
This gives a three-body helicity state with total 3-momentum zero
and the momentum of the pair in the positive $z$-direction:
\begin{equation}
|(q,0,0), \tilde{p} j m,\lambda_1 (\lambda_2 \lambda_3) \rho \rangle
= |({q},0,0),\lambda_1 \rangle \otimes |\tilde{p} j m, \lambda_2 \lambda_3
\;\rho \rangle \, .  \label{Eq9}
\end{equation}
Following the convention for rotation of states first introduced in
Eq.~(\ref{Eq1}), the state in which the momentum of the pair is in an
arbitrary direction
is obtained from (\ref{Eq9}) by applying the rotation $R_{\Phi,\Theta,0}$
\begin{equation}
|(q,\Theta,\Phi),\tilde{p} j m,\lambda_1 (\lambda_2\lambda_3) \rho \rangle  =
R_{\Phi,\Theta,0} |(q,0,0), \tilde{p} j m,\lambda_1 (\lambda_2 \lambda_3)
\rho \rangle\, .  \label{Eq10}
\end{equation}
Finally, the three-body helicity states with fixed total angular momentum
$J$ and
projection $M$ are obtained from the states (\ref{Eq10}) by the angular average
\begin{eqnarray}
|q J M,  \tilde{p} j m, \lambda_1 (\lambda_2 \lambda_3) \rho \rangle
 =&&\eta_J \! \int_{0}^{2\pi}\!\! d\Phi\int_{0}^{\pi}\!\! d\Theta \sin \Theta
{\cal D}^{(J)*}_{M,m-\lambda_1}(\Phi,\Theta,0)\;
|(q,\Theta,\Phi),\tilde{p},j m,\lambda_1 (\lambda_2\lambda_3) \rho \rangle
\nonumber\\
=&&\eta_J \eta_j
\int_{0}^{2\pi}\! d\Phi
\int_{0}^{\pi} \! d\Theta \sin \Theta
\int_{0}^{2\pi} d\phi
\int_0^{\pi} d\tilde{\theta} \sin\tilde{\theta}\;
{\cal D}^{(J)*}_{M,m-\lambda_1}(\Phi,\Theta,0)
\nonumber \\
&& \times
{\cal D}^{(j)*}_{m,\lambda_2-\lambda_3}(\phi,\tilde{\theta},0 )
R_{\Phi,\Theta,0}\,\left\{
| (q,\pi,\pi),\lambda_1 \rangle \!
\otimes \!
 Z_q R_{\phi,\tilde\theta,0}
|(\tilde{p},0,0), \lambda_2 \lambda_3 \;\rho\rangle \right\} \, ,\nonumber\\&&
\label{Eq11}
\end{eqnarray}
where $\eta_J$ is obtained from Eq.~(\ref{Eq7a}) by replacing $j$ with $J$ .
Note that
this expression contains the two-body partial wave states (\ref{Eq7}),
and if we denote
the rotation $R_{\Phi,\Theta,0}$ by $R_U$, and
$$\int dU= \int_{0}^{2\pi}\!\! d\Phi\int_{0}^{\pi}\!\! d\Theta \sin \Theta\, $$
$${\cal D}^{(J)*}_{M,m-\lambda_1}(\Phi,\Theta,0)=
{\cal D}^{(J)*}_{M,m-\lambda_1}(U)$$
we have
\begin{equation}
|q J M,  \tilde{p} j m, \lambda_1 (\lambda_2 \lambda_3) \rho \rangle
 =\eta_J \! \int dU {\cal D}^{(J)*}_{M,m-\lambda_1}(U)\, R_U \,
  \left\{| (q,\pi,\pi),\lambda_1 \rangle \!
\otimes \! Z_q | \tilde{p} j m, \lambda_2 \lambda_3 \;\rho \rangle \right\}
\, .  \label{Eq11a}
\end{equation}

Another useful form of Eq.~(\ref{Eq11}) is obtained by exploiting the fact
that a rotation
about the $z$-axis commutes with a boost in $z$-direction, so that the
operation
of the rotations on the (23) pair can be written
\begin{equation}
R_{\Phi,\Theta,0} Z_q R_{\phi,\tilde \theta,0} =
R_{\Phi,\Theta,0} R_{0,0,\phi} Z_q R_{0,\tilde \theta,0}
= R_{\Phi,\Theta,\phi} Z_q R_{0,\tilde \theta,0}
\, . \label{Eq12}
\end{equation}
On the other hand, the rotation of particle 1 can be written
\begin{equation}
R_{\Phi,\Theta,0} = R_{\Phi,\Theta,\phi} R_{0,0,-\phi} \to R_{\Phi,\Theta,\phi}
e^{-i\phi\lambda_1}
\, , \label{Eq13}
\end{equation}
where the last step is obtained by letting $R_{0,0,-\phi}$ operate on
$| (q,\pi,\pi),\lambda_1 \rangle$, and recalling that this state is an
eigenstate
of $J_z$ with projection $-\lambda_1$.  Finally, noting that
\begin{eqnarray}
{\cal D}^{(J)*}_{M,m-\lambda_1}(\Phi,\Theta,0)
{\cal D}^{(j)*}_{m,\lambda_2-\lambda_3}(\phi,\tilde{\theta},0 ) =&&
e^{i\lambda_1\phi}{\cal D}^{(J)*}_{M,m-\lambda_1}(\Phi,\Theta,\phi)
{\cal D}^{(j)*}_{m,\lambda_2-\lambda_3}(0,\tilde{\theta},0 )\nonumber\\
=&&e^{i\lambda_1\phi}{\cal D}^{(J)*}_{M,m-\lambda_1}(\Phi,\Theta,\phi)
\,d^j_{m,\lambda_2-\lambda_3}(\tilde{\theta}) \label{Eq13a}
\end{eqnarray}
shows that the factors of $e^{i\lambda_1\phi}$ cancel, and that
Eq.~(\ref{Eq11}) can be written
\begin{eqnarray}
|q J M,  \tilde{p} j m, \lambda_1 (\lambda_2 \lambda_3)  \rho \rangle
 = && \eta_J \eta_j \int dS
\;{\cal D}^{(J)*}_{M,m-\lambda_1}(S)
\int_0^{\pi} d\tilde{\theta} \sin\tilde{\theta}\;
d^{(j)}_{m,\lambda_2-\lambda_3}(\tilde{\theta})\nonumber\\
&&\qquad \times
 R_{S}\,| k_1^o \lambda_1^{\phantom{o}} (k_2^o \lambda_2^{\phantom{o}}
\overline{k_3^o}\lambda_3^{\phantom{o}})\rho \rangle \, ,
\label{Eq14}
\end{eqnarray}
where
\begin{equation}
| k_1^o \lambda_1^{\phantom{o}} (k_2^o \lambda_2^{\phantom{o}}
\overline{k_3^o}\lambda_3^{\phantom{o}})\rho \rangle  =
| (q,\pi,\pi),\lambda_1 \rangle
\otimes Z_q R_{0,\tilde\theta,0}
| (\tilde{p},0,0), \lambda_2 \lambda_3 \;\rho \rangle \label{Eq14a}
\end{equation}
is the three-body state in its {\it canonical configuration\/} in
the $xz$ plane with special
4-momenta $k_1^o,k_2^o,$ and $k_3^o$, as shown in Fig.~4, $R_{S} =
R_{\Phi,\Theta,\phi}$ is the rotation which carries the three  body system from
its canonical  configuration to the most
general orientation described by Euler angles $\Phi,\Theta,$ and $\phi$, and
\begin{eqnarray}
&&\int dS =\int_{0}^{2\pi} d\Phi\int_{0}^{\pi} d\Theta \sin \Theta
\int_{0}^{2\pi} d\phi \nonumber\\
&&{\cal D}^{(J)*}_{M,m-\lambda_1}(S) =
{\cal D}^{(J)*}_{M,m-\lambda_1}(\Phi,\Theta,\phi) \, . \label{Eq15}
\end{eqnarray}
Eq.~(\ref{Eq14a}) shows that the canonical three-body configuration is
constructed by starting from a two-body state in the two-body rest frame where
the relative momentum of the two particles is restricted to the $xz$ plane with
polar angle $\tilde\theta$, then boosting this state in the positive $z$
direction, and finally adding the spectator (particle 1) with momentum along
the  negative $z$-axis.  Since the most general rotation $R_S$ is
performed {\it after\/}  the boost $Z_q$, the Wigner rotations that accompany
the boost  are all rotations about the $y$-axis, which greatly simplifies the
calculation.

The results (\ref{Eq11a}) and (\ref{Eq14}) are equivalent, and either may
be used to evaluate matrix elements.

\subsection{Representation of the states}

In the previous subsection we showed how the states
\begin{equation}
| k_1 \lambda_1 ( k_2 \lambda_2 \overline{k}_3 \lambda_3)
\rho \rangle =
R_{S}\,| k_1^o \lambda_1^{\phantom{o}} (k_2^o \lambda_2^{\phantom{o}}
\overline{k_3^o}\lambda_3^{\phantom{o}})\rho \rangle \, , \label{Eq16}
\end{equation}
introduced abstractly in Eq.~(\ref{Eq16a}), are to be constructed.
These states can also be written as a direct product of the momentum space
plane wave states introduced in Eq.~(\ref{Eq17a}) and Dirac helicity spinors:
\begin{eqnarray}
| k_1 \lambda_1 ( k_2 \lambda_2 \overline{k}_3 \lambda_3)
\rho \rangle =&& e^{-i\pi(s_1+s_3)}\;| k_1 (k_2 \overline{k}_3)
\rangle \nonumber\\
&& \otimes R_S
\,\left(\left[R_{\pi,\pi,0}\, u(q,\lambda_1)\right]_\alpha \, Z_q
R_{0,\tilde{\theta},0}\left\{ u_\beta(\tilde{p},\lambda_2)  \left[R_{\pi,\pi,0}
\, u^\rho(\tilde{p},\lambda_3) \right]_\gamma\right\}\right) \, , \label{Eq17}
\end{eqnarray}
where $\alpha,\beta,$ and $\gamma$ are the Dirac indices of
particles 1,2, and 3, respectively, and all rotations are displayed
explicitly, so that all spinor states in Eq.~(\ref{Eq17}) are
``particle 1'' states (in the sense of Jacob and Wick).  Explicitly
\begin{equation}
u(p,\lambda) = \left(\begin{array}{c} \cosh(\eta/2) \\  \\
2\lambda \sinh(\eta/2)\end{array}\right) \chi(\lambda) \qquad
v(p,\lambda) = \left(\begin{array}{c} - 2\lambda \sinh(\eta/2)\\  \\
\cosh(\eta/2) \end{array}\right) \chi(\lambda)   \, , \label{Eq18}
\end{equation}
with
\begin{equation}
\chi({\scriptstyle{{1\over2}}}) = \left(\begin{array}{c} 1 \\
0 \end{array}\right) \qquad
\chi({\scriptstyle{-{1\over2}}}) = \left(\begin{array}{c} 0 \\
1 \end{array}\right)  \, , \label{Eq19}
\end{equation}
and
\begin{equation}
\tanh\eta={p\over E_p} \, . \label{Eq20}
\end{equation}
Since the helicity spinors depend only on the {\it magnitude\/} of the three
momentum ${\bf p}$, we have used the notation
$v(p,\lambda)=v(-{\bf p},\lambda)$, so that the correspondence given in
Eq.~(\ref{Eq20a}) now becomes
\begin{equation}
u^\rho(p,\lambda)=\cases{u(p,\lambda)
\qquad\hbox{if } \rho=+\cr
	v(p,\lambda) \qquad\hbox{if } \rho=- \; . \cr} \label{Eq21}
\end{equation}
All of these conventions are consistent with our previous work (check
Eq.~(A9) of Ref.~\cite{R1} with $i=1$).

Using the representation (\ref{Eq17}), and the orthogonality relations
(\ref{Eq22}) we obtain {\it generalized\/} orthogonality relations for the
three-particle helicity states (\ref{Eq16})
\begin{eqnarray}
\langle 1' (2'3') \rho' |1(23) \rho \rangle =&&
\langle k'_1 \lambda'_1 ( k'_2 \lambda'_2 \overline{k'_3} \lambda'_3)
\rho'
| k_1 \lambda_1 ( k_2 \lambda_2 \overline{k}_3 \lambda_3) \rho \rangle
\nonumber\\
=&& \langle k'_1 (k'_2 \overline{k'_3})
| k_1 (k_2 \overline{k}_3)\rangle \; \left[\overline{u}(q',\lambda'_1)\,
R_{\pi,\pi,0}^{-1} R_{S'}^{-1}  R_S^{\phantom{-1}}\!\!\!
R_{\pi,\pi,0}^{\phantom{-1}}\,
u(q,\lambda_1) \right] \nonumber\\
&&\times \left[\overline{u}(\tilde{p'},\lambda'_2)
\,R_{0,\tilde{\theta'},0}^{-1}
Z_{q'}^{-1}R_{S'}^{-1}  R_S^{\phantom{-1}}\!\!\! Z_q^{\phantom{-1}}\!\!\!
R_{0,\tilde{\theta},0}^{\phantom{-1}}\, u(\tilde{p},\lambda_2)
\right] \nonumber\\
&&\times \left[\overline{u}^{\rho'}(\tilde{p'},\lambda'_3)\,
R_{\pi,\pi,0}^{-1} R_{0,\tilde{\theta'},0}^{-1}
Z_{q'}^{-1}R_{S'}^{-1} \, R_S^{\phantom{-1}}\!\!\!
Z_q^{\phantom{-1}}\!\!\!  R_{0,\tilde{\theta},0}^{\phantom{-1}}
R_{\pi,\pi,0}^{\phantom{-1}} \, u^\rho(\tilde{p},\lambda_3)
\right] \nonumber\\
=&& 2 E_{k_1} \delta^{3}(k_1-k'_1)
2 E_{k_2} \delta^{3}(k_2-k'_2) \delta^4(P-P') \,\delta_{\lambda'_1\lambda_1}
\,\delta_{\lambda'_2\lambda_2}\,\nonumber\\
&&\qquad\qquad\times
\left[\overline{u}^{\rho'}(\tilde{p},\lambda'_3)\,
u^\rho(\tilde{p},\lambda_3) \right] \, .
\label{Eq23}
\end{eqnarray}
Note that the states are {\it not\/} orthogonal in $\rho$ space.  Using
Eq.~(\ref{Eq18}) gives
\begin{equation}
\left[\overline{u}^{\rho'}(\tilde{p},\lambda'_3)\,
u^\rho(\tilde{p},\lambda_3) \right] = \delta_{\lambda'_3\lambda_3}\,
{\cal O}_{\rho'\rho}(\tilde{p},\lambda_3) \, ,\label{Eq24}
\end{equation}
where, if $\rho=+$ is the first column and $\rho=-$ the second, the matrix
representation of ${\cal O}$ is
\begin{eqnarray}
{\cal O}_{\rho'\rho} ({p},\lambda)=&& \left({\begin{array}{cc}
1 & -2\lambda\sinh\eta\\
-2\lambda\sinh\eta & -1 \end{array}} \right) \nonumber\\
=&& (\tau_3)_{\rho'\rho} -2\lambda \sinh\eta\, (\tau_1)_{\rho'\rho}
 \, , \label{Eq24a}
\end{eqnarray}
where $\sinh \eta=p/m$.
It is also useful to express the covariant product
$2 E_{k_2} \delta^{3}(k_2-k'_2)$ in the rest frame of the two-body
subsystem, where ${\bf k}_2=\tilde{{\bf p}}$ depends on the polar angles
$\tilde\theta$ and $\phi$.
Hence the generalized orthogonality relation will be written
\begin{eqnarray}
\langle 1' (2'3')\rho' |1(23)\rho\rangle =&&
\langle k'_1 \lambda'_1 ( k'_2 \lambda'_2 \overline{k'_3} \lambda'_3)
\rho'
| k_1 \lambda_1 ( k_2 \lambda_2 \overline{k}_3 \lambda_3) \rho \rangle
\nonumber\\
=&& \delta_{\lambda'_1\lambda_1}
\,\delta_{\lambda'_2\lambda_2}\,\delta_{\lambda'_3\lambda_3}
\,{\cal O}_{\rho'\rho}(\tilde{p},\lambda_3) \nonumber\\
&&\qquad \times 2 E_{k_1} \delta^{3}(k_1-k'_1)
2 E_{\tilde{p}} \delta^{3}(\tilde{p}'-\tilde{p}) \delta^4(P'-P) \,  \, .
\label{Eq25}
\end{eqnarray}

Even though the states (\ref{Eq16}) are not orthogonal, the matrix
${\cal O}(\tilde{p},\lambda)$ has a simple property which
will enable us to
carry out the calculation much as it they were.  From
Eq.~(\ref{Eq24a}) we obtain
\begin{equation}
\left[{\cal O}(p,\lambda){\cal O}(p,\lambda)\right]_{\rho'\rho} =
\delta_{\rho'\rho}\,\cosh^2\eta =
\delta_{\rho'\rho}\,{E^2_p\over m^2} \, . \label{Eq25a}
\end{equation}
Using this we can show that the completeness relation for the states can be
written
\begin{eqnarray}
{\bf 1} \equiv &&
{\cal Q}_{\alpha'\alpha}\,
{\cal Q}_{\beta'\beta}\, \delta_{\gamma'\gamma}=
\int \frac{d^3k_1}{2 E_{k_1}}\frac{d^3\tilde{p}}{2 E_{\tilde{p}}}\,
{m^2\over E_{\tilde{p}}^2} \, d^4P \
\sum_{\lambda_1\lambda_2\lambda_3\,\rho'\rho}
|1(23)\rho'\rangle \,{\cal O}_{\rho'\rho}(\tilde{p},\lambda_3)\,
 \langle 1(23)\rho | \nonumber\\
=&& \int \frac{d^3k_1}{2 E_{k_1}}\frac{d^3\tilde{p}}{2 E_{\tilde{p}}} \,
{m^2\over E_{\tilde{p}}^2} \, d^4P \!
\sum_{\lambda_1\lambda_2\lambda_3\,\rho'\rho}
| k_1 \lambda_1 ( k_2 \lambda_2 \overline{k}_3 \lambda_3) \rho' \rangle
\,{\cal O}_{\rho'\rho}(\tilde{p},\lambda_3)\,
\langle k_1 \lambda_1 ( k_2 \lambda_2 \overline{k}_3 \lambda_3) \rho|
 \, ,
\label{Eq26}
\end{eqnarray}
where ${\cal Q}_1={\cal Q}_{\alpha'\alpha}$ is the positive energy
projection operator for particle 1 (with Dirac indices $\alpha$ and
$\alpha'$) introduced in Sec.~IIB.  Eq.~(\ref{Eq26}) tells us that the states
span only the positive energy sectors of particles 1 and 2
(which is sufficient) but they span the entire four-dimensional Dirac space
for the off-shell particle 3.

We will only describe the emergence of the factor
$\delta_{\gamma'\gamma}$ in the derivation of
the completeness relation (\ref{Eq26}).  To see how this factor emerges,
evaluate the sum over $\lambda_3,\rho,$ and $\rho'$ explicitly using
(\ref{Eq18})
\begin{eqnarray}
{m^2\over E_{\tilde{p}}^2}
\sum_{\lambda_3\,\rho'\rho} &&
u^{\rho'}_{\gamma'}(\tilde{p},\lambda_3)\,
{\cal O}_{\rho'\rho}(\tilde{p},\lambda_3)\,
\overline{u}^{\rho}_{\gamma}(\tilde{p},\lambda_3 )\nonumber\\
&&=
{m^2\over E_{\tilde{p}}^2} \left\{
\sum_{\lambda_3\,\rho} \rho\,
u^{\rho}_{\gamma'}(\tilde{p},\lambda_3 )\,
\overline{u}^{\rho}_{\gamma}(\tilde{p},\lambda_3 )-
\sinh\tilde{\eta}
\sum_{\lambda_3,\,\rho \not=\rho'} 2\lambda_3\,u^{\rho'}_{\gamma'}
(\tilde{p},\lambda_3 )\,
\overline{u}^{\rho}_{\gamma}(\tilde{p},\lambda_3 ) \right\}
\nonumber\\
&&= {m^2\over E_{\tilde{p}}^2}\left[ \begin{array}{cc} 1 &
-\tau_3\sinh\tilde{\eta}
\\ \tau_3\sinh\tilde{\eta} & 1 \end{array}\right]_{\gamma'\gamma} +
{m^2\sinh\tilde{\eta}\over E_{\tilde{p}}^2}\left[ \begin{array}{cc}
\sinh{\tilde{\eta}} & \tau_3
\\ -\tau_3 & \sinh\tilde{\eta} \end{array}\right]_{\gamma'\gamma}
\nonumber\\
&&= \delta_{\gamma'\gamma}\,{m^2\over E_{\tilde{p}}^2}\,
\cosh^2\!\tilde{\eta} =
\delta_{\gamma'\gamma}\,  .   \label{Eq27}
\end{eqnarray}
Subsequent operations by the rotations and boosts leave this factor
invariant. In the same way, the sums over $\lambda_1$ and
$\lambda_2$ give the projection operators ${\cal Q}_1$ and ${\cal
Q}_2$.

We will now use these relations to work out the generalized
orthogonality and completeness relations for the partial wave
amplitudes (\ref{Eq14}).

\subsection{Generalized orthogonality and completeness relations}

Using the generalized orthogonality relations (\ref{Eq25}), the definition
(\ref{Eq16}), and the notation
\begin{equation}
| J' 1'' (2''3'')\rho\rangle=
| q'' J' M', \tilde{p}'' j' m'; \lambda''_1 (\lambda''_2 \lambda''_3)
\rho \rangle \, ,  \label{Eq33a}
\end{equation}
we obtain
\begin{eqnarray}
\langle J' 1' (2'3')\rho'|J 1 (23)\rho\rangle =& &
\langle q' J' M', \tilde{p}' j' m'; \lambda'_1 (\lambda'_2 \lambda'_3) \rho' |
q J M, \tilde{p} j m; \lambda_1 (\lambda_2 \lambda_3) \rho \rangle \nonumber\\
=& & \eta_{J'} \eta_{j'}\eta_J \eta_j
\int dS'
\;{\cal D}^{(J')}_{M',m'-\lambda'_1}(S')
\int dS
\;{\cal D}^{(J)*}_{M,m-\lambda_1}(S)
\nonumber \\
& & \times
\int_0^{\pi} d\tilde{\theta}' \sin\tilde{\theta}'\;
d^{(j')}_{m',\lambda'_2-\lambda'_3}(\tilde{\theta}')
\int_0^{\pi} d\tilde{\theta} \sin\tilde{\theta}\;
d^{(j)}_{m,\lambda_2-\lambda_3}(\tilde{\theta})\,
\langle 1' (2'3')\rho' |1(23)\rho\rangle \nonumber\\
=&& \delta_{\lambda'_1\lambda_1}
\,\delta_{\lambda'_2\lambda_2}\,\delta_{\lambda'_3\lambda_3}
\, \eta_{J'} \eta_{j'}\eta_J \eta_j
\int dS'
\;{\cal D}^{(J')}_{M',m'-\lambda_1}(S')
\int dS
\;{\cal D}^{(J)*}_{M,m-\lambda_1}(S) \nonumber\\
&&\times
\int_0^{\pi} d\tilde{\theta}' \sin\tilde{\theta}'\;
d^{(j')}_{m',\lambda_2-\lambda_3}(\tilde{\theta}')
\int_0^{\pi} d\tilde{\theta} \sin\tilde{\theta}\;
d^{(j)}_{m,\lambda_2-\lambda_3}(\tilde{\theta}) \,\nonumber\\
&&\times{\cal O}_{\rho'\rho}(\tilde{p},\lambda_3)\; 2 E_{k_1}
\delta^{3}(k_1-k'_1)
\; 2 E_{\tilde{p}} \delta^{3}(\tilde{p}'-\tilde{p})\; \delta^4(P'-P)
\, . \label{Eq28}
\end{eqnarray}
Writing the $\delta^3$ functions in polar coordinates
\begin{eqnarray}
2 E_{k_1} \delta^{3}(k_1-k'_1)
\; 2 E_{\tilde{p}} \delta^{3}(\tilde{p}'-\tilde{p})
= && 2E_q \frac{\delta(q'-q)}{q^2} \delta(\cos \Theta' - \cos \Theta )
\delta(\Phi' - \Phi) \nonumber\\
&&\times 2E_{\tilde p}
\frac{\delta(\tilde p' - \tilde p)}{\tilde{p}^2}
\delta(\cos \tilde{\theta}' - \cos \tilde{\theta} )
\delta(\phi' - \phi)  \, ,
\label{Eq29}
\end{eqnarray}
allows us to integrate easily over $dS'$ and $d\tilde\theta'$, giving
$S'=S$ and $\tilde\theta'=\tilde\theta$.  The remaining integrals over
$dS$ and $d\tilde\theta$ can then be easily done using the orthogonality
properties of the ${\cal D}$ and $d$ functions.  We obtain
\begin{eqnarray}
\langle J' 1' (2'3')\rho'|J 1 (23)\rho\rangle =&&
\delta_{J' J} \delta_{M' M}
\delta_{j' j} \delta_{m' m}
\delta_{\lambda'_1 \lambda_1}
\delta_{\lambda'_2 \lambda_2}
\delta_{\lambda'_3 \lambda_3}\,
{\cal O}_{\rho'\rho}(\tilde{p},\lambda_3)\,
\nonumber\\ &&\quad\times
2 E_q \frac{ \delta(q-q') }{q^2}
2 E_{\tilde{p}} \frac{ \delta (\tilde{p}-\tilde{p}')}{\tilde{p}^2} \,
\delta^4(P'-P)\, .
\label{Eq30}
\end{eqnarray}

Using Eq.~(\ref{Eq26}) and the orthogonality of the ${\cal D}$ and $d$
functions, the completeness relation for the partial wave states can be
derived
\begin{eqnarray}
{\bf 1} \equiv &&
{\cal Q}_{\alpha'\alpha}\,
{\cal Q}_{\beta'\beta}\, \delta_{\gamma'\gamma}=
\int \frac{q^2 dq}{2 E_q}  \frac{\tilde{p}^2 d\tilde{p}}{2 E_{\tilde{p}}}
\,{m^2\over E_{\tilde{p}}^2} \, d^4P \!\!\!
\sum_{{JMjm\atop \lambda_1 \lambda_2 \lambda_3\,\rho'\rho} } \!\!\!\!
| J 1 (23)\rho'\rangle \,{\cal O}_{\rho'\rho}(\tilde{p},\lambda_3)\,
 \langle J 1 (23)\rho | \nonumber\\
=&&
\int \frac{q^2 dq}{2 E_q}  \frac{\tilde{p}^2 d\tilde{p}}
{2 E_{\tilde{p}}}
\,{m^2\over E_{\tilde{p}}^2} \, d^4P \!\!\!\!\!\!
\sum_{{JMjm\atop \lambda_1 \lambda_2 \lambda_3\,\rho'\rho} }
| q J M, \tilde{p} j m; \lambda_1 (\lambda_2 \lambda_3) \rho' \rangle
\,{\cal O}_{\rho'\rho}(\tilde{p},\lambda_3)\,
\langle q J M, \tilde{p} j m; \lambda_1 (\lambda_2 \lambda_3) \rho |
\, .  \nonumber\\ && \label{Eq31}
\end{eqnarray}
Note that this is consistent with (\ref{Eq30}).

\subsection{Reduction of the equations}

The partial-wave expanded three-body equations can now be obtained
directly from the operator equation (\ref{EqSpec2}).  Restoring the
projection operators  ${\cal Q}$, this equation is
\begin{equation}
{\cal Q}_1{\cal Q}_2\Gamma^1 = 2 {\cal Q}_2 M^1 G_3 {\cal Q}_2 \,
\left[{\cal Q}_1{\cal Q}_2\right]\, {\cal P}_{12}\,\left[{\cal Q}_1{\cal Q}_2\right]\,
{\cal Q}_1{\cal Q}_2 \Gamma^1 \, , \label{Eq32}
\end{equation}
where we have made frequent use of the fact that ${\cal Q}_i$
operates only in the space of particle $i$, and hence commutes with
all operators which operate only on particles $j\ne i$, and the
property ${\cal Q}_i{\cal Q}_i={\cal Q}_i$.   Replacing
the two terms $\left[{\cal Q}_1{\cal Q}_2\right]\otimes {\bf 1}$ with the
completeness relation  (\ref{Eq31}), and using the relation
\begin{equation}
{\cal Q}_1{\cal Q}_2 | J 1 (23)\rho\rangle = | J 1 (23)\rho\rangle\, ,
\label{Eq33}
\end{equation}
we obtain
\begin{eqnarray}
\langle J 1 (23)\rho | \Gamma^1 \rangle =&& 2
\sum_{J'M'j'm'}
\sum_{{\lambda''_1 \lambda''_2 \lambda''_3\,\rho_4\rho_3}
\atop {\lambda'_1 \lambda'_2 \lambda'_3\,\rho_2\rho_1} }
\int \frac{q''^2 dq''}{2 E_{q''}}  \frac{\tilde{p}''^2 d\tilde{p}''}
{2 E_{\tilde{p}''}}
\,{m^2\over E_{\tilde{p}''}^2}
\int \frac{q'^2 dq'}{2 E_{q'}}  \frac{\tilde{p}'^2 d\tilde{p}'}{2
E_{\tilde{p}'}}
\,{m^2\over E_{\tilde{p}'}^2} \nonumber\\
&&\times
\langle J 1 (23)\rho | M^1 G_3 | J 1'' (2''3'')\rho_4\rangle \,
{\cal O}_{\rho_4\rho_3}(\tilde{p}'',\lambda''_3)\, \nonumber\\
&&\times \langle J 1'' (2''3'')\rho_3 | {\cal P}_{12} | J' 1' (2'3')
\rho_2\rangle\,
{\cal O}_{\rho_2\rho_1}(\tilde{p}',\lambda'_3)\;
\langle J' 1' (2'3')\rho_1
|\Gamma^1 \rangle
\, .  \label{Eq34}
\end{eqnarray}
This result anticipates the overall conservation of the total four-momentum,
$P$, and the
conservation of the quantum numbers $J, M, j,$ and $m$ by the operator
$M^1 G_3$.

The matrix elements of the operator $M^1 G_3$ are most
easily evaluated using Eq.~(\ref{Eq11a}).  Recalling that
$M^1 G_3$ operates only in the two-body  subspace (23),
the matrix element is
\begin{eqnarray}
\langle J 1 (23)\rho  | M^1 G_3 | && J' 1'
(2'3')\rho_2\rangle
{\cal O}_{\rho_2\rho_1}(\tilde{p}',\lambda'_3) \nonumber\\
=&& \eta_J\eta_{J'} \int dU \int dU'
{\cal D}^{(J)}_{M,m-\lambda_1}(U)
{\cal D}^{(J')*}_{M',m'-\lambda'_1}(U') \,\nonumber\\
&&\times \langle (q,\pi,\pi),\lambda_1 |
 R^{-1}_U R_{U'}| (q,\pi,\pi),\lambda_1 \rangle \nonumber\\
&&\times \langle \tilde{p} j m, \lambda_2 \lambda_3 \;\rho
| R^{-1}_U Z_q^{-1}
  M^1 G_3  Z_{q'} R_{U'} | \tilde{p}' j' m', \lambda'_2
\lambda'_3 \;\rho_2  \rangle
{\cal O}_{\rho_2\rho_1}(\tilde{p}',\lambda'_3) \, .
\label{Eq35}
\end{eqnarray}
The orthogonality relation  for single particle states is
\begin{eqnarray}
\langle (q,\pi,\pi),\lambda_1 |
 R^{-1}_U R_{U'}| (q,\pi,\pi),\lambda_1 \rangle =&&
\langle (q,\Theta,\Phi), \lambda_1 | (q',\Theta',\Phi'),
\lambda'_1 \rangle  \nonumber\\
=&&
\delta_{\lambda_1 \lambda'_1} 2E_q \delta^{(3)}({q}-{q}')
\nonumber \\
= &&
\delta_{\lambda_1 \lambda'_1} 2E_q
\frac{\delta(q-q')}{q^2}
\delta(\cos \Theta - \cos \Theta' )\delta(\Phi - \Phi') \, . \label{Eq36}
\end{eqnarray}
Inserting this, integrating over $dU'$ (which fixes $U'=U$),  noting
that the  boost
operator $Z_q$ and the rotation operator $R_U$ commute with
$M^1G_3$,  and  carrying
out the $U$ integration using the orthogonality of the
${\cal D}$  functions  gives
\begin{eqnarray}
\langle J 1 (23)\rho  | M^1 G_3 | J' 1' (2'3')\rho_2\rangle
\,
&&{\cal O}_{\rho_2\rho_1}(\tilde{p}',\lambda'_3) \nonumber\\
=&& \delta_{J J'}\delta_{M M'} \delta_{j j'} \delta_{m m'}
\delta_{\lambda_1 \lambda'_1}
2 E_q \frac{ \delta(q-q') }{q^2} \nonumber\\
&&\times \;\langle j (23) \rho |
  M^1 G_3  |  j (2'3') \rho_2 \rangle \,
{\cal O}_{\rho_2\rho_1}(\tilde{p}',\lambda'_3) \, ,
\label{Eq37}
\end{eqnarray}
where we have used the following shortened notation
\begin{equation}
|j(2'3')\rho_2\rangle= | \tilde{p}' j m, \lambda'_2
\lambda'_3 \;\rho_2
\rangle
\label{Eq38}
\end{equation}
and removed from the two-body matrix element of
$M^1 G_3$ the factor of $\delta_{j j'} \delta_{m m'}$.
Now, including the projection operators, the propagator
$G^3_2$  in Dirac space is
\begin{eqnarray}
G^3_2 &&= G_3{\cal Q}_2 \to (m+ \rlap/ k_2)
\frac{(m+ \rlap/ k_3)}{m^2-k^2_3-i \epsilon} \nonumber\\
&&\to 2m\,{\cal Q}_2\,{m \over E_{{\tilde p}'}} \left[
u^{\rho'}( {\tilde p}',\lambda''_3) \, g^{\rho'}(q,{\tilde p}')
\, {\bar u}^{\rho'} ({\tilde p}',\lambda''_3) \right]\,  ,
\label{Eq38a}
\end{eqnarray}
where we inserted the decomposition (\ref{RhoDecomp}) with
$g^\rho(q,\tilde{p})$ given in Eq.~(\ref{Eq20b}).  The only dependence of the 
two-body helicity states $|j (2'3') \rho_2 \rangle$ on $\rho_2$ and $\lambda'_3$
comes from the factor
$u^{\rho_2}({\tilde p}',\lambda'_3)$, which leads to the following
property (where there is no sum over $\rho'$)
\begin{equation}
\left[ u^{\rho'}( {\tilde p}',\lambda''_3) {\bar u}^{\rho'}
({\tilde p}',\lambda''_3) \right]\; |  j (2'3') \rho_2 \rangle =
|  j (2'3'') \rho' \rangle\;\left[{\bar u}^{\rho'}
({\tilde p}',\lambda''_3) u^{\rho_2}( {\tilde p}',\lambda'_3) \right]
\,  .  \label{Eq40}
\end{equation}
Using Eqs.~(\ref{Eq38a}) and (\ref{Eq40}), and recalling Eq.~(\ref{Eq25a}),
the two-body matrix element in (\ref{Eq37}) is reduced as follows
\begin{eqnarray}
\sum_{\rho_2}\langle j (23)&& \rho |
  M^1 G_3  |  j (2'3') \rho_2 \rangle \,
{\cal O}_{\rho_2\rho_1}({\tilde p}',\lambda'_3)\nonumber\\
=&& 2m {m \over E_{{\tilde p}'}}\sum_{\rho'\rho_2\lambda''_3}
\langle j (23) \rho |   M^1 \left[
u^{\rho'}( {\tilde p}',\lambda''_3) \, g^{\rho'}(q,{\tilde p}')
\, {\bar u}^{\rho'} ({\tilde p}',\lambda''_3) \right] |  j (2'3')
\rho_2  \rangle \,
{\cal O}_{\rho_2\rho_1}({\tilde p}',\lambda'_3) \nonumber\\
=&& 2m {m \over E_{{\tilde p}'}}\sum_{\rho'\rho_2\lambda''_3}
\langle j (23) \rho |   M^1 |  j (2'3'') \rho' \rangle \, g^{\rho'}
(q,{\tilde p}')
\left[u^{\rho'}( {\tilde p}',\lambda''_3)
{\bar u}^{\rho_2} ({\tilde p}',\lambda'_3)\right]
{\cal O}_{\rho_2\rho_1}({\tilde p}',\lambda'_3) \nonumber\\
=&& 2m {m \over E_{{\tilde p}'}}\sum_{\rho'\rho_2\lambda''_3}
\langle j (23) \rho |   M^1 |  j (2'3'') \rho' \rangle \, g^{\rho'}
(q,{\tilde p}') \,
\delta_{\lambda''_3\lambda'_3} \,
{\cal O}_{\rho'\rho_2}({\tilde p}',\lambda'_3)
{\cal O}_{\rho_2\rho_1}({\tilde p}',\lambda'_3) \nonumber\\
=&& 2m {E_{{\tilde p}'} \over m} \;
\langle j (23) \rho |   M^1 |  j (2'3') \rho_1 \rangle \, g^{\rho_1}
(q,{\tilde p}')
 \, . \label{Eq39}
\end{eqnarray}
Inserting this result into Eq.~(\ref{Eq34}) gives finally
\begin{eqnarray}
\langle J 1 (23)\rho | \Gamma^1 \rangle =&& 2
\sum_{J'M'j'm'}
\sum_{{\lambda''_2 \lambda''_3\,\rho_3}
\atop {\lambda'_1 \lambda'_2 \lambda'_3\,\rho_2\rho_1} }
\int {q'^2 dq'}\frac{m}{E_{q'}} \int {\tilde{p}''^2 d\tilde{p}''
\over 2E_{\tilde{p}''}}
{m\over E_{\tilde{p}''}}
\int{\tilde{p}'^2 d\tilde{p}'\over 2E_{\tilde{p}'}}
{m^2 \over E^2_{\tilde{p}'}}
\nonumber\\  &&\times
\langle j (23)\rho | M^1| j (2''3'')\rho_3\rangle \,g^{\rho_3}
(q,{\tilde p}'')\,\phantom{{1\over2}}\nonumber\\\
&&\times
\langle J 1 (2''3'')\rho_3 | {\cal P}_{12} | J' 1' (2'3')\rho_2\rangle\,
{\cal O}_{\rho_2\rho_1}(\tilde{p}',\lambda'_3)\;
\langle J' 1' (2'3')\rho_1 |\Gamma^1 \rangle
\, .  \label{Eq41}
\end{eqnarray}
In Appendix A we show that
\begin{equation}
\langle j (23)\rho | M^1| j (2''3'')\rho_3\rangle =
{E_{\tilde{p}} E_{\tilde{p}''} \over (2\pi)^3 m^2}
M^{\rho\rho_3\;j}_{\lambda_2\lambda''_2,
\lambda_3\lambda''_3}(\tilde{p},\tilde{p}'';P_{23}) \, , \label{Eq42}
\end{equation}
where $M^{\rho\rho_3\;j}_{\lambda_2\lambda''_2,
\lambda_3\lambda''_3}(\tilde{p},\tilde{p}'';P_{23})$ are the  {\it
two-body  amplitudes
previously determined\/} from Ref.~\cite{R1}, Eq.~(2.88).

To obtain the final three-body equations, the matrix elements of the
permutation operator must be evaluated, which will be done in the following
section.

\section{The permutation operator ${\cal P}_{12}$}

In this section we will derive the matrix elements of the operator 
${\cal P}_{12}$ which interchanges the states of particles 1 and 2.  Using the
shorthand notation given in Eq.~(\ref{Eq33a}), the action of the permutation
operator is
\begin{equation}
{\cal P}_{12} |J 1 (23)\rho\rangle =  |J 2(13) \rho \rangle
\, , \label{Eq4.1}
\end{equation}
where the relation of the momenta of the individual
particles to the relative momenta $q$ and $\tilde{p}$, and to the
quantum
numbers $j$ and $m$, is  unambiguously determined by the order  in
which the single-particle names are written. For example, the state
$|J 2(13) \rho \rangle$ is one in which the second particle is the
spectator with momentum $q$, and particles 1 and 3 are the pair with
angular momentum $j$ and $m$, and particle 1 has CM momentum
variables $\tilde{p}$ and $\tilde{\theta}$.  More precisely, from
the result (\ref{Eq14}), the  state   $|J 1(23)\rho\rangle$  is
obtained by averaging over rotations $R_{S'}$ of a state with the
canonical configuration $k'^o_1,k'^o_2,k'^o_3$, shown in Fig.~5a,
and the state  $|J 2(13) \rho\rangle$ is
obtained by averaging over rotations $R_{S''}$ of a state with the
canonical configuration $k''^o_1,k''^o_2,k''^o_3$, shown in Fig.~5b.
Note that the two configurations are related by interchanging
particles 1  and 2, but the definition of the states requires that
${\bf k}'^o_1$ and ${\bf k}''^o_2$  both be in the negative
$z$-direction, and that ${\bf k}'^o_2$ and ${\bf k}''^o_1$  both lie
in the $xz$-plane with positive $x$ component.

%
%
\begin{figure}[t]
\begin{center}
\mbox{
   \epsfysize=1.5in
\epsfbox{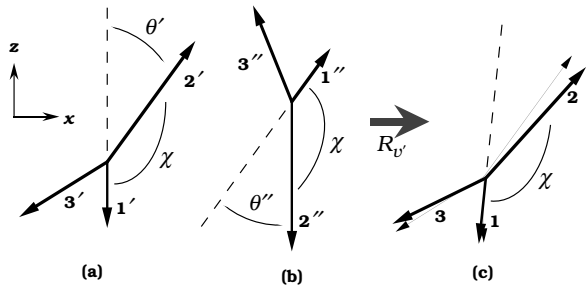}
}
\end{center}
\caption{(a) The momenta $k'^o_1,k'^o_2,k'^o_3$ in their canonical
configuration. (b) The canonical configuration of the momenta
$k''^o_1,k''^o_2,k''^o_3$ which result from the interchange of particles
1 and 2.  (c) Figure showing how $k''^o_1,k''^o_2,k''^o_3$ are rotated
into $k_1,k_2,k_3$ by the rotation $R_{V'}$ which equals $R_{V}$ only
when $k_1,k_2,k_3$ line up precisely with $k'^o_1,k'^o_2,k'^o_3$.}
\label{x5}
\end{figure}

Any three-body configuration with canonical
orientations in the $xz$-plane can be
completely characterized by three variables.  For the vectors in Fig.~5a,
these variables can be chosen to be $|{\bf k}'_1|=q'$,
$|\tilde{{\bf k}}'_2|=\tilde{p}'$, and the angle $\chi$ between
${\bf k}'_1$ and ${\bf k}'_2$ (recall that $|\tilde{{\bf k}}'_2|$ is
the vector ${\bf k}'_2$ in the c.m. of the pair).  For the
configuration in Fig.~5b, the corresponding variables are $|{\bf
k}''_2|=q''$, $|\tilde{{\bf k}}''_1|=\tilde{p}''$, and the same
angle $\chi$.  If the total c.m. energy of the three-body system is
fixed, then there is a constraint between these three variables,
leaving only two independent.  If $q'$ and $\tilde{p}'$ are
specified, the angle $\chi$ can be determined, and this was the
approach taken by one of us previously \cite{R5}.  However, the
final equations are more tractable if $q'$ and $\chi$ are specified,
and $\tilde{p}'$ is determined by the constraints, and this is the
approach we will take below.

Examination of the
two configurations shown in Figs.~5a and 5b shows that the rotation
$R_V=R_{\pi,\chi,0}$ (not to be confused with the rotation $R_{V'}$ discussed
below which carries the configuration shown in Fig.~5b into 5c) will bring them
into alignment, or
\begin{eqnarray}
R_Vk''^o_1&&=k'^o_1 \nonumber\\
R_Vk''^o_2&&=k'^o_2 \nonumber\\
R_Vk''^o_3&&=k'^o_3 \, .  \label{Eq4.2}
\end{eqnarray}
Furthermore, since the final momenta $k_1,k_2,$ and $k_3$ can be
obtained either by rotating the ($k'^o$)'s through $R_{S'}$, or the
($k''^o$)'s through $R_{S''}$ (because they are equal), we have the
relation %
\begin{equation}
k_1=R_{S'}k'^o_1=R_{S''}k''^o_1=R_{S''}R_Vk'^o_1 \label{Eq4.3}
\end{equation}
which implies
\begin{equation}
R_{S'}=R_{S''}R_V \, . \label{Eq4.4}
\end{equation}
This rotation $R_V$ will eventually emerge from the derivation
below.

\subsection{Initial reduction of the matrix element}

We now turn to the details of the  evaluation of the matrix element.
Using the form (\ref{Eq14}) for the three-body state,  the matrix
element of ${\cal P}_{12}$ can be written
\begin{eqnarray}
\langle J' 1' (2'3')\rho'| {\cal P}_{12} |J 1 (23)\rho\rangle =&&
\langle q' J' M',  \tilde{p}' j'
m', \lambda'_1 (\lambda'_2 \lambda'_3)  \rho' | {\cal P}_{12} |q J M,
\tilde{p} j
m, \lambda_1 (\lambda_2 \lambda_3)  \rho \rangle \nonumber\\
=&&
\langle q' J' M',  \tilde{p}' j'
m', \lambda'_1 (\lambda'_2 \lambda'_3)  \rho' | q'' J M,
\tilde{p}'' j
m, \lambda_2 (\lambda_1 \lambda_3)  \rho \rangle \nonumber\\
= && \eta_{J'} \eta_J \eta_{j'} \eta_j \int dS' \int dS'' \;
{\cal D}^{(J')}_{M',m'-\lambda'_1}(S')
{\cal D}^{(J)*}_{M,m-\lambda_2}(S'') \nonumber\\
&&\times \int_0^{\pi} d\tilde{\theta}' \sin\tilde{\theta}'
\int_0^{\pi} d\tilde{\theta}'' \sin\tilde{\theta}''\;
d^{(j')}_{m',\lambda'_2-\lambda'_3}(\tilde{\theta}')
d^{(j)}_{m,\lambda_1-\lambda_3}(\tilde{\theta}'')\nonumber\\
&&\times
\langle k'^o_1 \lambda'^{\phantom{o}}_1 (k'^o_2
\lambda'^{\phantom{o}}_2
\overline{k'^o_3}\lambda'^{\phantom{o}}_3)\rho'|\,
 R^{-1}_{S'}R_{S''}\,| k''^o_2 \lambda_2^{\phantom{o}} (k''^o_1
\lambda_1^{\phantom{o}}  \overline{k''^o_3}\lambda_3^{\phantom{o}})
\rho
\rangle \, . \label{Eq4.5}
\end{eqnarray}
Hence the matrix element depends only on the rotation
$R^{-1}_{S'S''}$.  This rotation will be equal to $R_V$ after
the constraints imposed by the evaluation of the
$\langle\; k'^o\;|\; k''^o \; \rangle$ matrix element have been
realized, but until then this rotation will be denoted
$R_{V'}=R_{\alpha,\chi',\beta}$. Hence
$R_{S''}=R_{S'}R_{S'}^{-1}R_{S''}=R_{S'}R_{V'}$, and using the group
properties of the rotation matrices
\begin{equation}
{\cal D}^{(J)*}_{M,m-\lambda_2}(S'') =
\sum_{\Lambda}
{\cal D}^{(J)*}_{M,\Lambda}(S')
{\cal D}^{(J)*}_{\Lambda,m-\lambda_2}(V') \, . \label{Eq4.6}
\end{equation}
The invariance of the group integration insures that
\begin{equation}
\int dS'' = \int dV' \, , \label{Eq4.7}
\end{equation}
and the orthogonality relation for the ${\cal D}$ functions,
\begin{equation}
\int dS'
{\cal D}^{(J')}_{M',m'-\lambda'_1}(S')
{\cal D}^{(J)*}_{M,\Lambda}(S')
=\delta_{J' J}
\delta_{M' M}
\delta_{m'-\lambda'_1,\Lambda}\, \frac{2\pi}{\eta_J^2} \, ,
\label{Eq4.8}
\end{equation}
allows the reduction of (\ref{Eq4.5}) to
\begin{eqnarray}
\langle J' 1' (2'3')\rho'| {\cal P}_{12} |J 1 (23)\rho\rangle = &&
2\pi\,\delta_{J'J}\delta_{M'M} \eta_{j'} \eta_j\int dV' \;
 {\cal D}^{(J)*}_{m'-\lambda'_1,m-\lambda_2}(V') \nonumber\\
&&\times \int_0^{\pi} d\tilde{\theta}' \sin\tilde{\theta}'
\int_0^{\pi} d\tilde{\theta}'' \sin\tilde{\theta}''\;
d^{(j')}_{m',\lambda'_2-\lambda'_3}(\tilde{\theta}')
d^{(j)}_{m,\lambda_1-\lambda_3}(\tilde{\theta}'')\nonumber\\
&&\times
\langle k'^o_1 \lambda'^{\phantom{o}}_1
(k'^o_2\lambda'^{\phantom{o}}_2
\overline{k'^o_3}\lambda'^{\phantom{o}}_3)\rho'|\,
 R_{V'}\,| k''^o_2 \lambda_2^{\phantom{o}} (k''^o_1
\lambda_1^{\phantom{o}}  \overline{k''^o_3}
\lambda_3^{\phantom{o}})\rho
\rangle \, . \label{Eq4.9}
\end{eqnarray}

We now define the vectors
\begin{eqnarray}
R_{V'}k''^o_1&&=k_1 \nonumber\\
R_{V'}k''^o_2&&=k_2 \nonumber\\
R_{V'}k''^o_3&&=k_3   \label{Eq4.10}
\end{eqnarray}
(where $k_1$ is not lined up along the negative $z$ axis
and equal to $k'^o_1$ until $R_{V'}=R_V$).  This rotation of the vectors $k_i''$
into the vectors $k_i$ is represented in Fig.~5c.  Guided by the discussion
leading up to Eq.~(\ref{Eq4.2}) and  the representation (\ref{Eq17}) for the
three-body states, the matrix element involving $R_{V'}$ is a product of a plane
wave  momentum space matrix element and Dirac space matrix
elements
\begin{eqnarray}
\langle k'^o_1 \lambda'^{\phantom{o}}_1
(k'^o_2 && \lambda'^{\phantom{o}}_2
\overline{k'^o_3}\lambda'^{\phantom{o}}_3)\rho'|\,
R_{V'}\,| k''^o_2 \lambda_2^{\phantom{o}} (k''^o_1
\lambda_1^{\phantom{o}}  \overline{k''^o_3}
\lambda_3^{\phantom{o}})\rho
\rangle\nonumber \\
&&=\langle k'^o_1 (k'^o_2 \overline{k'^o_3}) |\, R_{V'}\,
| k''^o_2  (k''^o_1 \overline{k''^o_3}) \rangle \times
{\cal U} \nonumber\\ &&=
2E_{k_1}  \delta^{(3)}(k'^o_1 - k_1)
2E_{k_2}  \delta^{(3)}(k'^o_2 - k_2)\delta^{(4)}(P'-P)\;
{\cal U} \, , \label{Eq4.11}
\end{eqnarray}
where
\begin{eqnarray}
{\cal U}=&& \left[ e^{i\pi s_1} \overline{u}(q',\lambda'_1)\,
R^{-1}_{\pi,\pi,0} R_{V'} Z_q R_{0,\tilde{\theta},0}\,
u(\tilde{p},\lambda_1)\right] \nonumber\\
&&\times
\left[ e^{-i\pi s_2} \overline{u}(\tilde{p}',\lambda'_2) \,
R^{-1}_{0,\tilde{\theta}',0}Z^{-1}_{q'} R_{V'}R_{\pi,\pi,0}\,
u(q,\lambda_2)\right] \nonumber\\
&&\times
\left[\overline{u}^{\rho'}(\tilde{p}',\lambda'_3)\,
R^{-1}_{\pi,\pi,0} R^{-1}_{0,\tilde{\theta}',0} Z^{-1}_{q'}
R_{V'}Z_q R_{0,\tilde{\theta},0} R_{\pi,\pi,0}\,
u^\rho(\tilde{p},\lambda_3) \right] \nonumber\\
=&& {\cal U}^{(1)}_{\lambda'_1\lambda_1}\,
{\cal U}^{(2)}_{\lambda'_2\lambda_2}\,
{\cal U}^{(3)}_{\lambda'_3\rho', \lambda_3\rho } =
\langle\!\langle
\tilde{p}', \lambda'_1(\lambda'_2\lambda'_3)  \rho'|
\tilde{p}, \lambda_2(\lambda_1\lambda_3) \rho
\rangle\!\rangle    \label{Eq4.11a}
\end{eqnarray}
will be referred to as the {\it reduced\/} matrix element
of the  permutation operator.  Note that $q=|{\bf k}_1|$ and $q'=|{\bf k}'_1|$
(as above).  The matrix element (\ref{Eq4.11a}) contains Wigner 
rotations which result from the fact that the helicities
$\{\lambda'_i\}$ and  $\{\lambda_i\}$ are defined in
different frames.

We first turn to the evaluation of the $\delta$ functions on
the  right  hand side of Eq.~(\ref{Eq4.11}).

\subsection{Evaluation of the $\delta$ functions}

In Appendix B it is shown that the two delta functions can
be written
\begin{eqnarray}
2 E_{k_1} \delta^{(3)}( k'^o_1 - k_1) 
2 E_{k_2} \delta^{(3)}(k'^o_2 -k_2) =
&&4 E_{\tilde{p}_0}E_{\tilde{p}'_0}
\delta (\beta)
\delta (\alpha-\pi)
\frac{ \delta \left( \tilde{p} - \tilde{p}_0
\right)} {\tilde{p}^2_0} \frac{ \delta \left(\tilde{p}' -
\tilde{p}_0\right)}  {\tilde{p}'^2_0} \nonumber\\
&&\times \delta \left(\cos\tilde{\theta} -\cos \tilde{\theta}_0 \right)
\delta \left(\cos \tilde{\theta}' - \cos \tilde{\theta}_0\right)  \, ,
\label{Eq4.12}
\end{eqnarray}
with
\begin{eqnarray}
\tilde{p}_0=\tilde{p}_0(q,q',\chi) &&\qquad
\tilde{p}'_0=\tilde{p}_0(q',q,\chi) \nonumber\\
\tilde{\theta}_0=\tilde{\theta}_0(q,q',\chi) &&\qquad
\tilde{\theta}'_0=\tilde{\theta}_0(q',q,\chi) \, , \label{Eq4.15x}
\end{eqnarray}
and
\begin{eqnarray}
&&\tilde{p}_0(q,q',\chi)=\sqrt{
\left[
\frac{(M_t-E_q)E_{q'} + q q' \cos \chi}{W_q}
\right]^2 - m^2}  \nonumber\\
&&\cos \left\{\tilde{\theta}_0(q,q',\chi) \right\}=
\frac{ W_q E_{q'} - (M_t - E_q)\,E_{\tilde{p}_0(q,q',\chi)} }
{ q\,\tilde{p}_0(q,q',\chi)}
\, . \label{Eq4.13}
\end{eqnarray}
The first two $\delta$ functions insure that the rotation
$R_{V'}=R_{\alpha,\chi', \beta}$ is now $R_{\pi,\chi',0}$, and
the delta functions in $\tilde{p}$
and $\tilde{p}'$ fix the angle $\chi'$ to $\chi$.  The angle $\chi$
will remain a variable,
since we prefer to express the ``allowed''  magnitudes of the
momenta $\tilde p$ and $\tilde p'$ as functions of $\chi$ rather
than the
other way round.

We now combine the expressions (\ref{Eq4.9}), (\ref{Eq4.11}), and
(\ref{Eq4.12}) and
insert the result into the three-body equation (\ref{Eq41}).  In doing this
we must be careful to change the arguments of the matrix element
(\ref{Eq4.11}), which is expressed in terms of 
$\langle p',q',\lambda_1'\lambda_2'\lambda_3' | p, q,\lambda_1\lambda_2\lambda_3
\rangle$, to $\langle p'',q,\lambda_1\lambda''_2\lambda''_3 | p',
q',\lambda'_1\lambda'_2\lambda'_3 \rangle$,
so as to agree with the labeling used in Eq.~(\ref{Eq41}).  Carrying out the
$d\tilde{p}'$ and
$d\tilde{p}$ integrations then gives
\begin{eqnarray}
\langle J 1 (23)\rho | \Gamma^1 \rangle &&=
\sum_{j'm'}\sqrt{2j+1}\sqrt{2j'+1}\!\!\!
\sum_{{\lambda''_2 \lambda''_3\,\rho_3}
\atop {\lambda'_1 \lambda'_2 \lambda'_3\,\rho_2\rho_1} }
\int {q'^2 dq'}{m\over E_{q'}} \int_0^\pi d\chi \sin\chi \;
d^{(J)}_{m-\lambda_1,m'-\lambda'_2}(\chi) \nonumber\\
&&\times\langle j (23)\rho | M^1| j (2''3'')\rho_3\rangle
\, g^{\rho_3}(q,{\tilde p}'')\,e^{i\pi(m-\lambda_1)}
d^{(j)}_{m,\lambda''_2-\lambda''_3}(\tilde{\theta}'')
d^{(j')}_{m',\lambda'_1-\lambda'_3}(\tilde{\theta}')
\nonumber\\
&&\times {m\over E_{\tilde{p}''}}
{m^2 \over E^2_{\tilde{p}'}}\,
\langle\!\langle {\tilde p}'', \lambda_1(\lambda''_2
\lambda''_3)\rho_3|{\tilde p}', \lambda'_2(\lambda'_1\lambda'_3)
\rho_2
\rangle\!\rangle\; {\cal O}_{\rho_2\rho_1}({\tilde p}',\lambda'_3)\;
\langle J' 1' (2'3')\rho_1 |\Gamma^1 \rangle   \label{Eq4.14}
\end{eqnarray}
where
\begin{eqnarray}
\tilde{p}'=&&\tilde{p}_0(q',q,\chi)\qquad
\tilde{p}''=\tilde{p}_0(q,q',\chi)  \nonumber\\
\tilde{\theta}'=&&\tilde\theta_0(q',q,\chi) \qquad
\tilde{\theta}''=\tilde\theta_0(q,q',\chi) \,  , \label{Eq4.15}
\end{eqnarray}
and $\langle\!\langle {\tilde p}'', \lambda_1(\lambda''_2
\lambda''_3)\rho_3|{\tilde p}',
\lambda'_2(\lambda'_1\lambda'_3)\rho_2
\rangle\!\rangle$ is the reduced
matrix element defined in Eq.~(\ref{Eq4.11a}).
This matrix element is calculated in the next subsection.

\subsection{Wigner rotations and the reduced matrix element}

%
%
\begin{figure}[t]
\begin{center}
\mbox{
   \epsfysize=1.5in
\epsfbox{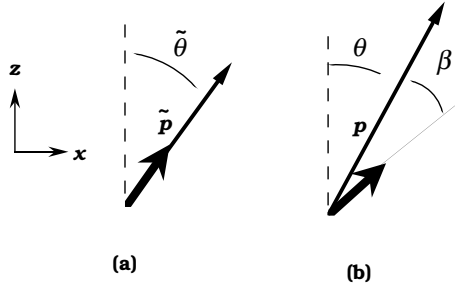}
}
\end{center}
\caption{(a) The canonical configuration of momentum
for the calculation of the Wigner rotation.  The helicity is
$+{1\over2}$ in this example. (b) The transformation of momentum and spin
after the boost in the $+\hat z$ direction.  The spin is now no longer
aligned with the momentum, but rotated by angle $\beta$
with respect to it.}
\label{x6}
\end{figure}

It will be sufficient to define a Wigner rotation only for the 
special case when a spinor with helicity $\lambda$ and 
three-momentum in the right half of the $xz$ plane is boosted in the
positive $z$-direction, as shown in Fig.~6.  The boost is denoted by
$Z_q$ [defined in Eq.~(\ref{B2})], the initial three-momentum of
the state by
$\tilde{{\bf p}}$ (with  magnitude $\tilde{p}$ and polar angle
$\tilde{\theta}$), the final three-momentum by ${\bf p}$ (with
magnitude $p=q'$ and polar angle $\theta=\pi-\chi$), so that the 
Wigner  rotation
${\cal R}(q,q',\chi)$ is defined by the relation
\begin{equation}
Z_q u(\tilde{{\bf p}},\lambda) =
Z_q R_{0,\tilde\theta,0} L_{\tilde{p}}
u(0,\lambda)= R_{0,\theta,0} L_p {\cal R}(q,q',\chi)
u(0,\lambda)\, , \label{Eq5.1}
\end{equation}
where the representation of the pure boosts $L_k$ (for $k= p$ or
$\tilde{p}$) in four-dimensional space-time is
\begin{equation}
L_k=\left(\begin{array}{cccc} \cosh\eta_k & & & \sinh\eta_k \\
&\phantom{\cosh} 1 & & \\ & & \phantom{\cosh}1 & \\
\sinh\eta_k & & & \cosh\eta_k
\end{array} \right) \, , \label{Eq5.2}
\end{equation}
with
\begin{equation}
\tanh\eta_k = {k\over E_k}\, . \label{Eq5.3}
\end{equation}
In Appendix C we show that ${\cal R}$ is a pure rotation about the
$y$-axis,
\begin{equation}
{\cal R}(q,q',\chi)=
R_{0,\beta,0}\, , \label{Eq5.4}
\end{equation}
and find the general equation for $\cos\beta$ as a function of
$q,q',$ and $\chi$.  Since $0\ge \beta \ge \pi$,
$\beta$ is uniquely determined by its cosine.
Using the result (\ref{Eq5.4}), we have
\begin{equation}
Z_q u(\tilde{{\bf p}},\lambda) =
\sum_\nu u({\bf p},\nu)\, d^{(1/2)}_{\nu\lambda}(\beta)\, .
\label{Eq5.5}
\end{equation}

We now are ready to evaluate each of the matrix elements in
Eq.~(\ref{Eq4.11a}), but we will make the substitution $\langle
p',q',\lambda_1'\lambda_2'\lambda_3' | p, q,\lambda_1\lambda_2\lambda_3
\rangle  \to \langle p'',q,\lambda_1\lambda''_2\lambda''_3 | p',
q',\lambda'_1\lambda'_2\lambda'_3 \rangle$ so as to agree with the labeling
used in Eq.~(\ref{Eq4.14}).  Noting that $k_1=k'^o_1$ implies that
${p}'=q$ and $\theta'+\chi=\pi$ (see Fig.~5) the
matrix element for particle 1 becomes
\begin{eqnarray}
{\cal U}^{(1)}_{\lambda_1 \lambda'_1}=&&\left[ e^{i\pi s_1}
\overline{u}(q,\lambda_1)\, R^{-1}_{\pi,\pi,0} R_{V}
Z_{q'} R_{0,\tilde{\theta}',0}\, u(\tilde{p}',\lambda'_1)\right]
\nonumber\\ =&&e^{i\pi s_1} \sum_\nu \left[
\overline{u}(q,\lambda_1)\, R^{-1}_{0,\pi,0}R_{0,\chi,0}
R_{0,{\theta}',0}\, u({p}',\nu)\right] \,
d^{(1/2)}_{\nu\lambda'_1}(\beta_1) \nonumber\\
=&& e^{i\pi s_1}\sum_\nu \left[
\overline{u}(q,\lambda_1) u(q,\nu)\right]
\, d^{(1/2)}_{\nu\lambda'_1}(\beta_1) \nonumber\\
=&& e^{i\pi s_1} d^{(1/2)}_{\lambda_1 \lambda'_1}(\beta_1)
\, , \label{Eq5.7}
\end{eqnarray}
where, using the function $\beta$ defined in Appendix C,
Eq.~(\ref{Cwr}),
\begin{equation}
\beta_1 = \beta(q',q,\chi) \,  ,  \label{Eq5.8}
\end{equation}
Similarly, to evaluate the matrix element for particle 2 use
$\theta''=\pi-\chi$, ${p}''=q'$, and
\begin{eqnarray}
R_{\alpha,\pi,0}=&&R_{0,\pi,-\alpha}\nonumber\\
R_{\pi,\alpha,0}=&&R_{0,-\alpha,\pi}\, , \label{Eq5.6a}
\end{eqnarray}
which gives
\begin{eqnarray}
{\cal U}^{(2)}_{\lambda''_2\lambda'_2}=&&
\left[ e^{-i\pi s_2} \overline{u}(\tilde{p}'',\lambda''_2) \,
R^{-1}_{0,\tilde{\theta}'',0}Z^{-1}_{q} R_{V}
R_{\pi,\pi,0}\, u(q',\lambda'_2)\right] \nonumber\\
=&& e^{-i\pi s_2}\sum_\nu \left[ \overline{u}(q',\nu) \,
R^{-1}_{0,{\theta}'',0} R_{\pi,\chi,0}R_{\pi,\pi,0}
\, u(q',\lambda'_2)\right]\,
d^{(1/2)}_{\nu\lambda''_2}(\beta_2)\nonumber\\
=&& e^{-i\pi s_2} \sum_\nu \left[ \overline{u}(q',\nu) \,
R_{0,0,-2\pi}\; u(q',\lambda'_2)\right]
\, d^{(1/2)}_{\nu\lambda''_2}(\beta_2) \nonumber\\
=&& e^{-i\pi( s_2-2\lambda'_2)}
d^{(1/2)}_{\lambda'_2 \lambda''_2}(\beta_2)
= e^{-i\pi( s_2-2\lambda'_2)}
d^{(1/2)}_{\lambda''_2 \lambda'_2}(-\beta_2)
  \, , \label{Eq5.9}
\end{eqnarray}
where
\begin{equation}
\beta_2 = \beta(q,q',\chi) \,  .  \label{Eq5.10}
\end{equation}

Calculation of the matrix element for particle 3 is
complicated by the fact that its physical four-momentum is
off-shell, while the four-momenta used in the definition of the
$u^\rho$ spinors are on-shell. However, as shown in
Eq.~(\ref{Eq20a}), the four-momentum of the {\it negative energy\/}
spinor is identical to the four-momentum of the on-shell particle in
the interacting pair, and an efficient way to
proceed is to first express both of the $u^\rho$ spinors in terms of
$u^-$.  Then it will turn out that the boosts of both $\rho$-spin
states can be evaluated in terms of quantities related to the
on-shell particle in the interacting pair.

To this end, note that
$u^+$ can be expanded in terms of $u^-$ and $\gamma^5\,u^-$
\begin{equation}
u^+(p,\lambda)=\gamma^5 \left[{E_p\over m} +2\lambda {p\over m}\,
\gamma^5 \right] u^-(p, \lambda)
\, , \label{Eq5.14}
\end{equation}
where the matrix
$$\gamma^5=\left(\begin{array}{cc} 0&1\\1&0\end{array}
\right)$$
commutes with all rotations and boosts.  Then, using the fact that
\begin{equation}
R_{0,\pi,0}=e^{-i\pi\gamma^5\alpha_2/2}=-i\gamma^5\alpha_2
\, , \label{Eq5.11}
\end{equation}
where $\alpha_2$ is the Dirac matrix, and using the explicit
form of the spinors given in Eq.~(\ref{Eq18}), we obtain
\begin{equation}
R_{\pi,\pi,0}\,u^-(p,\lambda)=e^{i\pi s_3}\gamma^5
u^+(p, -\lambda) \, . \label{Eq5.13}
\end{equation}
Combining this with Eq.~(\ref{Eq5.14}) gives a simple formula for
$R_{\pi,\pi,0}\,u^+(p,\lambda)$.  These two relations will be
summarized
\begin{equation}
R_{\pi,\pi,0}\,u^\rho(p,\lambda)=e^{i\pi s_3} \,N_\rho(\lambda)
\, u(p, -\lambda) \, , \label{Eq5.16}
\end{equation}
where $u^+=u$ is implied and
\begin{equation}
N_+(p,\lambda)= \left[{E_p\over m} +2\lambda  {p\over m}\, \gamma^5
\right]\, ,  \qquad N_-(p,\lambda)=\gamma^5\, .
\label{Eq5.17}
\end{equation}

These relations can now be used to evaluate
the matrix element ${\cal U}^{(3)}$
\begin{eqnarray}
{\cal U}^{(3)}_{\lambda''_3\rho',\lambda'_3\rho}
=&& \left[\overline{u}^{\rho'}(\tilde{p}'',\lambda''_3)\,
R^{-1}_{\pi,\pi,0} R^{-1}_{0,\tilde{\theta}'',0} Z^{-1}_{q}
R_{\pi,\chi, 0}Z_{q'} R_{0,\tilde{\theta}',0} R_{\pi,\pi,0}\,
u^{\rho}(\tilde{p}',\lambda'_3) \right] \nonumber\\
=&&\left[\overline{u}(\tilde{p}'',-\lambda''_3)\,
\tilde{N}_{\rho'}(\tilde{p}'',\lambda''_3) R^{-1}_{0,\tilde{\theta}'',0}
Z^{-1}_{q}  R_{\pi,\chi, 0}Z_{q'} R_{0,\tilde{\theta}',0}
N_{\rho}(\tilde{p}',\lambda'_3)\,u(\tilde{p}',-\lambda'_3) \right]
\nonumber\\ =&&\sum_{\nu\nu'}\left[\overline{u}(q',-\nu')\,
R^{-1}_{0,\pi-\chi,0}
\tilde{N}_{\rho'}(\tilde{p}'',\lambda''_3)\,N_\rho(\tilde{p}',\lambda'_3)
 R_{\pi,\chi, 0} R_{0,\pi-\chi,0}\,u(q,-\nu) \right]\nonumber\\
&&\qquad\times
d^{(1/2)}_{-\nu',-\lambda''_3}(\beta_2)\,d^{(1/2)}_{-\nu,-\lambda'_3}
(\beta_1)\,  ,\label{Eq5.18}
\end{eqnarray}
where $\tilde{N}_\rho=\gamma^0\,N_\rho\,\gamma^0$ and, because we
were able to write the states $R_{\pi,\pi,0}\,u^\rho$ in terms of the
positive energy on-shell spinors $u$ using Eq.~(\ref{Eq5.16}), the
matrix elements of particle 3 have been expressed in terms of the
Wigner rotations which already appeared in the treatment of particles
1 and 2.  Since the $N$ factors commute with the rotations, the
matrix element can be further simplified as follows
\begin{eqnarray}
{\cal U}^{(3)}_{\lambda''_3\rho',\lambda'_3\rho}
=&&-\sum_{\nu\nu'}e^{i\pi\nu'}d^{(1/2)}_{-\nu',-\lambda''_3}(\beta_2)
\,d^{(1/2)}_{-\nu,-\lambda'_3}(\beta_1)
\left[\overline{u}(q',-\nu')\,R_{0,-\chi,0}
\tilde{N}_{\rho'}(\tilde{p}'',\lambda''_3)\,N_\rho(\tilde{p}',\lambda'_3)
\,u(q,-\nu)
\right]\nonumber\\
=&&-\sum_{\nu\nu'}e^{i\pi\nu'}d^{(1/2)}_{-\nu',-\lambda''_3}(\beta_2)
\,d^{(1/2)}_{-\nu,-\lambda'_3}(\beta_1)\,
d^{(1/2)}_{-\nu',-\nu}(-\chi)\,
\left[A_{\rho'\rho}d_1
+ B_{\rho'\rho}d_5\right]\, ,
\label{Eq5.19}
\end{eqnarray}
where the matrix elements $d_1$ and $d_5$ are
\begin{eqnarray}
\left[\overline{u}(q',-\nu')\,R_{0,-\chi,0}\,u(q,-\nu)\right]
=&&d^{(1/2)}_{-\nu',-\nu}(-\chi)\,d_1=
d^{(1/2)}_{-\nu',-\nu}(-\chi)\,\left(c'c-4\nu'\nu s's\right)
\nonumber\\
\left[\overline{u}(q',-\nu')\,R_{0,-\chi,0}\gamma^5\,u(q,-\nu)\right]
=&&d^{(1/2)}_{-\nu',-\nu}(-\chi)\,d_5=
d^{(1/2)}_{-\nu',-\nu}(-\chi)\,\left(2\nu' s'c-2\nu c's\right)
\, ,
\label{Eq5.20}
\end{eqnarray}
with
\begin{eqnarray}
&&c=\cosh(\eta_q/2)\qquad c'=\cosh(\eta_{q'}/2)\nonumber\\
&&s=\sinh(\eta_q/2)\qquad s'=\sinh(\eta_{q'}/2)\, ,
\label{Eq5.21}
\end{eqnarray}
and
\begin{equation}
\sinh\eta_q ={q\over m} \qquad \sinh\eta_{q'} ={q'\over m}
\, .
\end{equation}
From the explicit form of the $N'$s, we obtain
\begin{eqnarray}
A_{\rho'\rho}=&&\left(\begin{array}{cc}
{\displaystyle{E_{\tilde{p}''}E_{\tilde{p}'}
\over m^2} - 4\lambda''_3\lambda'_3{\tilde{p}''\tilde{p}'
\over m^2 }}
& \qquad{\displaystyle -2\lambda''_3 {\tilde{p}''\over m} }\\
{\displaystyle-2\lambda'_3 {\tilde{p}'\over m}} &\qquad -1
\end{array}\right)
\nonumber\\
B_{\rho'\rho}= &&\left(\begin{array}{cc}
{\displaystyle 2\lambda'_3
{E_{\tilde{p}''} \tilde{p}'\over m^2} - 2\lambda''_3
{E_{\tilde{p}'} \tilde{p}''\over m^2}} & \qquad
{\displaystyle {E_{\tilde{p}''}
\over m} }\\ {\displaystyle-{E_{\tilde{p}'}\over m}} & \qquad 0
\end{array}\right)
\, . \label{Eq5.22}
\end{eqnarray}

The sum over $\nu$ and $\nu'$ in Eq.~(\ref{Eq5.19}) can now be
carried out if care is taken to remove all phases which depend
on  $\nu$ or $\nu'$.  There are four possibilities, all of which
occur. We may write the ``standard'' sum in a compact form
\begin{eqnarray}
\sum_{\nu\nu'}{\cal I}\equiv&& \sum_{\nu\nu'} e^{i\pi\nu'}
d^{(1/2)}_{-\nu',-\lambda''_3}(\beta_2)
d^{(1/2)}_{-\nu', -\nu}(-\chi)
d^{(1/2)}_{-\nu, -\lambda'_3}(\beta_1) \nonumber\\
=&& e^{i\pi\lambda'_3}\sum_{\nu\nu'}
d^{(1/2)}_{\lambda''_3 \nu'}(\beta_2)
d^{(1/2)}_{\nu'\nu}(-\chi)
d^{(1/2)}_{\nu\lambda'_3}(\beta_1) \nonumber\\
 =&& e^{i\pi \lambda'_3}
d^{(1/2)}_{\lambda''_3 \lambda'_3}(\beta_1+\beta_2-\chi)
= e^{i\pi \lambda'_3}
d(\beta_1+\beta_2-\chi)\, , \label{Eq5.29}
\end{eqnarray}
where symmetry properties of the $d$ functions have been used,
and for compactness the $\lambda''_3\lambda'_3$
indices are suppressed in the final result.  By similar
arguments, the remaining three sums give
\begin{eqnarray}
\sum_{\nu\nu'} 4\nu'\nu{\cal I} =&&
= e^{i\pi \lambda'_3}
d(\beta_1+\beta_2+\chi)\nonumber\\
\sum_{\nu\nu'} 2\nu'{\cal I} =&&
= e^{i\pi s_3}
d(-\beta_1+\beta_2+\chi)\nonumber\\
\sum_{\nu\nu'} 2\nu{\cal I} =&&
= e^{i\pi s_3}
d(-\beta_1+\beta_2-\chi)
\, . \label{Eq5.30}
\end{eqnarray}
Using these identities, the matrix element for particle 3
finally becomes
\begin{equation}
{\cal U}^{(3)}_{\lambda''_3\rho_3,\lambda'_3\rho_2} =
e^{i\pi \lambda'_3}
{\cal X}^{\rho_3\rho_2}_{\lambda''_3\lambda'_3}\, , \label{Eq5.31}
\end{equation}
where the matrix ${\cal X}^{\rho'\rho}$ is
\begin{equation}
{\cal X}^{\rho'\rho}_{\lambda''_3\lambda'_3}=
\left(\begin{array}{cc} -\left[D_1A+D_5B\right] &
-(-1)^{1/2-\lambda'_3}\left[{\displaystyle
{E_{\tilde{p}''}\over m}D_5
- (-1)^{\lambda''_3-\lambda'_3}{\tilde{p}''\over m} D_1}\right]
\\ & \\
(-1)^{1/2-\lambda'_3}\left[{\displaystyle{E_{\tilde{p}'}\over m}D_5
+{\tilde{p}'\over m} D_1}\right] & D_1
\end{array}\right) \, , \label{Eq5.32}
\end{equation}
with the notation
\begin{eqnarray}
D_1=&&d^{(1/2)}_{\lambda''_3 \lambda'_3}
(\beta_1+\beta_2-\chi)c'c - d^{(1/2)}_{\lambda''_3 \lambda'_3}
(\beta_1+\beta_2+\chi) s's\nonumber\\
D_5=&&d^{(1/2)}_{\lambda''_3 \lambda'_3}
(-\beta_1+\beta_2+\chi)s'c - d^{(1/2)}_{\lambda''_3 \lambda'_3}
(-\beta_1+\beta_2-\chi) c's\nonumber\\
A=&&A_{++}\nonumber\\
B=&&2\lambda'_3 B_{++}  \, .
\label{Eq5.33}
\end{eqnarray}

Combining the results (\ref{Eq5.7}), (\ref{Eq5.9}), and
(\ref{Eq5.31}) gives the following expression for the reduced
matrix element
\begin{equation}
\langle\!\langle
\tilde{p}'', \lambda_1(\lambda''_2\lambda''_3)  \rho_3|
\tilde{p}', \lambda'_2(\lambda'_1\lambda'_3) \rho_2
\rangle\!\rangle = -e^{i\pi\lambda'_3}\,
 d^{(1/2)}_{\lambda_1\lambda'_1}(\beta_1)\,
d^{(1/2)}_{\lambda''_2\lambda'_2}(-\beta_2)\,
{\cal X}^{\rho_3\rho_2}_{\lambda''_3\lambda'_3}  \, ,
\label{Eq5.34}
\end{equation}

\subsection{Symmetries of the permutation operator}

The reduced matrix element (\ref{Eq5.34}) satisfies a symmetry
condition which can be obtained from the following property of
scalar products
\begin{equation}
\langle\!\langle
\tilde{p}'', \lambda_1(\lambda''_2\lambda''_3)  \rho_3|
\tilde{p}', \lambda'_2(\lambda'_1\lambda'_3) \rho_2
\rangle\!\rangle = \langle\!\langle
\tilde{p}', \lambda'_2(\lambda'_1\lambda'_3)  \rho_2|
\tilde{p}'', \lambda_1(\lambda''_2\lambda''_3) \rho_3
\rangle\!\rangle^* \, .
\label{Eq5.35}
\end{equation}
Because the permutation operator is hermitian and the initial and
final states are composed of identical nucleons, this
equation tells us that the matrix elements (\ref{Eq4.9}) must
be identical under the substitution $q\leftrightarrow q'$
(which also implies $\tilde{p}''\leftrightarrow \tilde{p}'$,
$\tilde{\theta}''\leftrightarrow \tilde{\theta}'$, and
$\beta_1\leftrightarrow \beta_2$) and
\begin{equation}
\begin{array}{c}
j\leftrightarrow j'\\
m\leftrightarrow m'\\
\rho_3\leftrightarrow \rho_2
\end{array}\qquad
\begin{array}{c}
\lambda_1\leftrightarrow\lambda'_2\\
\lambda''_2\leftrightarrow\lambda'_1\\
\lambda''_3\leftrightarrow\lambda'_3
\end{array}\, . \label{Eq5.36}
\end{equation}
Examination of the matrix elements shows that this implies
\begin{eqnarray}
(-1)^{m-\lambda_1+\lambda'_3}
&&d^{(J)}_{m-\lambda_1,m'-\lambda'_2}(\chi)\,
d^{(1/2)}_{\lambda_1\lambda'_1}(\beta_1)\,
d^{(1/2)}_{\lambda''_2\lambda'_2}(-\beta_2)\,
{\cal X}^{\rho_3\rho_2}_{\lambda''_3\lambda'_3} \nonumber\\
&&=
(-1)^{m'-\lambda'_2+\lambda''_3}
d^{(J)}_{m'-\lambda'_2,m-\lambda_1}(\chi)\,
d^{(1/2)}_{\lambda'_2\lambda''_2}(\beta_2)\,
d^{(1/2)}_{\lambda'_1\lambda_1}(-\beta_1)\,
{\cal X}^{\rho_2\rho_3}_{\lambda'_3\lambda''_3}
\, ,
\label{Eq5.37}
\end{eqnarray}
which reduces to the condition
\begin{equation}
(-1)^{\lambda'_3-\lambda''_3}
{\cal X}^{\rho_3\rho_2}_{\lambda''_3\lambda'_3} =
{\cal X}^{\rho_2\rho_3}_{\lambda'_3\lambda''_3}
\, .
\label{Eq5.38}
\end{equation}
However, the transformation (\ref{Eq5.36}) gives
$A\leftrightarrow A$, $B\leftrightarrow
-(-1)^{\lambda''_3-\lambda'_3}B$,  $D_1\leftrightarrow
(-1)^{\lambda''_3-\lambda'_3}D_1$, and $D_5 \leftrightarrow
-D_5$, showing that the result (\ref{Eq5.32}) satisfies the
symmetry condition (\ref{Eq5.38}).

Another symmetry of the matrix elements of the permutation 
operator follows from the fact that ${\cal P}_{12}$ commutes with the
parity operator ${\cal P}$, which leads to the identity 
\begin{equation}
{\cal P}_{12}= {\cal P}_{12} {\cal P}^2 = {\cal P} {\cal P}_{12} {\cal P} \, .
\end{equation}
The action of the parity operator on the states defined in Eq.~(\ref{Eq11})
can be worked out, giving 
\begin{equation}
{\cal P} |q\, J M,\tilde{p} j
m, \lambda_1 (\lambda_2 \lambda_3)  \rho \rangle
= (-1)^{j-1} (-1)^{J-j-1/2}\rho 
|q\, J M,\tilde{p}\, j\;
-\!m, -\!\lambda_1 (-\!\lambda_2 -\!\lambda_3)  \rho \rangle \, .
\end{equation}
Hence the matrix elements of ${\cal P}_{12}$ must satisfy the identity
\begin{eqnarray}
\lefteqn{\langle q'\, J M',  \tilde{p}' j'
m', \lambda'_1 (\lambda'_2 \lambda'_3)  \rho' | {\cal P}_{12} |q J M,
\tilde{p}\, j m, \lambda_1 (\lambda_2 \lambda_3)  \rho \rangle
} & &\nonumber \\
& = &
\langle q'\, J M',  \tilde{p}' j'
m', \lambda'_1 (\lambda'_2 \lambda'_3)  \rho' | 
{\cal P} {\cal P}_{12} {\cal P}
|q\, J M,\tilde{p} j
m, \lambda_1 (\lambda_2 \lambda_3)  \rho \rangle 
 \nonumber \\
& = &(-1)^{2J-1} \rho \rho'
\langle q'\, J M',  \tilde{p}'\, j'\; -\!m', -\!\lambda'_1 
(-\!\lambda'_2 -\!\lambda'_3)  \rho' | {\cal P}_{12} |q \,J M,
\tilde{p}\, j\;-\!m, -\!\lambda_1 (-\!\lambda_2 -\!\lambda_3)  \rho \rangle
\, .\label{parity}
\end{eqnarray}
For the triton, where $J=1/2$, this means that under the
substitutions
\begin{eqnarray}
& \lambda_1' \leftrightarrow -\lambda_1' \qquad
& \lambda_1 \leftrightarrow -\lambda_1 \nonumber \\
& \lambda_2' \leftrightarrow -\lambda_2' \qquad
& \lambda_2 \leftrightarrow -\lambda_2 \nonumber \\
& \lambda_3' \leftrightarrow -\lambda_3' \qquad
& \lambda_3 \leftrightarrow -\lambda_3 \nonumber \\
& m' \leftrightarrow -m' \qquad 
& m \leftrightarrow -m \label{eqsubst}
\end{eqnarray}
we should recover the same matrix element multiplied by
a factor of $\rho \rho'$.

To verify that the matrix elements of ${\cal P}_{12}$ satisfy 
this symmetry, return to the full expression given
in Eq.~(\ref{Eqxx}), and use the identity
\begin{equation}
d^{(j)}_{\lambda,\lambda'}(\theta)=(-1)^{\lambda-\lambda'}
d^{(j)}_{-\lambda,-\lambda'}(\theta)
\end{equation}
to obtain the condition
\begin{equation}
{\cal N}_{\lambda''_3\lambda'_3}^{\rho''\rho'}
=(-1)^{\lambda''_3-\lambda'_3}\rho''\rho'
{\cal N}_{-\!\lambda''_3\,-\!\lambda'_3}^{\rho''\rho'}\, , \label{Ncond2}
\end{equation}
where ${\cal N}_{\lambda''_3\lambda'_3}^{\rho''\rho'}$ is defined in
Eq.~(\ref{Eq5.40}) below.  Examination of this equation confirms that
Eq.~(\ref{Ncond2}) is indeed satisfied.

In the next section we present our final results for the three
body equations.

\section{Final equations}

In this final section we collect the previous results together, and
explain how it is that integration over the the spectator momentum, $q'$,
is limited to a finite interval.  Then we describe the changes in the
equations which are required by isospin and the conservation of
parity.  Finally, we describe how the three-body channels are
classified and counted.   

\subsection{Spectator equations in angular momentum space}

Using Eqs.~(\ref{Eq5.32}) and (\ref{Eq24a}) gives the
following compact result
\begin{equation}
\sum_{\rho_2}{\cal X}^{\rho_3\rho_2}_{\lambda''_3\lambda'_3}
{\cal
O}_{\rho_2\rho_1}(\tilde{p}',\lambda'_3)=-{E_{\tilde{p}'}
\over m}\,{\cal N}^{\rho_3\rho_1}_{\lambda''_3\lambda'_3} \, ,
\label{Eq5.39}
\end{equation}
where the matrix ${\cal N}^{\rho'\rho}$ is
\begin{equation}
{\cal N}^{\rho'\rho}_{\lambda''_3\lambda'_3}=
\left(\begin{array}{cc} {\displaystyle
{E_{\tilde{p}''}\over m}D_1
- 4\lambda''_3\lambda'_3{\tilde{p}''\over m} D_5 }
&\qquad\quad -2\lambda'_3\left[D_1B+D_5A\right]
\\ & \\
-2\lambda'_3 D_5 & {\displaystyle
{E_{\tilde{p}'}\over m}D_1 + {\tilde{p}'\over m} D_5 }
\end{array}\right) \, , \label{Eq5.40}
\end{equation}
and $A$, $B$, $D_1$ and $D_5$ were defined in
Eq.~(\ref{Eq5.33}).

Combining Eqs.~(\ref{Eq5.34}) and (\ref{Eq5.39}) and
substituting into Eq.~(\ref{Eq4.14}) gives
\begin{eqnarray}
\langle J 1 (23)\rho | \Gamma^1 \rangle
&&=
\sum_{j'm'}\sqrt{2j+1}\sqrt{2j'+1}\!\!\!
\sum_{{\lambda''_2 \lambda''_3\,\rho''}
\atop {\lambda'_1 \lambda'_2 \lambda'_3\,\rho'} }
\int_0^{q_{\hbox{{\tiny crit}}}} {q'^2 dq'}{m\over
E_{q'}}
\int_0^\pi d\chi \sin\chi \;
d^{(J)}_{m-\lambda_1,m'-\lambda'_2}(\chi) \nonumber\\
&&\times
\langle j (23)\rho | M^1| j (2''3'')\rho''\rangle
\,{m\over E_{\tilde{p}''}}\, g^{\rho''}(q,{\tilde p}'')\,
d^{(j)}_{m,\lambda''_2-\lambda''_3}(\tilde{\theta}'')
\,d^{(j')}_{m',\lambda'_1-\lambda'_3}(\tilde{\theta}')\nonumber\\
&&\times (-1)^{m-\lambda_1+ \lambda'_3}
\,d^{(1/2)}_{\lambda_1\lambda'_1}(\beta_1)\,
d^{(1/2)}_{\lambda''_2\lambda'_2}(-\beta_2)\,
{\cal N}^{\rho''\rho'}_{\lambda''_3\lambda'_3}
\,{m \over E_{\tilde{p}'}}\,
\langle J' 1' (2'3')\rho' |\Gamma^1 \rangle \, .  \label{Eq6.1}
\end{eqnarray}
This is identical to the final result given in
Sec.~I, Eqs.~(\ref{Eq6.1x}) and (\ref{Eqxx}).  It is a 
two-dimensional integral equation depending on the variables $q'$ and
$\chi$, where the integration over the angle $\chi$ runs from 0 to
$\pi$, independent of the value of $q'$, and the integration over
$q'$ has been limited to the finite interval
$[0,q_{\hbox{{\scriptsize crit}}}]$, as discussed in the next
subsection.  The momenta
$\tilde{p}'$ and $\tilde{p}''$, the angles $\tilde{\theta}'$ and
$\tilde{\theta}''$, and the Wigner rotation angles $\beta_1$ and
$\beta_2$  all depend  on
$q, q'$, and $\chi$, and are defined in
Eqs.~(\ref{Eq4.15}), (\ref{Eq5.8}) and (\ref{Eq5.10}).   The
matrix ${\cal N}$ is defined in Eq.~(\ref{Eq5.40}).

\subsection{Removal of the space-like region}

The physical reason for restricting the $q'$ integration in
Eq.~(\ref{Eq6.1}) to the finite interval $0\le q'\le
q_{\hbox{{\scriptsize crit}}}$ will be discussed now.

As given in Eq.~(\ref{EqWq}), the invariant mass of
the two-body subsystem decreases with increasing momentum of the
spectator, $q$, and at the value
\begin{equation}
q=q_{\hbox{{\scriptsize crit}}}= {M_t^2-m^2\over 2M_t}\simeq
{4\over3}m
\label{Eq6.2}
\end{equation}
the mass of the two-body subsystem is zero.   This means it is
recoiling with the speed of light, and under such circumstances
the relativistic effects are clearly enormous! Furthermore, as
$q$ increases beyond the critical value, we pass from a region
where the two-body states are time-like into a region where they
are space-like.  The two-body scattering calculations are
carried out in the rest frame of the two-body system, which does
not exist for space-like states, and, more generally, it is
unlikely that an effective theory designed to describe time-like
scattering would  be useful in the space-like region.
Furthermore, since the space-like two-body states appear only at
rather high momentum (above 1200 MeV) where the amplitudes are
expected to be very small anyway, it would be sensible to simply
neglect the region $q\ge q_{\hbox{{\scriptsize crit}}}$, and
set the three-body amplitudes to zero in this region.  As it
turns out, the three-body amplitudes {\it go to zero
automatically at the critical value of $q$\/}, permitting us to
impose the condition that they are zero for $q\ge
q_{\hbox{{\scriptsize crit}}}$ {\it without making the three
body amplitudes discontinuous in $q$\/}.

To see that the Faddeev amplitudes $\langle J 1 (23)\rho |
\Gamma^1  \rangle
\to0$ as $q\to q_{\hbox{{\scriptsize crit}}}$, note that the
function
$\tilde{p}_0(q,q',\chi)$ [defined in Eq.~(\ref{Eq4.13})]
approaches infinity as $q\to q_{\hbox{{\scriptsize crit}}}$ (as
long as
$q'>0$, which is true over the entire region of the $q'$
integration except at the boundary where the integrand is
zero).  Specifically,
%
\begin{equation}
\tilde{p}'' 
{\mathrel{\mathop{\kern0pt\longrightarrow
}\limits_{q\to q_{\hbox{{\scriptsize crit}}}}}}
{q_{{\scriptsize \hbox{crit}}}\,\left(
E_{q'} +q'\cos\chi \right) \over W_q }
= {C\over W_q} \, , \label{Eq6.3}
\end{equation}
and in this limit
\begin{eqnarray}
{m\over E_{\tilde{p}''}}g^+(\tilde{p}''){\cal N}^{+\rho}
&&\longrightarrow {m\over 2\tilde{p}''^2}\,\left(\tilde{p}''K_{+\rho}\right)
\quad\longrightarrow  \quad (W_q)^1
\nonumber\\
{m\over E_{\tilde{p}''}}g^-(\tilde{p}''){\cal N}^{-\rho}
&&\longrightarrow -{1\over W_q} \left({W_q\over mC}\right) K_{-\rho} \longrightarrow
(W_q)^0\, ,  \label{Eq6.4}
\end{eqnarray}
where $C$ and $K_{\pm\rho}$ are functions which are finite in
the limit as $q\longrightarrow q_{\hbox{{\scriptsize crit}}}$.  Note that
the possible $1/W_q$ singularity from the negative energy
part of the propagator is canceled by the
$m/E_{\tilde{p}''}$ factor.  Hence the Faddeev amplitudes
go like
\begin{equation}
\langle J 1 (23)\rho |\Gamma^1  \rangle 
{\mathrel{\mathop{\kern0pt\longrightarrow
}\limits_{W_q\longrightarrow 0} }}
c_{\rho+}(W_q)^{(n_{\rho+}+1)} +
c_{\rho-}\,(W_q)^{n_{\rho-}} \, , \label{Eq6.5}
\end{equation}
where $n_{\rho+}$ and $n_{\rho-}$ are powers with which
the two-body amplitudes (\ref{Eq42}) fall with momentum as
$\tilde{p}''\longrightarrow\infty$:
\begin{equation}
\langle j (23)\rho | M^1| j (2''3'')\rho_3\rangle
{\mathrel{\mathop{\kern0pt\longrightarrow
}\limits_{\tilde{p}''\longrightarrow\infty} }}
\left({1\over \tilde{p}''}\right)^{n_{\rho\rho_3}}
\, . \label{Eq6.6}
\end{equation}
We conclude that the Faddeev amplitudes not only go to zero as
$q\to q_{\hbox{{\scriptsize crit}}}$, but that they approach
this limit smoothly.

\subsection{Isospin}

Since the main application of the three-body equations for 
spin $1/2$ particles will be the three-nucleon system, we
have to incorporate the isospin degree of freedom. This can
be  done separately from the other degrees of freedom, as
described in this subsection.  We will assume that isospin
is conserved by the equations.

To lay the foundation we return to the discussion in 
Sec.~II.  The exchange operators ${\cal P}_{ij}$ are a product
of a part  which acts {\it only\/} in isospin
space, and a part which acts on all other
coordinates, denoted by $\widetilde{{\cal P}}_{ij}$.  
If the $ij$ pair has isospin $T_{ij}$, the action of 
${\cal P}_{ij}$ on the isospin part of the wave function will
be denoted simply by its eigenvalue $(-1)^{T_{ij}-1}$.   
The phase $\zeta$ which occurs in Eq.~(\ref{SpecSymm1}) is
$-1$ for fermions and is therefore a product of the phase
$(-1)^{T_{ij}-1}$ from the exchange of the isospin variables, and the
phase $u$, resulting from the operation of
$\widetilde{{\cal P}}_{ij}$.  Hence
\begin{equation}
\zeta=-1=u\;(-1)^{T_{ij}-1} \, . \label{I1}
\end{equation}
Even though $\zeta$  is always
$-1$ for fermions, there are in general two possible values of $u$
corresponding to the two possible isospin channels, 
and Eq.~(\ref{SpecSymm1}) generalizes to 
\begin{eqnarray}
\widetilde{{\cal P}}_{32} M^1_{22} && = u\, M^1_{32} =
M^1_{33}\widetilde{{\cal P}}_{32} \nonumber\\ 
\widetilde{{\cal P}}_{23}
M^1_{33} && = u\, M^1_{23} = M^1_{22}
\widetilde{{\cal P}}_{23} \,  .  \label{I8}
\end{eqnarray}

The vectors $|\Gamma^1_2\rangle$ are also vectors in isospin space.  Taking
matrix elements of Eq.~(\ref{EqSpec2}), and inserting $1=\sum_T |T\rangle
\langle T |$, gives 
\begin{equation}
\langle T|\Gamma^1_{2}\rangle= 2\sum_{T'} \langle T| {\cal P}_{12}
| T'\rangle\; M_{22}^{1\,T}
G_2^1 \widetilde{{\cal P}}_{12} \langle T'| \Gamma^1_{2} \rangle
\, , \label{I4}
\end{equation}
where $|T\rangle $ are the isospin wave functions discussed below, and 
$\langle T| {\cal P}_{12}| T'\rangle$ is the matrix
element of the permutation operator in isospin space.
The calculation of this matrix element is familiar from
the nonrelativistic theory, but for completeness we will
briefly present it here.

In more detail, the states in isospin space will be denoted
\begin{equation}
|T\rangle=| ((t_2 t_3) T t_1) {\cal T}{\cal T}_z \rangle \,,
\end{equation}
where $t_i$ is the isospin of particle $i$, $T$ is the isospin
of the pair, and  $\cal T$ and ${\cal T}_z$ are the total
three-body isospin and its projection.  As the notation
suggests, $t_2$ and $t_3$ are first coupled to $T$, and then
$T$ and $t_1$ are coupled to ${\cal T}$.
These states form a
complete, orthonormal basis
\begin{eqnarray}
&&\langle ((t_2 t_3) T t_1) {\cal T}{\cal T}_z |   
((t'_2 t'_3) T' t'_1) {\cal T'}{\cal T}'_z \rangle =
\delta_{t_1 t'_1}
\delta_{t_2 t'_2}
\delta_{t_3 t'_3}
\delta_{T T'}
\delta_{{\cal T}{\cal T}'}
\delta_{{\cal T}_z{\cal T}'_z}\nonumber\\
&&\sum_{t_1 t_2 t_3 T {\cal T} {\cal T}_z}
|  ((t_2 t_3) T t_1) {\cal T}{\cal T}_z \rangle \langle
((t_2 t_3) T t_1) {\cal T}{\cal T}_z |
= {\bf 1} \, .
\end{eqnarray}

The effect of ${\cal P}_{12}$ is to interchange particles 1 and 2:
\begin{equation}
{\cal P}_{12}  | ((t_2 t_3) T' t_1) {\cal T}
{\cal T}_z\rangle
=| ((t_1 t_3) T' t_2) {\cal T}{\cal T}_z\rangle
\, .
\end{equation}
The matrix element of ${\cal P}_{12}$ in isospin space 
reduces therefore to a simple recoupling coefficient,
\begin{eqnarray}
\langle T| {\cal P}_{12}| T'\rangle=&&
\langle ((t_2 t_3) T t_1) {\cal T}{\cal T}_z| 
{\cal P}_{12} | ((t_2 t_3) T' t_1) {\cal T}
{\cal T}_z\rangle \nonumber\\
 =&& 
\langle ((t_2 t_3) T t_1) {\cal T}{\cal T}_z| 
((t_1 t_3) T' t_2) {\cal T}{\cal
T}_z\rangle \nonumber \\
=& & 
 - \sqrt{2T+1} \sqrt{2T'+1}
\left\{ \begin{array}{ccc} t_2 & t_3 & T \\ t_1 & {\cal T}
& T'
\end{array}  \right\} \, .
\end{eqnarray}

In the next subsection we complete the reduction of Eq.~(\ref{I4}) by
inserting a complete set of good parity eigenstates.

\subsection{Parity eigenstates}

The two-body scattering amplitudes which drive the three-body equations
are separated into channels which are eigenstates of ${\cal P}^1$ (the
parity operator on the 23 subspace) and isospin.  Isospin was just
discussed in the previous subsection, and need not be revisited again
until the next subsection below where we explain how isospin (or
exchange symmetry) plays a role in the description and counting of
the channels.  The role of the conservation of parity, which has not
yet been taken into account, will be discussed in this subsection.  

The helicity states $|J 1(23)\rho \rangle$ are neither eigenstates of the
full parity operator, ${\cal P}$, nor of the two-body parity operator,
${\cal P}^1$.  Since the two-body scattering amplitudes which emerge
from the two-body calculations
are eigenstates of ${\cal P}^1$, the three-body equations must be
re-expressed in terms of these states. This is not difficult because the
eigenstates are merely linear combinations of the states we have already
obtained.

First we return to the two-body helicity states
$|jm(\lambda_2 \lambda_3) \rho \rangle$, where
the relative momentum, $\tilde{p}$ is suppressed because it will play  no
role in the discussion which follows.  If we apply the operator
${\cal P}^1$ (referred to simply as ${\cal P}$ in Ref.~\cite{R1}) to this
state we get 
\begin{equation}
{\cal P}^1 |jm (\lambda_2 \lambda_3) \rho \rangle =
\rho \epsilon \;|jm (-\!\lambda_2 \;-\!\lambda_3) \rho
\rangle \, ,\label{p2}
\end{equation}
where $\epsilon=(-1)^{j-1}$.  It is easy to see that the state
\begin{eqnarray}
|j^{\;r}(m \lambda) \rho \rangle \equiv&&
\frac{1}{\sqrt{2}}\left(1+ r{\cal P}^1\right)\, |jm
(\lambda_2 \lambda_3) \rho \rangle
\nonumber\\ =&&\frac{1}{\sqrt{2}}
\left\{ |jm (\lambda_2 \lambda_3) \rho \rangle +
r \rho \epsilon  |jm (-\!\lambda_2\; -\!\lambda_3) \rho
\rangle \right\}  \, ,
\label{eqpar2}
\end{eqnarray}
with $\lambda \equiv \lambda_2-\lambda_3$, is a normalized eigenstate
of ${\cal P}^1$  
\begin{equation}
{\cal P}^1|j^{\;r}(m \lambda) \rho \rangle = r\,|j^{\;r}(m \lambda)
\rho \rangle \, .
\end{equation}

If we replace the individual particle helicities $\lambda_2$ and
$\lambda_3$ on the rhs Eq.~(\ref{eqpar2})  by
$-\lambda_2$ and $-\lambda_3$, we obtain the same state, apart from a
phase factor.  We should therefore
include in our new basis (\ref{eqpar2}) only states that are not related
to each other by changing the sign of both helicities. We choose the
convention
$\lambda_2=+\frac{1}{2}$
and label the states by the difference
$\lambda$.  With this convention $\lambda$ can be 0 or 1, and 
the parity $r$ can be 
$+$ and $-$, and we have again 4 independent
states, just as before when each of the individual helicities were
allowed to be $\pm \frac{1}{2}$.
The selection rule
\begin{equation}
|\lambda_2-\lambda_3| \leq j \label{EqZZ6}
\end{equation}
excludes $\lambda=1$ for states with $j=0$.

For the construction of three-body parity eigenstates we can proceed
in precisely the same way, treating the two-body subsystem as one
elementary particle with spin $j$, helicity $m$, and (now well defined)
intrinsic parity $r$.  The parity operator which
acts in the three-body space will be
denoted ${\cal P}$, and should be distinguished from ${\cal P}^1$.  
To carry out this construction, we first introduce the three-body 
states
\begin{eqnarray}
|J j^{\;r}\lambda_1 (m \lambda ) \rho\rangle =&& \frac{1}{\sqrt{2}}
\left\{ |J,jm; \lambda_1  (\lambda_2\lambda_3)\rho\rangle +
r \rho \epsilon  |J,jm; \lambda_1
(-\!\lambda_2\;-\!\lambda_3)\rho\rangle  \right\}\nonumber\\
=&&\frac{1}{\sqrt{2}}\left(1+r{\cal P}^1\right) |J,jm;
\lambda_1  (\lambda_2\lambda_3)\rho\rangle\, ,
\end{eqnarray}
where $|J,jm; \lambda_1
(\lambda_2\lambda_3)\rho\rangle=|J1(23)\rho\rangle$ are the same three
body states introduced in Sec.~IV, Eq.~(\ref{Eq33a}), but with some of 
the notation restored for clarity.  The parity
operation ${\cal P}$ on these states yields
\begin{equation}
{\cal P} |J j^{\;r}\lambda_1(m \lambda ) \rho\rangle =
 \eta_1 r (-1)^{J-j-s_1} |J j^{\;r}\;-\!\lambda_1 (-m \lambda)  \rho
\rangle
\, , \label{p3}
\end{equation}
and the three-body eigenstates of parity, with eigenvalue $\Pi=\pm$ are
therefore 
\begin{eqnarray}
|J^\Pi j^{\;r} (m \lambda) \rho \rangle \equiv&&
\frac{1}{\sqrt{2}}\left\{
| J j^{\;r} \lambda_1(m \lambda) \rho \rangle
+ \Pi \eta_1 r (-1)^{J-j-s_1}
| J j^{\;r}\;-\!\lambda_1 (-m\lambda) \rho \rangle \right\} \nonumber\\
=&&\frac{1}{\sqrt{2}}\left(1+\Pi{\cal P}\right)
| J j^{\;r} \lambda_1(m \lambda) \rho \rangle\, .
\end{eqnarray}
In this case we adopt the convention
$\lambda_1 = +\frac{1}{2}$ and let $m$ vary, subject to the
condition that 
\begin{equation}
|m-\lambda_1|=|m-\frac{1}{2}| \leq J \, .\label{EqZZ7}
\end{equation}

Nucleon one is always in a positive-energy state, and therefore $\eta_1
=1$ and $s_1 = \frac{1}{2}$.
The triton is characterized by $J^\Pi={\frac{1}{2}}^+$.
Three-body states with these quantum numbers are
\begin{equation}
|{\textstyle{\frac{1}{2}}^+},j^{\;r} (m \lambda) \rho \rangle =
\frac{1}{\sqrt{2}}
\left\{ |{\textstyle{\frac{1}{2}}}\, j^{\;r}\, {\textstyle{\frac{1}{2}}}
(m\lambda) \rho \rangle - r \epsilon  | {\textstyle{\frac{1}{2}}}
\,j^{\;r}\; -\!{\textstyle{\frac{1}{2}}} (-m \lambda) \rho \rangle
\right\}
\, .
\end{equation}
Expanding these three-body parity eigenstates in terms of
the original three-body helicity states (\ref{Eq33a}) gives
\begin{eqnarray}
|{\textstyle{\frac{1}{2}}}^+\, j^{\;r} (m \lambda) \rho \rangle =&&
{\frac{1}{{2}}} \left\{
|{\textstyle{\frac{1}{2}}}, j m; \lambda_1 (\lambda_2
\lambda_3) \rho\rangle + r\rho\epsilon 
|{\textstyle{\frac{1}{2}}}, j m; \lambda_1 (-\!\lambda_2\;
-\!\lambda_3)
\rho\rangle \right.  \nonumber \\
&&\left. -r\epsilon 
|{\textstyle{\frac{1}{2}}},  j\, -\!m;\, -\!\lambda_1
(\lambda_2\lambda_3)\rho\rangle -\rho
|{\textstyle{\frac{1}{2}}},j\, -\!m;\, -\!\lambda_1
(-\!\lambda_2\; -\!\lambda_3)\rho\rangle \right\}
\nonumber\\ =&&{\frac{1}{{2}}} \left(1+{\cal P}\right) 
\left(1+r{\cal P}^1\right) 
|{\textstyle{\frac{1}{2}}}, j m; \lambda_1 (\lambda_2
\lambda_3) \rho\rangle \, . \label{final3}
\end{eqnarray}

We now return to Eq.~(\ref{I4}) and take the remaining spin-momentum
matrix elements of the operators using the basis of good parity states
Eq.~(\ref{final3}).  For simplicity, we represent the states by 
$|{\textstyle{\frac{1}{2}}}^+\, j^{\;r} (m \lambda) \rho \rangle =
| j^{\;r} (m \lambda) \rho \rangle$, and the direct product states by
$|T\rangle \otimes | j^{\;r} (m \lambda) \rho \rangle= | T j^{\;r} (m
\lambda) \rho \rangle$ .  Then, in our abbreviated notation,
\begin{eqnarray}
\langle T j^{\;r} (m \lambda) \rho |\Gamma^1_T\rangle =&&
\sum_{{j'r'}\atop{m'T'}}
\sum_{{\lambda''\,\rho''}
\atop {\lambda'\,\rho'}}
\int_0^{q_{\hbox{{\tiny crit}}}} {q'^2 dq'}{m\over
E_{q'}}
\int_0^\pi d\chi \sin\chi \;
 \nonumber\\
&&\quad\times \langle T j^{\;r} (m
\lambda) \rho | M^{1\,T} |T j^{\;r} (m \lambda'') \rho''\rangle 
\,{m\over E_{\tilde{p}''}}\, g^{\rho''}(q,{\tilde p}'')\nonumber\\
&&\quad\times \overline{{\cal P}}_{12}^{\rho''\rho'}
[T j^{\;r} (m \lambda'') \rho'' ,T' j'^{\;r'} 
(m' \lambda') \rho' ] \,{m \over E_{\tilde{p}'}}\,
\langle   T' j'^{\;r'} (m'\lambda') \rho' | \Gamma^1_T \rangle
\, , \label{Ifinal}
\end{eqnarray}
where $\langle T j^{\;r} (m \lambda) \rho 
| M^{1\,T} |T j^{\;r} (m \lambda'') \rho''\rangle=
M_{\rho\rho''}^{\lambda\lambda''}(T j^{\;r})$ is the two-body
scattering amplitude for the $j$th partial wave with parity $r$ and
isospin $T$, and 
\begin{eqnarray}
\overline{{\cal P}}_{12}^{\rho''\rho'}
[T j^{\;r} (m \lambda'') \rho'' ,T' j'^{\;r'} 
(m' \lambda') \rho' ]=&&  \langle T| {\cal P}_{12}| T'\rangle
\times \langle  j^{\;r} (m \lambda'') \rho'' |
\widetilde{{\cal P}}_{12} | j'^{\;r'} (m' \lambda') \rho'
\rangle \nonumber\\
=&& \langle T| {\cal P}_{12}| T'\rangle \times 
\langle\!\langle \overline{{\cal P}_{12}} \rangle\!\rangle \, .
\label{I6}
\end{eqnarray}
Eq.~(\ref{Ifinal}) is our final result.  It expresses the three-body
equations in terms of the physical states with definite parity and
isospin, driven by two-body amplitudes
$M_{\rho\rho''}^{\lambda\lambda''}(T j^{\;r})$ which have been
previously calculated as described in Ref.~\cite{R1}.

The new matrix element $\langle\!\langle \overline{{\cal P}_{12}}
\rangle\!\rangle$ is readily obtained from the original matrix
elements Eq.~(\ref{Eq4.5}), which are, in the notation of this
section,
\begin{equation}
\langle\!\langle {\cal P}_{12} \rangle\!\rangle =
\langle{\textstyle{\frac{1}{2}}}, j' m'; \lambda'_1
(\lambda'_2\lambda'_3) \rho'|{\cal P}_{12}| {\textstyle{\frac{1}{2}}},
j m; \lambda_1 (\lambda_2\lambda_3) \rho\rangle \, .
\end{equation}
From the definition (\ref{final3}) we have
\begin{eqnarray}
\langle\!\langle \overline{{\cal P}_{12}} \rangle\!\rangle =&&
{\frac{1}{{4}}} 
\langle{\textstyle{\frac{1}{2}}}, j' m'; \lambda'_1
(\lambda'_2\lambda'_3) \rho'|
\left(1+r'{\cal P}^1\right) \left(1+{\cal P}\right)  {\cal P}_{12}
\left(1+{\cal P}\right) 
\left(1+r{\cal P}^1\right) 
| {\textstyle{\frac{1}{2}}},j m; \lambda_1 (\lambda_2\lambda_3)
\rho\rangle \nonumber\\
=&&{\frac{1}{{2}}} 
\langle{\textstyle{\frac{1}{2}}}, j' m'; \lambda'_1
(\lambda'_2\lambda'_3) \rho'|
\left(1+r'{\cal P}^1\right)  {\cal P}_{12}
\left(1+{\cal P}\right) 
\left(1+r{\cal P}^1\right) 
| {\textstyle{\frac{1}{2}}},j m; \lambda_1 (\lambda_2\lambda_3)
\rho\rangle \, ,
\end{eqnarray}
where we used the fact that ${\cal P}$ commutes with ${\cal P}_{12}$.  Using
Eqs.~(\ref{p2}), (\ref{p3}), and (\ref{parity}), we obtain
\begin{eqnarray}
\langle\!\langle \overline{{\cal P}_{12}} \rangle\!\rangle =&& {\frac{1}{{2}}}
\biggl\{ 
\langle{\textstyle{\frac{1}{2}}}, j' m'; \lambda'_1
(\lambda'_2\lambda'_3) \rho'|{\cal P}_{12}| {\textstyle{\frac{1}{2}}},
j m; \lambda_1 (\lambda_2\lambda_3) \rho\rangle \nonumber\\
&&+ r'\rho'\epsilon'
\langle{\textstyle{\frac{1}{2}}}, j' m'; \lambda'_1
(-\!\lambda'_2\,-\!\lambda'_3)
\rho'|{\cal P}_{12}| {\textstyle{\frac{1}{2}}}, j m; \lambda_1
(\lambda_2\lambda_3)\rho\rangle  \nonumber\\
&&+r\rho\,\epsilon
\langle{\textstyle{\frac{1}{2}}}, j' m'; \lambda'_1
(\lambda'_2\lambda'_3) \rho'|{\cal P}_{12}| {\textstyle{\frac{1}{2}}}, j m;
\lambda_1 (-\!\lambda_2\,-\!\lambda_3) \rho\rangle \nonumber\\
&&+ r'r\rho'\rho\,\epsilon'\epsilon
\langle{\textstyle{\frac{1}{2}}}, j' m'; \lambda'_1
(-\!\lambda'_2\,-\!\lambda'_3)
\rho'|{\cal P}_{12}| {\textstyle{\frac{1}{2}}}, j m; \lambda_1
(-\!\lambda_2\,-\!\lambda_3)\rho\rangle  \nonumber\\
&&-\rho
\langle{\textstyle{\frac{1}{2}}}, j' m'; \lambda'_1
(\lambda'_2\lambda'_3) \rho'|{\cal P}_{12}| {\textstyle{\frac{1}{2}}}, j\, -\!m;
-\!\lambda_1 (-\!\lambda_2\,-\!\lambda_3) \rho\rangle \nonumber\\
&&- r'\rho'\rho\,\epsilon'
\langle{\textstyle{\frac{1}{2}}}, j' m'; \lambda'_1
(-\!\lambda'_2\, -\!\lambda'_3)
\rho'|{\cal P}_{12}| {\textstyle{\frac{1}{2}}}, j \,-\!m; -\!\lambda_1
(-\!\lambda_2\, -\!\lambda_3)\rho\rangle  \nonumber\\
&&-r\epsilon
\langle{\textstyle{\frac{1}{2}}}, j' m'; \lambda'_1
(\lambda'_2\lambda'_3) \rho'|{\cal P}_{12}| {\textstyle{\frac{1}{2}}}, j\,
-\!m;
-\!\lambda_1 (\lambda_2\lambda_3) \rho\rangle \nonumber\\
&&-r'r\rho'\epsilon'\epsilon
\langle{\textstyle{\frac{1}{2}}}, j' m'; \lambda'_1
(-\!\lambda'_2\,-\!\lambda'_3)
\rho'|{\cal P}_{12}| {\textstyle{\frac{1}{2}}}, j\,-\! m; -\!\lambda_1
(\lambda_2\lambda_3) \rho\rangle\biggr\} \, , \label{EqZZ8}
\end{eqnarray}
This is the correct form of the permutation operator to be used with the
physical, good parity states.

\subsection{Three-body channels}

We conclude this paper by counting and classifying the channels which
contribute to the final three-body equations.

In order to clarify the following discussion we restore some of the
notation which we previously suppressed, and denote the two-body helicity
states (\ref{eqpar2}) with good parity, $|j^{\;r}(m \lambda) \rho
\rangle $, by $|\tilde{p}_0\,j^{\;r}(m \lambda) \rho \rangle $,
where  
\begin{equation}
\tilde{p}_0 = E_{\tilde{p}_0} - {1\over2}W_{23} \label{rhxx}
\end{equation}
is the difference in the energies of the two particles in
the two-body rest frame (with particle 2 on shell) and $W_{23}$
the rest frame energy of the two-body system.  Note
that $\tilde{p}_0$ is in general not zero because particle three is
off-shell, and that we continue to suppress explicit reference to the
magnitude of the three component of the relative momentum, $\tilde{p}$,
because it will play  no role in the discussion which follows.  This
state satisfies the relation
\begin{equation}
{\cal Q}_2\,|\tilde{p}_0\, j^{\;r}(m \lambda) \rho \rangle =
|\tilde{p}_0\,j^{\;r}(m \lambda) \rho \rangle \, , \label{EqZZ1}
\end{equation}
where ${\cal Q}_2$ is the projection operator introduced in Sec.~II
which places particle 2 on shell, and is equivalent to the
identity when operating on states where particle 2 is already on
shell.  Note that the state with relative energy $-\tilde{p}_0$ has
particle 3 on shell and hence
\begin{equation}
{\cal Q}_3\,|-\!\tilde{p}_0\,j^{\;r}(m \lambda) \rho\rangle
= |-\!\tilde{p}_0\,j^{\;r}(m \lambda) \rho\rangle \, .
\end{equation}

In terms of the states (\ref{eqpar2}), the interchange of space and spin
coordinates (everything but isospin) has the following effect:
\begin{equation}
\widetilde{{\cal P}}_{23}\, |\tilde{p}_0\,j^{\;r}(m \lambda) \rho
\rangle =
(r\rho)^\lambda\,\epsilon^{(1-\lambda)} \; |-\!\tilde{p}_0\,j^{\;r}(m
\lambda)
\rho\rangle
\, .
\label{EqZZ2}
\end{equation}
This equation follows from the definition (\ref{eqpar2}) and the relation
\begin{equation}
\widetilde{{\cal P}}_{23}\, |\tilde{p}_0\,j m (\lambda_2\lambda_3) \rho
\rangle =
\epsilon \; |-\!\tilde{p}_0\,j m (\lambda_3 \lambda_2)
\rho\rangle \, ,
\end{equation}
which (except for notational changes) is Eq.~(2.97) of Ref.~\cite{R1}.

Using Eq.~(\ref{EqZZ2}), we can extend the discussion of the previous
subsection and introduce states with {\it both\/} good parity and good
exchange symmetry.  Introduce the states
\begin{equation}
|j^{\;ru}(m \lambda) \rho \rangle \equiv=
\frac{1}{\sqrt{2}}\left(1+u\widetilde{{\cal P}}_{23}\right)
\,|\tilde{p}_0\,j^{\;r}(m \lambda) \rho \rangle  \, ,
\label{eqpar2xx}
\end{equation}
These are normalized eigenstates of both ${\cal P}^1$ and
${\cal P}_{23}$ 
\begin{eqnarray}
{\cal P}^1|j^{\;ru}(m \lambda) \rho \rangle =&& r\,|j^{\;ru}(m \lambda)
\rho \rangle \nonumber\\
\widetilde{{\cal P}}_{23}|j^{\;ru}(m \lambda) \rho \rangle =&&
u\,|j^{\;ru}(m
\lambda) \rho \rangle \, .\label{EqZZ4}
\end{eqnarray}
Since $u=(-1)^T$, which can be written $T=(1-u)/2$, these states are
also the correct spin-momentum states to use with isospin.

The counting and classifying of three-body states depends in part on the 
number and classification of the two-body scattering states.  Since both 
${\cal P}^1$ and  $\widetilde{{\cal P}}_{23}$ are conserved by the
two-body equations, two-body states can be classified by different
possible values of the quantum numbers $r$ and $u$.  There are four
combinations:

\vspace*{0.2cm}
$\begin{array}{lll}
\hbox{\rm singlet}&\qquad r=-\epsilon &\qquad u=\phantom{-}\epsilon \cr 
\hbox{\rm triplet}&\qquad r=-\epsilon &\qquad u=-\epsilon\cr
\hbox{\rm coupled}&\qquad r=\phantom{-}\epsilon &\qquad
u=\phantom{-}\epsilon\cr
\hbox{\rm virtual}&\qquad r=\phantom{-}\epsilon &\qquad u=-\epsilon\, .\cr
\end{array}$
\vspace*{0.2cm}

\noindent The last set of states, referred to as {\it virtual states\/} in
Ref.~\cite{R1}, do not contribute to physical two-body scattering.  This
is because, in the positive $\rho$-spin sector, their parity assignment
would require that $j=\ell\pm1$, which in turn requires a total spin
$S=1$.  These assignments are consistent with an exchange symmetry of
$u=-\epsilon$ only if these states are {\it odd under change of sign of the
relative energy variable\/} $\tilde{p}_0$, which insures that they are
zero on shell.  However, because the two-body quantum numbers $r$ and $u$
are {\it not conserved\/} in three-body scattering, they can contribute to
relativistic three-body scattering and to the three-body bound state. 
In the calculations completed thus far \cite{SG} we have neglected
these states, but we expect them to give a small contribution of
purely relativistic origin.  

Neglecting the virtual states, and recalling the selection rule
(\ref{EqZZ6}) leads to the following counting rules:
 
\vspace*{0.2cm}
$\begin{array}{ll}
j=0:&\qquad
\phantom{+}(\lambda=0)\times(\rho=\pm1)\times(r=\pm1)\times(u=-1)=
\;4\;\; \hbox{\rm states}\cr
j>0:& \phantom{+}\qquad
(\lambda=0,1)\times(\rho=\pm1)\times(r=\pm\epsilon)\times(u=\epsilon )\cr
&\qquad
+(\lambda=0,1)\times(\rho=\pm1)\times(r=-\epsilon)\times(u=-\epsilon)=\;12\;\;
\hbox{\rm states} \cr
\end{array}$
\vspace*{0.2cm}

\noindent The total number of two-body states with angular momenta
$j\le j_{max}$ is therefore $n_2=4+12j_{max}$.

%
%
\begin{table}
\caption{Possible quantum numbers for three-body states with
$J^\Pi = {\frac{1}{2}}^+$ (for the triton).  Virtual two body states
have been neglected. }
\begin{tabular}{ c c c c c c c}
 $j$  & $\lambda $ & $r$ & $u$ & $\rho $ &  $m$ & number of states \\
\tableline
 0 & $0$ & $\pm$ & $-$ & $\pm$ &  0 &  4  \\
$\geq 1$ & 0,1 & $\pm$ & $\epsilon$ & $\pm$  &  0,1 & 16 \\
$\geq 1$ & 0,1 & $-\epsilon$ & $-\epsilon$ & $\pm$  &  0,1 & 8 
\end{tabular} \label{Tstates}
\end{table}

In numerical calculations of the three-nucleon bound state it has become
customary to truncate the partial wave series according to the maximal
included total pair angular momentum $j$. Table
\ref{Tstates} shows how many different three-body states exist for a
given $j$. The pattern is simple: applying the selection rule
(\ref{EqZZ7}) for each $j>0$ gives 24 possible states corresponding
to 12 two-body states with either $m=0$ or 1, or
$2\times12=24$ states.  For $j=0$ only
$m=0$ is allowed, and hence there are only 4 different states.  For each
combination of quantum numbers there is one particular pair isospin
consistent with the Pauli principle. Since we have used exchange
symmetry to count the states, the inclusion of isospin does not lead to
any further increase in the number of channels.  The total number of
states up to a given maximal value
$j_{max}$ is therefore $n_3 = 4+24 j_{max}$.

\section{Acknowledgments}

One of us (FG) acknowledges the hospitality of Peter Sauer and the
University of Hannover, where the initial idea to pursue this program 
began. Early work was partially supported by a grant from NATO.
The major part of this work could not have been done without the support
of the Department of Energy
under grant \#DE-FG05-88ER40435 which is gratefully acknowledged. In
addition, one of us (AS) thanks JNICT for support under contract
\# praxis/2/2.1/FIS/223/94, \# CERN/P/FIS/1101/96, and 
\# BCC/4394/94, and and another (MF) thanks the DOE for support under 
grant \#DE-FG06-90ER40561.

\appendix

\section{Operator form of the two-body equations}

In this Appendix we present the operator form of the
two-body equations, and their subsequent reduction to
partial waves.  Using the notation of Sec.~IIB, the
two-body equations for the scattering amplitude are
\begin{equation}
M= V-VG_2{\cal Q}_1 M \, , \label{A1}
\end{equation}
where $V$ is the symmetrized kernel and $M$ is the
two-body scattering amplitude describing the
scattering of particles 1 and 2.  Note that particle
1 in on-shell in the intermediate state, in agreement
with the conventions of Ref.~\cite{R1} (see Sec.~IIB
in that reference).  In the three-body language, the
$M$ in Eq.~(\ref{A1}) is $M^3$, because the
``spectator'' (if it were present) would be particle
3.  To obtain a closed set of equations for
(\ref{A1}), multiply by ${\cal Q}_1$, giving
\begin{equation}
\left[{\cal Q}_1M\right]= \left[{\cal Q}_1V\right]
-\left[{\cal Q}_1 VG_2{\cal Q}_1\right] \,
\left[{\cal Q}_1 M \right]
\, , \label{A2}
\end{equation}
which shows that particle 1 is on-shell throughout the
interaction.

The two-body partial wave equations can be obtained
from Eq.~(\ref{A2}) by inserting a complete set of the
two-body angular momentum states defined in
Eq.~(\ref{Eq7}) [with the 23 pair relabeled 12].  The
completeness and generalized orthogonality relations
for the two-body states, implied by the work in
Sec.~IIC, is
\begin{eqnarray}
\langle j' (1'2') \rho'|j (12)\rho\rangle =&&
\delta_{j' j} \delta_{m' m}
\delta_{\lambda'_1 \lambda_1}
\delta_{\lambda'_2 \lambda_2}\,
{\cal O}_{\rho'\rho}(\tilde{p},\lambda_2)\,
2 E_{\tilde{p}} \frac{ \delta
(\tilde{p}-\tilde{p}')}{\tilde{p}^2} \,
\delta^4(P_{12}'-P_{12})\nonumber\\
{\bf 1} \equiv
{\cal Q}_{\alpha'\alpha}\,
\delta_{\beta'\beta}=&&\int
\frac{\tilde{p}^2 d\tilde{p}}{2 E_{\tilde{p}}}
\,{m^2\over E_{\tilde{p}}^2} \, d^4P \!\!\!
\sum_{{jm\,\rho'\rho\atop \lambda_1 \lambda_2 \lambda_3} }
\!\!\!\! | j (12)\rho'\rangle \,{\cal
O}_{\rho'\rho}(\tilde{p},\lambda_2)\,
 \langle j (12)\rho | \, , \label{A3}
\end{eqnarray}
where the shorthand notation defined in
Eq.~(\ref{Eq38}) has been used.  Substituting the
completeness relation into Eq.~(\ref{A2}) gives
\begin{eqnarray}
\langle j (12)\rho |M&& |j (1'2')\rho' \rangle =
\langle j (12)\rho |V |j (1'2')\rho' \rangle
\nonumber\\ -&&
\sum_{{\rho_1\rho_2}
\atop {\lambda''_1 \lambda''_2} }
\int{k^2 dk\over 2E_k}
{m^2 \over E^2_k}
\,\langle j (12)\rho | V G| j (1''2'')\rho_1\rangle
{\cal O}_{\rho_1\rho_2}(k,\lambda''_2)\;
\langle j  (1''2'')\rho_2 |M |j (1'2')\rho' \rangle
\, .  \label{A4}
\end{eqnarray}
The evaluation of the matrix element of $VG$ is identical
to the evaluation of $M^1G_3$ carried out in
Eq.~(\ref{Eq39}).  Substituting this (\ref{A4}) gives
\begin{eqnarray}
\langle j (12)\rho |M&& |j (1'2')\rho' \rangle =
\langle j (12)\rho |V |j (1'2')\rho' \rangle
\nonumber\\ -&&
\sum_{\rho''\,\lambda''_1 \lambda''_2}
\int  k^2 d k
\,\langle j (12)\rho | V | j (1''2'')\rho''\rangle
{m^2 \over E^2_k} g^{\rho''}(0,k)
\langle j  (1''2'')\rho'' |M |j (1'2')\rho' \rangle
\, .  \label{A5}
\end{eqnarray}
Multiplying both sides of the equation by
$(2\pi)^3m^2/E_{\tilde{p}}E_{\tilde{p}'}$, and introducing
the amplitudes
\begin{eqnarray}
{(2\pi)^3m^2\over E_{\tilde{p}}E_{\tilde{p}'}}\langle j
(12)\rho |M | j (1'2')\rho' \rangle &&=
M^{\rho\rho'\; j}
_{\lambda_1\lambda_1',\lambda_2\lambda_2'}
(\tilde{p},\tilde{p}';P_{12}) \nonumber\\
{(2\pi)^3m^2\over E_{\tilde{p}}E_{\tilde{p}'}}\langle j
(12)\rho |V | j (1'2')\rho' \rangle &&=
V^{\rho\rho'\; j}
_{\lambda_1\lambda_1',\lambda_2\lambda_2'}
(\tilde{p},\tilde{p}';P_{12}) \nonumber\\ {k^2\over
(2\pi)^3}g^\rho(0,k)=g^\rho(k) \, ,  \label{A6}
\end{eqnarray}
gives the equation
\begin{eqnarray}
M^{\rho\rho'\; j}
_{\lambda_1\lambda_1',\lambda_2\lambda_2'}
(\tilde{p},\tilde{p}';&&P_{12})  =
V^{\rho\rho'\; j}
_{\lambda_1\lambda_1',\lambda_2\lambda_2'}
(\tilde{p},\tilde{p}';P_{12})  \nonumber\\
&&-
\sum_{\rho''\,\lambda''_1 \lambda''_2}
\int d k
\,V^{\rho\rho''\; j}
_{\lambda_1\lambda''_1,\lambda_2\lambda''_2}
(\tilde{p},k;P_{12})
g^{\rho''}(k)
M^{\rho''\rho'\; j}
_{\lambda''_1\lambda_1',\lambda''_2\lambda_2'}
(k,\tilde{p}';P_{12})
\, .  \label{A7}
\end{eqnarray}
This is identical to Eq.~(2.88) of Ref.~\cite{R1},
establishing the relationship between this paper and
previous work on the two-body problem.

\section{Reduction of the $\delta$ functions in ${\cal P}_{12}$}

In this Appendix we derive Eq.~(\ref{Eq4.12}) for the
delta functions
\begin{equation}
2 E_{k_1} \delta^{(3)}( k'^o_1 - k_1)  2
E_{k_2} \delta^{(3)}(k'^o_2 -k_2) \label{B0}
\end{equation}
which appear in the
matrix element of the permutation operator,
Eq.~(\ref{Eq4.11}).

\subsection{Radial $\delta$ functions}

First find the vectors $k^{'o}_1$ and $k^{'o}_2$ shown in
Fig.~5a, and $k^{''o}_1$ and $k^{''o}_2$ shown in
Fig.~5b. The vectors $k^{'o}_1$ and $k^{''o}_2$ are
\begin{equation}
k^{'o}_1 =  \left(  \begin{array}{c} E_{q'}\\ 0 \\ 0 \\ -q'
\end{array} \right)  \, , \qquad k^{''o}_2 =
\left( \begin{array}{c} E_{q}\\ 0 \\ 0 \\ -q \end{array}
\right)  \, , \label{B00}
\end{equation}
where we anticipate that $\{k''\}\to\{k\}$ and hence use 
$|{\bf k}''_2|=q$ instead of $|{\bf k}''_2|=q''$. This agrees with
the notation in Eq.~(\ref{Eq4.11a}).  The vectors
$k^{'o}_2$ and $k^{''o}_1$ can be found by boosting the vectors
$\tilde{k}^{'o}_2$ and
$\tilde{k}^{''o}_1$ defined in the two-body rest frame:

\begin{equation}
\tilde{k}^{'o}_2 = \left( \begin{array}{c}
E_{\tilde p'}\\ \tilde{p}' \sin\tilde\theta' \\
0 \\ \tilde{p}' \cos\tilde\theta' \end{array}  \right)
\, , \qquad
\tilde{k}^{''o}_1 = \left( \begin{array}{c}
E_{\tilde p''} \\ \tilde{p}'' \sin\tilde\theta'' \\
0 \\ \tilde{p}'' \cos\tilde\theta''   \end{array}  \right)
\, . \label{B1}
\end{equation}
The boost in the $z$-direction for the configuration
shown in Fig.~5a is
\begin{equation}
Z_{q'} =
\left(  \begin{array}{cccc}
C'  & 0   &  0  &  S'  \\
0       &  1  &  0  &   0   \\
0       &  0  &  1  &   0   \\
S' & 0 & 0 & C'
\end{array}  \right)
\, , \label{B2}
\end{equation}
where
\begin{equation}
C'= \frac{M_t - E_{q'}}{W_{q'}}={\sqrt{W_{q'}^2+q'^2}\over W_{q'}}
\, ,\qquad  S' = \frac{q'}{W_{q'}} \,  .\label{AAA1}
\end{equation}
The boost for the configuration shown in Fig.~5b is
obtained from (\ref{B2}) by replacing $q'$ by $q$.  Hence
\begin{eqnarray}
k^{'o}_2 & = & Z_{q'} \tilde{k}^{'o}_2 =
\left(\begin{array}{c} C'E_{\tilde{p}'} + S'\tilde{p}'
\cos\tilde\theta' \\ \tilde{p}'\sin\tilde\theta' \\ 0 \\
S' E_{\tilde{p}'} + C'\tilde{p}' \cos\tilde\theta'
\end{array}  \right)
\,  \nonumber\\
k^{''o}_1 & = &
Z_{q} \tilde{k}^{''o}_1 =
\left(\begin{array}{c}
C E_{\tilde{p}''} + S \tilde{p}''  \cos\tilde\theta'' \\
\tilde{p}'' \sin\tilde\theta'' \\
0 \\ S E_{\tilde{p}''} +
C\tilde{p}'' \cos\tilde\theta''\end{array}  \right)
\, . \label{B2a}
\end{eqnarray}

Now the vectors which appear in the delta functions
(\ref{B0}) are ${\bf k}'^o_1$, ${\bf k}'^o_2$,
${\bf k}_1$, and ${\bf k}_2$.  The second
two of these are obtained by applying the rotation
$R_{V'}$  to ${\bf k}^{''o}_1$ and ${\bf k}^{''o}_2$, as
illustrated in Fig.~5c.  Since  the rotation $R_{V'}$
does not change the length of the three-vectors, the
condition $k^{'o}_2 = k_2$ imposed by one of the radial
delta functions becomes
\begin{eqnarray}
q  =&&  \left\{ \tilde{p}'^2 \sin^2\tilde\theta' +
S'^2 E^2_{\tilde{p}'}+C'^2 \tilde{p}'^2
\cos^2\tilde\theta'
 + 2 C'S' \tilde{p}' E_{\tilde{p}'}
\cos\tilde\theta' \right\}^{1/2} \nonumber\\
=&&\left\{
\left[C'E_{\tilde{p}'}+S'\tilde{p}'\cos\tilde\theta'
\right]^2 -m^2 \right\}^{1/2}  \, .
\end{eqnarray}
Requiring that $|\cos\tilde\theta'|<1$ gives the unique
solution
\begin{equation}
\cos\tilde\theta' = \cos\tilde\theta'_0 =
\frac{ E_{q} -C' E_{\tilde{p}'}}{S'\tilde{p}'} ={W_{q'}E_q - 
(M_t-E_{q'})E_{\tilde{p}'} \over q'\tilde{p}'} 
\, . \label{B3}
\end{equation}
The radial delta function can therefore be rewritten
$$
\delta(k^{'o}_2 - k_2) =
\delta(\cos\tilde\theta' -\cos\tilde\theta'_0)
\left|  \frac{d k_2}{d\cos\tilde\theta'}
\right|^{-1}_{\cos\tilde\theta' = \cos\tilde\theta'_0}
\, .
$$
The derivative is
$$
\left|\frac{d k_2}{d\cos\tilde\theta'}
\right|_{\cos\tilde\theta' = \cos\tilde\theta'_0}  =
\frac{S' E_{q}  \tilde{p}'}{q}
\, ,
$$
giving finally
\begin{equation}
\frac{\delta(k^{'o}_2-k_2)}{k^2_2}
=\frac{1}{S' q E_{q}\tilde{p}'}
\delta( \cos\tilde\theta' -  \cos\tilde\theta'_0)
= \frac{W_{q'}}{qq' E_{q}\tilde{p}'}
\delta( \cos\tilde\theta' -  \cos\tilde\theta'_0)\, ,
\label{B4}
\end{equation}
where $\cos\tilde\theta'_0$ was defined in Eq.~(\ref{B3}).

The result for the other radial delta function follows
immediately by interchanging primed and unprimed
variables.

\subsection{Angular $\delta$ functions}

Evaluation of the angular delta functions requires
explicit consideration of the rotation
$R_{V'}=R_{\alpha,\chi',\beta}$ which rotates the
vectors $k^{''o}_1$ and $k^{''o}_2$ into
$k_1$ and $k_2$ as discussed in Sec.~VA and
illustrated in Fig.~5c.  First we consider the delta
function
\begin{equation}
\delta^{(2)}(k^{'o}_2-k_2)= \delta(\cos\theta^{'o}_2
-\cos\theta_2)\delta(\phi^{'o}_2-\phi_2)\, ,
\label{B5}
\end{equation}
where $(\theta_2,\phi_2)$ and
$(\theta^{'o}_2,\phi^{'o}_2)$ are the polar and azimuthal
angles of ${\bf k}_2$ and ${\bf k}^{'o}_2$,
respectively. The three-vector ${\bf k}^{'o}_2$ was given
in Eq.~(\ref{B2a}), and ${\bf k}_2$ is
\begin{equation}
{\bf k}_2 = R_{\alpha,\chi',\beta} {\bf k}^{''o}_2
=R_{\alpha,\chi',\beta}
\left(\begin{array}{c} 0 \\ 0 \\ -q\end{array} \right)
=-q\left(  \begin{array}{c}\sin\chi' \cos\alpha \\
\sin\chi' \sin\alpha \\ \cos\chi'  \end{array} \right)
 \, .  \label{B6}
\end{equation}
Since $k_{2y}=k^{'o}_{2y}=0$ and $k_{2x}=k^{'o}_{2x}\ge
0$, the azimuthal part of Eq.~(\ref{B5}) becomes
\begin{equation}
\delta(\phi^{'o}_2-\phi_2)=\delta(\pi-\alpha) \, .
\label{B7}
\end{equation}
The polar part of the delta function is
\begin{equation}
\delta(\cos\theta^{'o}_2-\cos\theta_2)=\delta\left(
{1\over q}\left[S' E_{\tilde{p}'} + C'\tilde{p}'
\cos\tilde\theta'\right]+\cos\chi'\right) \, . \label{B8}
\end{equation}
If this delta function is used to eliminate the $\chi'$,
integration, one is left with an integration over
$\tilde{p}'$ with upper and lower limits depending on the
two external momenta and on the second integration
variable.  This makes numerical solutions of the
resulting equations awkward.  Because $\chi'$
is the angle between the momenta $\bf q$ and ${\bf q}'$,
and is therefore symmetric under interchange of initial
and final states, it is more convenient to retain
$\chi'$ as the independent variable and eliminate instead
the integration over $\tilde {p}'$.  The limits on the
$\chi'$ integration turn out to be independent of the
other variables, running over the expected range from 0 to
$\pi$.  Using Eq.~(\ref{B3}) to replace $\cos\tilde{\theta}'$, the
delta function becomes
\begin{equation}
\delta(\cos\theta^{'o}_2-\cos\theta_2)=\delta\left(
{ C'E_{q} - E_{\tilde{p}'}\over S'q}+\cos\chi\right) \, ,
\label{B9}
\end{equation}
with the solution
\begin{equation}
E_{\tilde{p}'}= E_{\tilde{p}'_0} = C'E_{q}+S'q\cos\chi=
{(M_t-E_{q'})E_q + qq'\cos\chi\over W_{q'}} \, , \label{B10}
\end{equation}
and, since the ${\bf k}^{'o}_2={\bf k}_2$ is now
satisfied, we have replaced $\chi'$ by $\chi$, as
discussed in Sec.~VA.  It is easy to show that
Eq.~(\ref{B9}) implies that  $E_{\tilde{p}'}\ge m$ for all
values of $\chi$, so that the delta function (\ref{B9})
places no additional restrictions on the $\chi$
integration.  Hence
\begin{equation}
\delta(\cos\theta^{'o}_2-\cos\theta_2)=
\delta (\tilde{p}' -\tilde{p}'_0)
\left|{d\cos\theta^{'o}_2 \over d\tilde{p}'}\right|^{-1}
_{\tilde{p}' =\tilde{p}'_0}
 = { S'qE_{\tilde{p}'_0}\over \tilde{p}'_0}\,
\delta (\tilde{p}' -\tilde{p}'_0)
\, .
\label{B11}
\end{equation}

The final angular delta function is
\begin{equation}
\delta^{(2)}(k^{'o}_1-k_1)= \delta(\cos\theta^{'o}_1
-\cos\theta_1)\delta(\phi^{'o}_1-\phi_1)\, ,
\label{B12}
\end{equation}
where $(\theta_1,\phi_1)$ and
$(\theta^{'o}_1,\phi^{'o}_1)$ are the polar and azimuthal
angles of ${\bf k}_1$ and ${\bf k}^{'o}_1$,
respectively.  The vector ${\bf k}^{'o}_1$ is given in
Eq.~(\ref{B00}); the vector ${\bf k}_1$ is
\begin{equation}
{\bf k}_1 = R_{\pi,\chi,\beta} {\bf k}^{''o}_1
=R_{\pi,\chi,\beta} \left(\begin{array}{c}
v_x \\ 0 \\ v_z
\end{array}\right)
=R_{\pi,\chi,0}\left(\begin{array}{c}
v_x\cos\beta \\ v_x\sin\beta \\ v_z \end{array} \right)
= -\left(\begin{array}{c}
v_x\cos\beta \cos\chi + v_z \sin\chi
\\ v_x\sin\beta \\ v_x\cos\beta\sin\chi-v_z\cos\chi
\end{array} \right)
 \, .  \label{B13}
\end{equation}
where
\begin{eqnarray}
v_x&&=\tilde{p}'' \sin\tilde\theta''\nonumber\\
v_z&&=SE_{\tilde{p}''} +C \tilde{p}''\cos\tilde\theta''
={CE_{q'}-E_{\tilde{p}''}\over S}\,  . \label{B14}
\end{eqnarray}
Setting ${\bf k}_1= {\bf k}^{'o}_1$ gives three equations
\begin{eqnarray}
0&&=v_x\cos\beta \cos\chi + v_z \sin\chi=q'\sin\theta_1\cos\phi_1
\nonumber\\
0&&=v_x\sin\beta=q'\sin\theta_1\sin\phi_1 \nonumber\\
q&&=v_x\cos\beta\sin\chi-v_z\cos\chi=-q'\cos\theta_1\, . \label{B15}
\end{eqnarray}

We will first use the left hand set of Eqs.~(\ref{B15}) to obtain the
values of $\beta$ and $\tilde{p}''$ which are fixed by the delta
functions (\ref{B12}).   Then we will use the right hand set to find
the Jacobian of the transformation from the variables
$\cos\theta_2,\phi_2$ to the variables $\tilde{p}'',\beta$.

The angle $\beta$ must be 0 or $\pi$.  Allowing for either possibility,
the first and third of Eqs.~(\ref{B15}) give
\begin{equation}
q'\cos\chi=-v_z\, .\label{B16}
\end{equation}
Substituting this result back into the third equation gives
\begin{equation}
q'\sin^2\chi=v_x\cos\beta\sin\chi=\tilde{p}''
\sin\tilde\theta''\cos\beta\sin\chi\, .
\end{equation}
Since the $\sin$ of all angles under consideration is positive,
this equation shows that $\cos\beta$ is also and that
therefore $\beta=0$.  Finally, from Eq.~(\ref{B16}) we obtain
\begin{equation}
\cos\chi+{CE_{q'}-E_{\tilde{p}''}\over q'S}=0
\end{equation}
which is the condition (\ref{B9}), with $q$ and $q'$ interchanged,
showing that $E_{\tilde{p}''}$ satisfies (\ref{B10}) (with  $q$ and
$q'$ interchanged).  We have shown that the second angular delta
function is
\begin{equation}
\delta^{(2)}(k^{'o}_1-k_1)=
\delta(\beta)\delta(\tilde{p}''-\tilde{p}_0)J\, , \label{B17}
\end{equation}
where $\tilde{p}_0$ is $\tilde{p}'_0$ with $q$ and $q'$ interchanged
and $J$ is the jacobian of the transformation from the variables
$\cos\theta_2,\phi_2$ to the variables $\tilde{p}'',\beta$.

We return to the right hand set of Eqs.~(\ref{B15}) to calculate
this jacobian.  Unlike the previous cases it is necessary to calculate
a full jacobian because the variables are all coupled unless we go to
the limit $\sin\theta_1=0$, which gives singular results.  Postponing
this limit until the end, we first eliminate $v_z$ from the
Eqs.~(\ref{B15}) and obtain
\begin{eqnarray}
v_x\sin\beta&&=q'\sin\theta_1\sin\phi_1 \nonumber\\
v_x\cos\beta&&=q'[\sin\theta_1\cos\phi_1\cos\chi-\cos\theta_1\sin\chi]
\,  . \label{B18}
\end{eqnarray}
Differentiating both of these equations with respect to $\cos\theta_1$
and $\phi_1$, computing the jacobian, and {\it then\/} taking the
limits $\theta_1=\pi, \phi_1=0,$ and $\beta=0$ gives
\begin{equation}
\left|{\partial v_x\over\partial \tilde{p}''}\right|\;J= \left|\matrix{
{\partial\beta\over\partial\phi_1} & {\partial
v_x\over\partial\phi_1}\cr  {\partial\beta\over\partial\cos\theta_1}
& {\partial v_x\over\partial\cos\theta_1}}\right|=
{q'^2\over v_x}\cos\chi \, .\label{B19}
\end{equation}
The radial delta functions have fixed $\cos\tilde{\theta}''$ in terms
of $\tilde{p}''\to\tilde{p}_0 $ [Eq.~(\ref{B3}) with $q'\to q$ and
$\tilde{p}'\to\tilde{p}''\to \tilde{p}_0$], and using this relation we
find that
\begin{equation}
{\partial v_x\over \partial\tilde{p}''}=-{q' W_{q}\cos\chi\over
qE_{\tilde{p}_0}\sin\tilde{\theta}''}\, ,
\end{equation}
and hence the jacobian is
\begin{equation}
J= {qq'E_{\tilde{p}_0}\over\tilde{p}_0\,W_{q}}={Sq'
E_{\tilde{p}_0}\over\tilde{p}_0}\, . \label{B20}
\end{equation}

Combining Eq.~(\ref{B4}), its companion, Eqs.~(\ref{B11}), 
(\ref{B17}) and (\ref{B20}), and anticipating the fact that the delta
functions fix the rotation so that $\tilde{p}''\to\tilde{p}$ and
$\tilde{\theta}''\to\tilde{\theta}$, we obtain our final result
\begin{eqnarray}
2 E_{k_1} \delta^{(3)}( k'^o_1 - k_1)  2
E_{k_2} \delta^{(3)}(k'^o_2 -k_2)  & = &
4 E_{\tilde{p}_0} E_{\tilde{p}'_0}
\delta(\alpha-\pi)
\delta(\beta)
\frac{\delta(\tilde{p}' - \tilde{p}'_0)}{\tilde{p}'^2_0}
\frac{\delta(\tilde{p} - \tilde{p}_0)}{\tilde{p}^2_0}
\nonumber \\
& & \times
\delta(\cos\tilde\theta' - \cos\tilde\theta'_0)
\delta(\cos\tilde\theta - \cos\tilde\theta_0)
\,.
\end{eqnarray}

\section{Wigner Rotations}

In this Appendix the Wigner rotation angle for the standard boost
(\ref{Eq5.1}) that occurs in the matrix elements of the permutation
operator is derived.

Consider a spin $1/2$ particle with mass $m$, helicity
$\lambda$, and three-momentum
$\tilde{{\bf p}}$ which lies in the $xz$ plane (with the
$x$-component positive, by convention). Under the boost $Z_q$ in
the $+z$ direction the state transforms like
\begin{equation}
Z_q |\tilde{p},\lambda \rangle = R(Q) |p, \lambda \rangle =
\sum_{\nu} |p, \nu \rangle 
d^{(1/2)}_{\nu,\lambda}(\beta) \, ,
\end{equation}
where $Z_q$ will by used to denote {\it both\/} the boost (\ref{B2}) in
four-dimensional space-time, {\it and\/} its representation on the
space of states.  Hence $p=Z_q\;\tilde{p}$.  In agreement with the
notation used in Sec.~IV, the magnitude $|{\bf p}|=q'$, and angle
between ${\bf p}$ and the $+z$ axis is 
$\theta=\pi-\chi$.  We will show that $Q$ is a rotation about the $y$
axis, and find the rotation angle, $\beta$, in terms of $q$, $q'$, and
$\chi$.

As previously described, the helicity states are constructed from rest
states by first boosting in the $+z$ direction, and then rotating
through the proper angle.  For the states $|p,\lambda\rangle$ and
$|\tilde{p},\lambda\rangle$, this construction gives
\begin{eqnarray}
|\tilde{p},\lambda \rangle =&& e^{-iJ_y\tilde{\theta}}\,L_{\tilde{p}}
\,|\hat m,\lambda\rangle\nonumber\\
|{p},\lambda \rangle =&& e^{-iJ_y{\theta}}\,L_{q'}
\,|\hat m,\lambda\rangle\, ,
\end{eqnarray}
where $L_k$ was defined in Eq.~(\ref{Eq5.2}) and $\hat m= (m, {\bf 0})$
is the four-momentum vector of a particle of mass $m$ at rest. Hence the
Wigner rotation operator is given by
\begin{equation}
R(Q)=\left(e^{-iJ_y{\theta}}\,L_{q'}\right)^{-1} \; Z_q\;
e^{-iJ_y\tilde{\theta}}\,L_{\tilde{p}} \, .
\end{equation}
Assuming that $R(Q)$ is a pure rotation about the $y$-axis, the
equation reduces to the following set of equations involving $\beta$
\begin{equation}
e^{-iJ_y{\theta}}\,L_{q'}\,e^{-iJ_y{\beta}} =
Z_q\,e^{-iJ_y\tilde{\theta}}\,L_{\tilde{p}} \label{AA1}
\end{equation}

We will solve these equations using the Dirac representation for the
operators.  In this representation the pure boosts in the $z$-direction
are 
\begin{equation}
L_{q'} = e^{\alpha_z {\eta_{q'} /2}} =c'+s'\alpha_z\, ,\label{AA3}
\end{equation}
where $\tanh{\eta_{q'}}=q'/E_{q'}$, $c'=\cosh(\eta_{q'}/2)$, and
$s'=\sinh(\eta_{q'}/2)$ [these relations were previously defined in
Eq.~(\ref{Eq5.21})]. The boost $Z_q$ has a structure similar to
(\ref{AA3}) but with $\eta_{q'}\to n_q$ where 
\begin{equation}
C={M_t-E_q\over W_q}=\cosh n_q \qquad S={q\over W_q}=\sinh n_q
\end{equation}  
as in Eq.~(\ref{AAA1}), but with $q'\to q$.  To simplify notation in
what follows, we denote the hyperbolic functions of $n_q/2$ by
\begin{equation}
c_n=\cosh (n_q / 2) \qquad s_n=\sinh(n_q / 2)\, .
\end{equation}
The pure rotations about the
$y$-axis are 
\begin{equation}
e^{-iJ_y{\theta}}=e^{-i{\theta}\gamma^5\alpha_y/2}=
\cos(\theta/2)-i\sin(\theta/2)\gamma^5\alpha_y \, .
\end{equation}
Hence Eq.~(\ref{AA1}) becomes
\begin{eqnarray}
\left[\cos (\theta/ 2) - i \gamma^5\alpha_y \sin ({\theta/ 2})
\right]&&
\left[
c' + s'\alpha_z \right]
\left[
\cos ({\beta / 2}) - i\gamma^5\alpha_y \sin ({\beta / 2})
\right]\nonumber\\
&&\qquad=
\left[c_n + \alpha_z\,s_n \right]
\left[
\cos (\tilde{\theta} / 2) - i \gamma^5\alpha_y \sin ({\tilde{\theta}
/ 2})\right]
\left[ c_p + \alpha_z s_p\right]\, .\label{AA2}
\end{eqnarray}
Using $\{\gamma^5\alpha_y,\alpha_z\}=0$, Eq.~(\ref{AA2}) becomes
\begin{eqnarray}
&&\left\{\cos [(\theta+\beta)/2]-i\gamma^5\alpha_y \sin [(\theta+\beta)/
2]\right\}\;c'+\left\{\cos[(\theta-\beta)/2] -i\gamma^5\alpha_y 
\sin [(\theta-\beta)/2]\right\}\;s'\alpha_z\nonumber\\
&&\qquad=\left\{c_nc_p+s_ns_p +\alpha_z(s_nc_p+c_ns_p)\right\}\;
\cos (\tilde{\theta} / 2)\nonumber\\
&&\qquad\quad -i\left\{c_nc_p-s_ns_p + \alpha_z(s_nc_p-c_ns_p)\right\}
\gamma^5\alpha_y  \sin (\tilde{\theta} / 2) \, .
\end{eqnarray}
Equating the coefficients of the independent operators on each
side of this equation gives four coupled equations   
\begin{eqnarray}
c'\,\cos [{ ({\theta + \beta}) / 2}] 
& = & (c_nc_p+s_ns_p) \,\cos (\tilde{\theta} / 2)\nonumber\\
c'\,\sin [(\theta+\beta)/2]  
& = & (c_nc_p-s_ns_p)\,\sin (\tilde{\theta} / 2)\nonumber\\
s'\,\sin [(\theta-\beta)/2]
& = & (c_ns_p-s_nc_p)\,\sin (\tilde{\theta} / 2)\nonumber\\
s'\,\cos[(\theta-\beta)/2]
& = & (c_ns_p+s_nc_p)\,\cos (\tilde{\theta} / 2) \,.
\end{eqnarray}
These are not all independent; squaring each of these and then adding
the first two and subtracting the second two gives an identity (1=1).
Hence only three are independent, and given the quantities
$q$, $q'$, and $\chi=\pi-\theta$, these three independent equations can
be regarded as equations for the unknown quantities
$\tilde{\theta}$, $\tilde{p}$, and $\beta$.

We are only interested in an equation for $\beta$.  This is obtained by
multiplying the first equation by the last equation, and adding it to
the product of the second equation and the third.  The result is:
\begin{equation}
\cos \beta = {C \sinh\eta_p +
S\cosh\eta_p\cos\tilde{\theta} \over \sinh\eta'}\,.
\end{equation}
To eliminate $\cos\tilde\theta$, we find an equation for it by summing
the  squares of each of the equations.  This gives
\begin{eqnarray}
\cos \tilde\theta =&& {\cosh\eta'  -C\cosh\eta_p 
\over S\sinh\eta_p }\nonumber\\
=&& {E_{q'}-C E_{\tilde{p}}\over S\tilde{p}}\,.
\end{eqnarray}
Note that this is identical to Eq.~(\ref{B3}) [with the primed and
unprimed variables exchanged], showing that the calculations are
consistent.  Substituting for $\cos\tilde\theta$ gives
\begin{eqnarray}
\cos \beta =&& {\cosh\eta_p\cosh\eta'-C \over \sinh\eta_p\sinh\eta'}
\nonumber\\
=&&{W_qE_{\tilde{p}}E_{q'}-m^2(M_t-E_q)\over \tilde{p}q'W_q}\nonumber\\
=&&\frac{
q' (M_t-E_q)  + q E_{q'} \cos\chi}
{\sqrt{ q'^2 W_q^2 + q^2 E_{q'}^2
      + 2 q q' E_{q'}( M_t-E_q ) \cos\chi
      + \left( q q' \cos\chi \right)^2 }} \, . \label{Cwr}
\end{eqnarray}
This is the formula for $\cos\left[\beta(q,q',\chi)\right]$.

\end{document}